\documentclass[
pteplogo
]{ptephy_v1}

\usepackage{amsmath} 
\usepackage{amsthm} 

\newcommand{\eqb}{\begin{equation}}
\newcommand{\eqe}{\end{equation}}
\newcommand{\eqbnon}{\begin{equation*}}
\newcommand{\eqenon}{\end{equation*}}

\newcommand{\eqab}{\begin{eqnarray}}
\newcommand{\eqae}{\end{eqnarray}}
\newcommand{\eqabnon}{\begin{eqnarray*}}
\newcommand{\eqaenon}{\end{eqnarray*}}

\newcommand{\seqb}{\begin{subequations}}
\newcommand{\seqe}{\end{subequations}}

\newcommand{\defeq}{:=}
\newcommand{\defeqr}{=:}
\newcommand{\us}[1]{^{\rm #1}} 
\newcommand{\ls}[1]{_{\rm #1}} 

\newcommand{\pd}[2]{\dfrac{\partial #1}{\partial #2}}
\newcommand{\od}[2]{\dfrac{{\rm d} #1}{{\rm d} #2}}
\newcommand{\diff}[1]{{\rm d}#1}

\newcommand{\delay}{\Delta t\ls{obs}}
\newcommand{\ratio}{\mathcal{R}\ls{obs}}
\newcommand{\tdf}{\mathcal{D}}
\newcommand{\atc}{\mathcal{C}}

\begin{document}

\title{
How to Measure Black Hole's Mass, Spin and Direction of Spin Axis in Kerr Lens Effect 1: 
test case with simple source emission near BH
}


\author{
Hiromi Saida
}
\affil{
Department of Physics, Daido University, Minami-ku 457-8530 Nagoya, Japan
\email{saida@daido-it.ac.jp}
}



\begin{abstract}
We propose a theoretical principle to measure the mass, spin and direction of spin axis of Kerr black holes (BHs) through observing 2 quantities of the spinning strong gravitational lens effect of BHs. 
Those observable quantities are generated by 2 light rays emitted at the same time by a source near the BH: the primary and secondary rays that reach a distant observer, respectively, the earliest and secondary temporally. 
The time delay between detection times and the ratio of observed specific fluxes of those rays are the observable quantities. 
Rigorously, our proposal is applicable to a single burst-like (short duration) isotropic emission by the source. 
An extension of our principle to cases of complicated emissions may be constructed by summing up appropriately the result of this paper, which will be treated in future works. 
\end{abstract}

\subjectindex{E31(Black holes)}

\maketitle

\section{Introduction}
\label{sec:intro}

What is the meaning of \emph{direct measurement} of BHs? 
To consider this question, we emphasize 2 theoretical facts of BHs~\cite{ref:frolov+1.1998} in the framework of general relativity: 
a BH is a very strong gravitating spacetime region that traps anything inside it, and a BH is completely characterized by 2 parameters due to the uniqueness theorem: its mass and spin angular-momentum. 
(We expect that BHs have no electric charge in astrophysical situations.) 
According to these theoretical facts, let us define direct measurement of BHs as follows: \emph{It is to measure the mass and spin angular-momentum of BHs by a direct observation of some strong gravitational (general relativistic) effect produced by the BH.}

The first example of direct BH measurement by this definition has been succeeded by the advanced-LIGO that detected the gravitational wave of a BH-BH binary coalescence~\cite{ref:abbott+many.2016}.
Gravitational wave astronomy for the direct BH measurement seems to be ready to start in the near future. 
On the other hand, no example of direct BH measurement by electro-magnetic wave observations has been achieved yet. 
Developments of electro-magnetic observations not only in technology but also in theory are expected to make collaborations of gravitational waves and electro-magnetic waves in the \emph{direct} observational study of BHs.

We consider the observation of light rays (massless particles) coming from a source near a single BH (not a BH binary), and propose a theoretical principle of direct BH measurement. 
Our principle makes use of some observable quantities of the strong gravitational lens effect produced by the mass and spin of a single Kerr BH (Kerr BH lens effect).

Concerning the Kerr BH lens effect for performing direct BH measurement, many ideas have been investigated so far. 
For example, Cunningham and Bardeen~\cite{ref:cunningham+1.1973} considered a star on a circular orbit on the equatorial plane of an extreme Kerr BH, and calculated the brightness of primary and secondary images detected by a distant observer. 
Holz and Wheeler~\cite{ref:holz+1.2002} proposed the retro-macho, which is the multi-image of a star (the Sun in their original paper) located between the observer and the lensing BH. 
Bozza and Mancini~\cite{ref:bozza+1.2004} have applied the idea of retro-macho to the star S2 orbiting the massive BH candidate at the center of our galaxy. 
Fukumura et al.~\cite{ref:fukumura+1.2008,ref:fukumura+2.2009} proposed the BH echoes in order to discuss the so-called quasi-periodic oscillations, in which they considered X-ray sources near a Kerr BH and calculated the observable spectrum produced by multiple rays emitted randomly by the sources. 
Ho\'{r}ak and Karas~\cite{ref:horak+1.2006} studied the effects of time delay of multiple rays on the polarization magnitude. 
James et al.~\cite{ref:james+3.2015} considered an observer near a BH, and calculated the observed images of the BH and distant stars seen by the observer. 
Apart from these examples, the famous observable quantities of the Kerr BH lens effect may be the broadening of iron line emission by an accretion disk around BH (see Kojima~\cite{ref:kojima.1991}, Fanton et al.~\cite{ref:fanton+3.1997}, and references therein) and the BH shadow (see Luminet~\cite{ref:luminet.1979}, Takahashi~\cite{ref:takahashi.2004}, and references therein).

Although these interesting ideas have not succeeded yet in observations, improving these ideas is necessary in order to perform direct BH measurement by electro-magnetic observations. 
In this paper, we consider a similar case with Cunningham and Bardeen~\cite{ref:cunningham+1.1973}, but do not restrict our discussions to the extreme Kerr BH and circular orbit on equatorial plane. 
Our theoretical proposal for direct BH measurement is made under assumptions: (i) the environment around the BH is transparent for photons, at least in the frequency band of observation, (ii) a single source is moving on any orbit around the Kerr BH, (iii) the source emits a single burst-like (short duration) isotropic emission, and (iv) a distant observer detects the primary and secondary rays that reach the observer, respectively, the earliest and secondary temporally. 
Conditions (i) and (iv) are the same conditions as Cunningham and Bardeen~\cite{ref:cunningham+1.1973}, but conditions (ii) and (iii) are not. 
Then, if our 4 conditions are applied to a BH, our theoretical principle can let us measure not only the BH's mass and spin but also the direction angle of the spin axis seen by the observer.

In this paper, we investigate 2 observable quantities: the delay between detection times of the primary and secondary rays, and the ratio of observed specific fluxes of the 2 rays. 
By performing some numerical estimations of these observable quantities, we discuss the potential detectability of them by present or near future telescope capability. 
Thorough numerical studies on the feasibility of our principle and on the behavior of our observable quantities will be reported in another paper, since the numerical calculations will need a powerful computer.

Note that, in some of the existing ideas of Kerr BH lens effects, e.g. the BH echo~\cite{ref:fukumura+2.2009} and broadening of iron line~\cite{ref:kojima.1991,ref:fanton+3.1997}, the light source radiates photons in some temporally and spatially successive, random, or continuous fashion. 
Such complicated emissions will be constructed by summing appropriately some burst-like emissions. 
Therefore, this paper concentrates on the case with single burst-like emission as the preparation for cases with complicated source emissions. 
The Kerr BH lens effects for such complicated source emissions are the issue of our next papers.

In Sect.\ref{sec:principle}, our target phenomenon and observable quantities are specified, and the theoretical principle of direct BH measurement is proposed. 
In Sect.\ref{sec:formalism}, the formulas and our numerical procedure for calculating the observable quantities are explained so that readers can check our proposal quantitatively. 
However, readers who do not need the details of formulas can skip Sect.\ref{sec:formalism}. 
Then, Sect.\ref{sec:result} shows some numerical estimations of our observable quantities to discuss the potential detectability of them. 
Finally, Sect.\ref{sec:sd} is for a summary and discussions. 
We use the units $c=1$ and $G=1$, and measure all quantities in the dimension of length. 
However, when MKS units are convenient for a line of thought, we use the units and show $c$ and $G$ explicitly.

\section{Principle of measuring BH's mass, spin and spin axis' direction angle}
\label{sec:principle}

\subsection{Basic assumptions}
\label{sec:principle.assumption}

Our target phenomenon is a single burst-like (short duration) isotropic emission by a source that is much smaller than its host BH. 
This is modeled as a point-like source that moves near the BH and emits light rays in all directions within a duration much shorter than a typical dynamical time scale of the system composed of the source and the BH (e.g. the Keplerian time scale of the source around BH). 
Further, we assume that the environment around the BH is transparent, at least in the frequency band of observation. 
Note that the details of the intrinsic structure of the source is outside the scope of this paper, and we assume such a simple emission occurs in the transparent environment around BH. 
We focus on how the gravitational lens effect of the Kerr BH appears in observable quantities of light rays that are emitted by the source and received by a rest observer distant from the BH.

On these assumptions, let us note that an example of a transparent environment has been recognized for the massive BH candidate at the center of our galaxy, Sagittarius A$^\ast$ (Sgr A$^\ast$)~\cite{ref:bower+5.2004,ref:doeleman+27.2008,ref:shen+4.2005}. 
Radio observational data of Sgr A$^\ast$ for a rather wide band range indicate the disappearance of plasma scattering effects at high frequency ($> 230$ GHz) radio waves, which implies a transparent environment around the BH for such high frequency radio photons~\cite{ref:bower+5.2004,ref:doeleman+27.2008,ref:shen+4.2005}. 
Further, it is also reported that intermediate mass BH candidates seem to exist in the central region of our galaxy~\cite{ref:oka+3.2016}. 
Such an intermediate mass BH candidate does not form a binary with some other body, but it is recognized by a significant (large acceleration) motion of gas cloud around a dark compact region, which is consistent with assuming an intermediate mass BH in the dark compact region~\cite{ref:oka+3.2016}. 
We may expect that such an intermediate mass BH, if it exists, wears a transparent and inactive environment.

Here note that the estimated mass of Sgr A$^\ast$ is about $4\times 10^6 M_{\odot}$~\cite{ref:boehle+13.2016}, and that of the intermediate mass BH is about $10^5 M_{\odot}$~\cite{ref:oka+3.2016}, where $M_{\odot} \simeq 2\times 10^{30}$ kg is the mass of the Sun. 
Therefore, gas clouds and celestial bodies whose mass is about $M_{\odot}$ are much smaller than those BHs. 
Then, let us consider the case that a gas cloud or a dark celestial body of stellar mass falls toward the BH and emits a short duration emission in all directions (due, e.g., to some shock formation or disruption event).
Such a case may be effectively described by our assumptions or by some appropriate summation of our simple source emission. 
Anyway, the detail of the source's structure is outside the scope of this paper and will be properly treated in another paper. 
Our interest, in this paper, is in the theoretical properties of the Kerr BH lens effect under our assumptions described in the first paragraph.

Let us make a comment on the terminology. 
The term ``strong gravitational lens effect'' is already used in, e.g., the survey of galaxy distribution, dark matter distribution, and so on. 
The meaning of this term in those research fields is the lens effect of a cluster of masses (not necessarily BHs) that produces the multi-image of a source (e.g. galaxy) behind the cluster. 
This usage of ``strong gravitational lens effect'' includes the effect of mass, but not the effect of a BH's spin. 
However, we are interested in the lens effect of a spinning BH, which shall include the effect not only of the mass but also of the spin of the BH. 
Therefore, we use the term \emph{Kerr BH lens effect} (or \emph{Kerr lens effect}) for the lens effect produced by the mass and spin of a Kerr BH.

\subsection{Kerr spacetime, notation in this paper}
\label{sec:principle.kerr}

To describe Kerr spacetime, we use Boyer-Lindquist (BL) coordinates $(t,r,\theta,\varphi)$ throughout this paper. 
The metric components $g_{\mu\nu}$ are read from the line element of spacetime,
\seqb
\label{eq:principle.metric}
\eqb
\label{eq:principle.metric-a}
 \diff{s^2}
 \,=\, g_{\mu\nu}\,\diff{x^\mu}\,\diff{x^\nu}
 \,=\,
  - \dfrac{\Sigma \Delta}{Z} \,\diff{t}^2
  + \dfrac{Z}{\Sigma} \sin^2\theta \,\Bigl( \omega \diff{t} - \diff{\varphi} \Bigr)^2
  + \dfrac{\Sigma}{\Delta} \,\diff{r}^2 + \Sigma \,\diff{\theta}^2 \,,
\eqe
where
\eqb
\label{eq:principle.metric-b}
\begin{aligned}
 \omega(r,\theta)
 &= \dfrac{2 M a r}{Z(r,\theta)}
 &,&
 &Z(r,\theta)
 &= (r^2+a^2)\,\Sigma(r,\theta) + 2Mra^2\sin^2\theta
\\
 \Delta(r)
 &= r^2 + a^2 - 2Mr
 &,&
 &\Sigma(r,\theta)
 &= r^2 + a^2 \cos^2\theta
\end{aligned}
\eqe
\seqe
and the time coordinate $t$, the BH's mass $M$ and spin parameter $a \defeq J/M$ (where $J$ is the spin angular-momentum) are measured in the dimension of length. 
Instead of $a$, we use the dimensionless spin parameter $\chi \defeq a/M$, when it is convenient for the line of thought. 
In some figures shown later, we project 4-dimensional null geodesics onto 3-dimensional screen spanned by
\eqb
\label{eq:principle.xyz}
 ( x , y , z ) \defeq
 ( r\sin\theta\cos\varphi , r\sin\theta\sin\varphi , r\cos\theta ) \,.
\eqe
In this 3D screen, the $z$-axis corresponds to the axis of the BH's spin.

The radius of the BH horizon is 
$r\ls{BH} = M \bigl[\, 1 + \sqrt{1-\chi^2} \,\bigr]$, 
and the ergosurface is 
$r\ls{erg}(\theta) = M \bigl[\, 1 + \sqrt{1-\chi^2\cos^2\theta}\,\bigr]$. 
The BH's spin is a rigid rotation, due to the rigidity theorem, with angular velocity
$\omega\ls{BH}
 = \omega(r\ls{BH},\pi/2)
 = \chi/(2 r\ls{BH})$. 
Without loss of generality, we can set $0 \le \chi < 1$, where the upper bound is due to the existence of the BH horizon, and the non-negativity of $\chi$ corresponds to the direction of BH spin given by $\omega\ls{BH} \ge 0$. 
We do not need to consider the opposite direction of spin because of the axisymmetry of Kerr spacetime.

The \emph{zero-angular-momentum-observer} (ZAMO) is defined as the one whose 4-velocity $u\ls{zamo}^\mu$ is perpendicular to the spacelike hypersurface at $t =$ constant, 
\eqb
\label{eq:principle.uzamo}
 u\ls{zamo}^\mu =
 \Bigl( \sqrt{\dfrac{Z}{\Sigma \Delta}} \,,\, 0 \,,\, 0 \,,\,
        \omega\sqrt{\dfrac{Z}{\Sigma \Delta}} \Bigr) \,.
\eqe
Although ZAMO has, by definition, no angular-momentum with respect to the spacelike hypersurface at $t =$ constant, its 4-velocity, however, has a $\varphi$-component. 
This is understood as the effect of the rotation of spacetime itself, which is called the \emph{frame dragging effect} of Kerr BH. 
The angular velocity of ZAMO measured in BL coordinates, $u\ls{zamo}^\varphi/u\ls{zamo}^t = \omega$, is regarded as the angular velocity of the frame dragging effect measured in BL coordinates.

The timelike and spacelike Killing vectors are, respectively in BL coordinates, 
$\xi\ls{(t)}^\mu = (1,0,0,0)$ and 
$\xi\ls{(\varphi)}^\mu = (0,0,0,1)$. 
The metric $g_{\mu\nu}$ satisfies the Killing equations, $\mathcal{L}_{\xi\ls{(t)}}g_{\mu\nu} = 0 \,,\, \mathcal{L}_{\xi\ls{(\varphi)}}g_{\mu\nu} = 0$, where $\mathcal{L}$ denotes the Lie derivative. 
And, Kerr spacetime is of Petrov type~D, which possesses 2 principal null vectors (a kind of eigenvector of Weyl curvature tensor), $m^\mu$ and $n^\mu$. 
These produce a Killing tensor that is, in BL coordinates,  
$\Xi_{\mu\nu} = (m_\mu n_\nu + m_\nu n_\mu)\Sigma + r^2 g_{\mu\nu}$, 
where 
$m_\mu = ( 1 , -\Sigma/\Delta , 0 , -a\sin^2\theta )$ and 
$n_\nu = ( \Delta/(2\Sigma) , 1/2 , 0 , [\Delta/(2\Sigma)] a \sin^2\theta )$. 
The Killing tensor equation $\Xi_{(\mu\nu ; \alpha)}=0$ holds, where the semicolon ``\,;\,'' in the subscript denotes the covariant derivative with respect to $g_{\mu\nu}$. 
These Killing fields, $\xi\ls{(t)}$, $\xi\ls{(\varphi)}$, and $\Xi$, express some symmetries of Kerr spacetime that produce some constants of motion of geodesics, and are used in Sect.\ref{sec:formalism} to derive the formulas of observable quantities.

\subsection{Observable quantities}
\label{sec:principle.observable}

\begin{figure}[t]
 \begin{center}
 \includegraphics[scale=0.50]{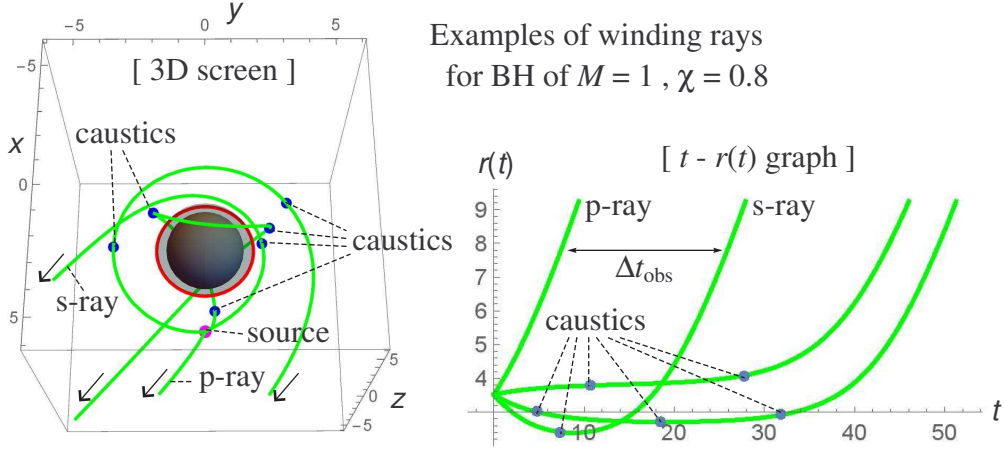}
 \end{center}
\caption{Examples of winding rays (green curves) for a BH of $(M,\chi) = (1\,,\,0.8)$. 
All rays shown here reach the same observer at $(r\ls{obs}\,,\theta\ls{obs}\,,\varphi\ls{obs}) = (100\, r\ls{BH}\,,15\pi/31\,,7\pi/30)$. 
Left panel is the 3D screen spanned by the variables~\eqref{eq:principle.xyz}. 
The ergoregion is colored gray, the equator of the ergosurface is the red circle, and the BH horizon is the black sphere. 
The source is at $(r\ls{emi}\,,\theta\ls{emi}\,,\varphi\ls{emi}) = (2.2\, r\ls{BH}\,,\pi/2\,,0)$. 
The right-hand panel is the $t$-$r(t)$ graph, from which the time delay $\delay$ is read. 
In calculating the flux ratio $\ratio$, the caustics that each ray passes through are found as by-products (see Sect.\ref{sec:formalism}).
}
\label{fig:rays.example}
\end{figure}

By the Kerr lens effect, the light ray winds around the BH before reaching the distant rest observer. 
An arbitrary number of windings around the BH are possible by tuning the values of the impact parameters of the light ray. 
In this paper, we focus on the 2 light rays that reach the distant observer earliest and secondary temporally. 
We call these rays, respectively, the \emph{primary ray (p-ray)} and \emph{secondary ray (s-ray)}. 
Some numerical examples of the orbits of these rays (solutions of null geodesic equations) are shown in Fig.\ref{fig:rays.example}, which depicts not only the p-ray and the s-ray but also other 2 rays of higher winding number that reach the same rest observer. 
(The detail is explained in Sect.\ref{sec:formalism}.)

As explained in Sect.\ref{sec:intro}, the observable quantities that we focus on are the delay $\delay$ between the detection time of the p-ray $t\ls{obs(p)}$ and that of the s-ray $t\ls{obs(s)}$, and the ratio $\ratio$ of the observed specific flux of the s-ray $F\ls{obs(s)}$ to that of the p-ray $F\ls{obs(p)}$:
\eqb
\label{eq:principle.observable}
 \delay \defeq t\ls{obs (s)} - t\ls{obs (p)} \quad,\quad
 \ratio \defeq \dfrac{F\ls{obs (s)}}{F\ls{obs (p)}} \,.
\eqe
The dimensions of $F\ls{obs}$ are W/(m$^2$\,Hz) in MKS units, where 1/m$^2$ denotes \emph{per unit spatial area perpendicular to the observer} and 1/Hz denotes \emph{per unit observed frequency} which is the meaning of ``specific''. 
Under the assumptions introduced in Sect.\ref{sec:principle.assumption}, the time delay $\delay$ and specific flux ratio $\ratio$ are regarded as functions of some parameters.

The time delay $\delay$ is regarded as a function of following parameters,
\eqb
\label{eq:principle.parameter-delay}
\begin{array}{rcl}
 M \,,\, \chi
 &:&\text{Parameters characterizing Kerr spacetime}
\\
 (\theta\ls{obs}\,,\,\varphi\ls{obs})
 &:&\text{Angular coordinates of the distant observer}
\\
 x^\mu\ls{emi} = (0\,,\,r\ls{emi}\,,\,\theta\ls{emi}\,,\,0)
 &:&\text{Spacetime position of the source at emission} \,,
\end{array}
\eqe
where the emission time is set to 0 ($t\ls{emi}=0$) and the distant limit of the observer ($r\ls{obs}\to\infty$) is supposed as mentioned in Sect.\ref{sec:principle.assumption}. 
The parameters $(\theta\ls{obs},\varphi\ls{obs})$ and $x^\mu\ls{emi}$ determine the relative location among the BH, source and distant observer.

Note that, by the axisymmetry of Kerr spacetime, we can rotate the BL coordinates in the $\varphi$-direction so as to make the new azimuthal angles of the observer and source $\varphi'\ls{obs}=0$ and $\varphi'\ls{emi} = -\varphi\ls{obs}$. 
However, for the convenience of our numerical calculation, we fix the source's azimuthal coordinate at $\varphi\ls{emi} = 0$ and let the observer's azimuthal coordinate $\varphi\ls{obs}$ be a free parameter. 
Also, it is worth emphasizing that \emph{the zenithal angle $\theta\ls{obs}$ is the direction angle of the BH's spin axis seen from the observer (the angle between the line of sight and the spin axis)}.

Once the relative location is fixed, the null geodesic equation can be solved and the orbits of the p-ray and the s-ray are specified. 
Then, the detection times $t\ls{obs (p)}$ and $t\ls{obs (s)}$ are determined, and the time delay $\delay$ can be read from the numerical result as shown in the right-hand panel of Fig.\ref{fig:rays.example}. 
This $\delay$ may be roughly estimated as $\delay \sim \alpha 2 \pi r\ls{emi}$, where $0 \lesssim \alpha \lesssim 1$, and $\alpha \sim 1$ for $\varphi\ls{obs} \sim 0$ (the emission event is in front of the BH seen from the observer) and $\alpha \sim 0$ for $\varphi\ls{obs} \sim \pi$ (the emission event is behind the BH seen from the observer). 
For the example in Fig.\ref{fig:rays.example}, $(M,\chi,r\ls{emi}) = (1\,,\,0.8\,,\,2.2\,r\ls{BH})$, this estimation formula gives $\delay \sim 4.4 \pi M (1+\sqrt{1-\chi^2}) \simeq 22$ which is roughly consistent with the value of $\delay$ read from the right-hand panel of Fig.\ref{fig:rays.example}.

Next, the observed specific fluxes of the p-ray and the s-ray, $F\ls{obs (p)}$ and $F\ls{obs (s)}$, are regarded as functions of the parameters in list \eqref{eq:principle.parameter-delay} and
\eqb
\label{eq:principle.parameter-ratio}
\begin{array}{rcl}
 u^\mu\ls{emi}
 &:&\text{4-velocity of the source at emission}
\\
 \nu\ls{obs}
 &:&\text{Observed frequency of the p-ray or s-ray}
\\
 I\ls{emi}(\nu\ls{emi})
 &:&\text{Intrinsic specific intensity of the source at emission}
\\
 &&\text{($\nu\ls{emi}$ is the emission frequency measured by the source)}
 \,,
\end{array}
\eqe
where, because the isotropic emission is assumed as mentioned in Sect.\ref{sec:principle.assumption}, the emission intensity $I\ls{emi}(\nu\ls{emi})$ depends only on the emission frequency $\nu\ls{emi}$ and not on the emission direction. 
The dimensions of $I\ls{emi}(\nu\ls{emi})$ are W/(m$^2$\,Hz\,\,ste-rad) in MKS units, where 1/m$^2$ denotes \emph{per unit spatial area perpendicular to the source}, 1/ste-rad denotes \emph{per unit solid-angle seen from the source} and 1/Hz denotes \emph{per unit frequency measured at the source}. 
Further, $F\ls{obs}$'s dependence on $u\ls{emi}^\mu$ is due to the kinetic (special relativistic) Doppler effect on the light ray. 
Since the kinetic Doppler effect is determined by the relative velocity between the source and ray, the strength of the kinetic Doppler effect is different between the p-ray and the s-ray. 
However, since the p-ray and the s-ray are emitted by the same source at the same time, the strength of gravitational Doppler effects on those rays are the same because the gravitational Doppler effect is determined by the gravitational potential at the emission.

There are 2 basic types of observation of the specific flux ratio $\ratio$:
\begin{description}
\item[Type LD (line detection): ]
Detect the p-ray and the s-ray at the same observation frequency $\nu\ls{obs}$. 
In this case, generally, the emission frequency of the p-ray $\nu\ls{emi (p)}$ and that of the s-ray $\nu\ls{emi (s)}$ are different ($\nu\ls{emi (p)} \neq \nu\ls{emi (s)}$), because the strength of the kinetic Doppler effect on the p-ray is generally different from that on the s-ray. 
\item[Type LE (line emission): ]
Detect the p-ray and the s-ray at different observation frequencies, respectively, $\nu\ls{obs (p)}$ and $\nu\ls{obs (s)}$, so that the difference between $\nu\ls{obs (p)}$ and $\nu\ls{obs (s)}$ offsets the difference between the kinetic Doppler effects on those rays. 
This means that the emission frequencies of the p-ray and the s-ray coincide, $\nu\ls{emi}(\nu\ls{obs (p)}) = \nu\ls{emi}(\nu\ls{obs (s)})$. 
This case corresponds to a line emission by the source. 
\end{description}
It will be shown in Sect.\ref{sec:formalism} that, while the ratio $\ratio\us{(LD)}$ of type LD depends on the functional form of $I\ls{emi}(\nu\ls{emi})$, the ratio $\ratio\us{(LE)}$ of type LE, however, does not depend on $I\ls{emi}(\nu\ls{emi})$.

\subsection{Principle of measuring $(M,\chi,\theta\ls{obs})$ from $(\delay,\ratio)$}
\label{sec:principle.principle}

We are studying a method of measuring the value of $(M,\chi,\theta\ls{obs})$ by observing quantities $(\delay,\ratio)$ which are functions of many parameters in lists \eqref{eq:principle.parameter-delay} and \eqref{eq:principle.parameter-ratio}. 
Our theoretical principle of measuring $(M,\chi,\theta\ls{obs})$ is simple, consisting of the following 3 steps:
\begin{description}
\item[Step 1 (BH measurement): ]
This step consists of theoretical and observational tasks.
\begin{description}
\item[Task of theory: ]
Theoretically calculate the values of $(\delay,\ratio)$ for various values of the parameters \eqref{eq:principle.parameter-delay} and \eqref{eq:principle.parameter-ratio}. 
\item[Task of observation: ]
Using an appropriate telescope, observe the values of $(\delay,\ratio)$ for 1 BH candidate as many times as possible. 
\end{description}
\item[Step 2 (BH measurement): ]
From the results of step~1, construct a table such as Table~\ref{table:measurement}.
In each row of this table, all sets of the values of parameters \eqref{eq:principle.parameter-delay} and \eqref{eq:principle.parameter-ratio} in the right-hand slot predict the same value of observable quantities $(\delay,\ratio)$ in the left-hand slot. 
\item[Step 3 (BH measurement): ]
It is a fact that the parameters $(M,\chi)$ and the direction angle $\theta\ls{obs}$ will not change for all observational data $(\delay,\ratio)$, because the target BH candidate is the same for all data. 
Hence, if we find only 1 set of values of $(M,\chi,\theta\ls{obs})$ that is shared by all rows in Table~\ref{table:measurement}, then the common values should be the true values of the mass, spin parameter, and direction angle of the BH candidate under observation. 
\end{description}

\begin{table}[t]
\begin{center}
\begin{tabular}{|c||c|}
 \hline
 Obs. data & Parameter values resulting in the obs. data in left slot
\\
 $(\delay,\ratio)$
 &
 $( M , \chi , \theta\ls{obs} \,;\, P)$
 \,,\,
 $P=$ [the others of \eqref{eq:principle.parameter-delay}
       and \eqref{eq:principle.parameter-ratio}\,]
\\
 \hline\hline
 $( \Delta t_1 \,,\, \mathcal{R}_1 )$
 &
 $\bigl( M_1, \chi_1 , \theta_1 \,;\, P_1 \bigr)$ \,,\,
 $\bigl( \boxed{M\ls{true}, \chi\ls{true} , \theta\ls{true}}
         \,;\, P^{\prime}_1 \bigr)$ \,,\, $\cdots$
\\
 &
 $\bigl( M^{\prime\prime}_1, \chi^{\prime\prime}_1 , \theta^{\prime\prime}_1
         \,;\, P^{\prime\prime}_1 \bigr)$ \,,\,
 $\bigl( M^{\prime\prime\prime}_1, \chi^{\prime\prime\prime}_1 , \theta^{\prime\prime\prime}_1
         \,;\, P^{\prime\prime\prime}_1 \bigr)$ \,,\,
 $\cdots$
\\
 \hline
 $( \Delta t_2 \,,\, \mathcal{R}_2 )$
 &
 $\bigl( \boxed{M\ls{true}, \chi\ls{true} , \theta\ls{true}}
         \,;\, P_2 \bigr)$ \,,\,
 $\bigl( M^{\prime}_2, \chi^{\prime}_2 , \theta^{\prime}_2\,;\, P^{\prime}_2 \bigr)$ \,,\, $\cdots$
\\
 &
 $\bigl( M^{\prime\prime}_2, \chi^{\prime\prime}_2 , \theta^{\prime\prime}_2
         \,;\, P^{\prime\prime}_2 \bigr)$ \,,\,
 $\bigl( M^{\prime\prime\prime}_2, \chi^{\prime\prime\prime}_2 , \theta^{\prime\prime\prime}_2
         \,;\, P^{\prime\prime\prime}_2 \bigr)$ \,,\,
 $\cdots$
\\
 \hline
 $( \Delta t_3 \,,\, \mathcal{R}_3 )$
 &
 $\bigl( M_3, \chi_3 , \theta_3 \,;\, P_3 \bigr)$ \,,\,
 $\bigl( M^{\prime}_3 , \chi^{\prime}_3 , \theta^{\prime}_3
         \,;\, P^{\prime}_3 \bigr)$ \,,\, $\cdots$
\\
 &
 $\bigl( \boxed{M\ls{true}, \chi\ls{true} , \theta\ls{true}}
         \,;\, P^{\prime\prime}_3 \bigr)$ \,,\,
 $\bigl( M^{\prime\prime\prime}_3, \chi^{\prime\prime\prime}_3 , \theta^{\prime\prime\prime}_3
         \,;\, P^{\prime\prime\prime}_3 \bigr)$ \,,\,
 $\cdots$
\\
 \hline
 $( \Delta t_4 \,,\, \mathcal{R}_4 )$
 &
 $\bigl( M_4, \chi_4 , \theta_4 \,;\, P_4 \bigr)$ \,,\,
 $\bigl( M^{\prime}_4 , \chi^{\prime}_4 , \theta^{\prime}_4
         \,;\, P^{\prime}_4 \bigr)$ \,,\, $\cdots$
\\
 &
 $\bigl( M^{\prime\prime}_4, \chi^{\prime\prime}_4 , \theta^{\prime\prime}_4
         \,;\, P^{\prime\prime}_4 \bigr)$ \,,\,
 $\bigl( \boxed{M\ls{true}, \chi\ls{true} , \theta\ls{true}}
         \,;\, P^{\prime\prime\prime}_4 \bigr)$ \,,\,
 $\cdots$
\\
 \hline
 $\vdots$ & \vdots
\\
 \hline
\end{tabular}
\end{center}
\caption{Our principle of direct BH measurement. 
If this table is obtained at step~2 of our principle, then we can conclude that the values of the BH's parameters are $(M\ls{true} , \chi\ls{true} , \theta\ls{true})$, which is the only set of values shared by all rows in this table.
}
\label{table:measurement}
\end{table}

The above method consisting of 3 steps is our principle for measuring the value of the BH's parameters $(M,\chi)$ and the direction angle $\theta\ls{obs}$ of the BH's spin axis. 
Because the value of $(M,\chi)$ is determined by the observable quantities of the Kerr BH lens effect (one of the general relativistic effects), our method is regarded as one principle of \emph{direct BH measurement} under the definition given at the beginning of Sect.\ref{sec:intro}.

Note that, in the task of observation, we do not need the repetition of observation under the same relative location among the BH, source, and observer. 
In general, the value of $(\delay,\ratio)$ for 1 relative location differs from that for the other relative location. 
The point is that the number of rows in Table~\ref{table:measurement} becomes large by observing the value of $(\delay,\ratio)$ as many times as possible for various relative locations. 
The larger the number of rows in Table~\ref{table:measurement}, the more certain the true value $(M\ls{true},\chi\ls{true},\theta\ls{true})$.

Also note that, when the values of parameters \eqref{eq:principle.parameter-delay} and \eqref{eq:principle.parameter-ratio} are given, the flux ratio $\ratio\us{(LD)}$ of the type LD observation differs from $\ratio\us{(LE)}$ of the type LE observation in general (but the time delay $\delay$ is common to both types of observation). 
Therefore, in the task of theory, we need to perform 2 calculations of $\ratio\us{(LD)}$ and $\ratio\us{(LE)}$ for each given value of parameters \eqref{eq:principle.parameter-delay} and \eqref{eq:principle.parameter-ratio}. 
Then, one of them should be selected (or their combination should be constructed) according to the real observational method used in the task of observation.

\section{Formalism to calculate $\delay$ and $\ratio$}
\label{sec:formalism}

Under the condition that the parameters \eqref{eq:principle.parameter-delay} and \eqref{eq:principle.parameter-ratio} are given, we construct the formalism for calculating $(\delay,\ratio)$.

\subsection{Numerical setup of the observer}
\label{sec:formalism.observer}

We consider the rest observer, $(r\ls{obs} , \theta\ls{obs} , \varphi\ls{obs}) =$ constant, whose velocity is $u\ls{obs}^\mu = (1/\sqrt{|g_{tt}|} , 0 , 0 , 0)$ satisfying $u\ls{obs \mu} u\ls{obs}^\mu = -1$. 
As mentioned in Sect.\ref{sec:principle.assumption}, we consider the distant observer, $r\ls{obs} \to \infty$, whose velocity becomes
\eqb
\label{eq:formalism.uobs}
 u\ls{obs}^\mu = (1,0,0,0) = \xi\ls{(t)}^\mu \,.
\eqe
The velocity of our observer coincides with the timelike Killing vector, $u\ls{obs} = \xi\ls{(t)}$.

Although our ideal setup is the observer at $r\ls{obs}\to\infty$, our numerical calculation will be performed with putting the rest observer at $r\ls{obs} = 100\,r\ls{BH}$ which gives $100/99\, (\simeq 1.01) \le 1/\sqrt{|g_{tt}|} \le 50/49\, (\simeq 1.02)$. 
On the other hand, we will use Eq.\eqref{eq:formalism.uobs} for $u\ls{obs}^\mu$ in our numerical calculation. 
This implies that our numerical results of observable quantities, shown in Sect.\ref{sec:result}, include an intrinsic error of $1\% \sim 2 \%$. 
However, this error does not affect the numerical results showing the potential detectability of our observable quantities.

\subsection{Time delay $\delay$ : null geodesic equations}
\label{sec:formalism.delay}

Since we assume the duration of emission is short (burst-like), the p-ray and the s-ray are both regarded as 1 pulse made of photons. 
The 4-dimensional orbit of a light ray is a null geodesic of spacetime. 
To emphasize the behavior of light ray as a pulse, we use the term ``light ray''. 
To emphasize the 4-dimensional orbit of a light ray, we use the term ``null geodesic''. 
Then, we can say that any light ray propagates on a null geodesic.

Let $h$ denote an affine parameter measured in the length dimension, and let $x\ls{ng}^\mu(h)$ be the spacetime position of a light ray on the null geodesic. 
The wave vector of a light ray is given by $K^\mu = \diff x\ls{ng}^\mu(h)/\diff h$, which must satisfy the null condition $K^\mu K_\mu = 0$. 
Due to the stationarity and axisymmetry of Kerr spacetime, any light ray possesses 3 constants of motion $(\nu\ls{obs},l_\varphi,l\ls{nor})$ given by $K^\mu$ and Killing fields, 
\eqb
\label{eq:formalism.constant}
\begin{aligned}
 \nu\ls{obs}
 &= - K_\mu \xi\ls{(t)}^\mu = - K_t > 0
 \quad,\quad
 l_\varphi
 = K_\mu \xi\ls{(\varphi)}^\mu = K_\varphi
\\
 l\ls{nor}^2
 &= K^\mu K^\nu \Xi_{\mu\nu}
  = \Bigl( \nu\ls{obs} a \sin\theta - \dfrac{l_\varphi}{\sin\theta} \Bigr)^2
    + \Sigma^2 \bigl(K^\theta\bigr)^2 \ge 0 \,,
\end{aligned}
\eqe
which are conserved along the null geodesic, $K^\mu Q_{;\mu} = 0$, where $Q = \nu\ls{obs} \,,\, l_\varphi \,,\, l\ls{nor}$.

Our setup \eqref{eq:formalism.uobs} of $u\ls{obs}^\mu$ means that the quantity $\nu\ls{obs}$ is calculated from $u\ls{obs}^\mu$ as $\nu\ls{obs} = -K_\mu u\ls{obs}^\mu$. 
This $\nu\ls{obs}$ is the dimensionless frequency of a light ray observed by a rest observer at infinity ($r\ls{obs}\to\infty$), where ``dimensionless'' denotes that $\nu\ls{obs}$ is normalized by a fiducial frequency. 
The quantity $l_\varphi$ is the $\varphi$-component (toroidal component) of the orbital angular-momentum of a light ray, and the quantity $l\ls{nor}$ relates to the norm of the orbital angular-momentum of a light ray. 
These components, $l_\varphi$ and $l\ls{nor}$, are measured per unit energy of the ray, since the dimension of them is the length in natural units ($c=1 \,,\, G=1$).

Note that the fiducial frequency for the normalization of $\nu\ls{obs}$ can be chosen arbitrarily. 
For example, in MKS units, if we choose $300$GHz as the fiducial frequency, then the unit of length size in our numerical calculation becomes $c/(300$GHz$) = 10^{-3}$m. 
If, conversely, we set the unit of length size to $r\ls{BH}$, then the fiducial frequency becomes $c/r\ls{BH}$.

Hereafter, for simplicity of calculations, we normalize the null geodesic by $\nu\ls{obs}$ as
\seqb
\label{eq:formalism.normalize}
\eqb
\label{eq:formalism.k}
 k^\mu \defeq \dfrac{1}{\nu\ls{obs}} K^\mu = \od{x\ls{ng}^\mu(\eta)}{\eta} \,,
\eqe
where $\eta \defeq \nu\ls{obs}\, h$ is the dimensionless affine parameter. 
Further we define the \emph{toroidal impact parameter} $b$ and the \emph{``normic'' impact parameter} $q$ by
\eqb
\label{eq:formalism.impactparameter}
 b \defeq \dfrac{l_\varphi}{\nu\ls{obs}} \quad,\quad q \defeq \dfrac{l\ls{nor}}{\nu\ls{obs}} \,.
\eqe
\seqe

The spacetime position of a light ray, $x\ls{ng}^\mu(\eta)$, is determined by the null geodesic equations, $k^\alpha k^\mu_{\phantom{\mu};\alpha} = 0$, whose components in BL coordinates, together with the null condition $k_\mu k^\mu = 0$, can be arranged as
\eqb
\label{eq:formalism.geodesic}
\begin{aligned}
 k^t
 &= \dfrac{1}{\Sigma(r,\theta)}\Bigl[ (r^2 + a^2) X(r) - a Y(\theta) \Bigr]
 &,&
 &\bigl( k^r \bigr)^2
 + \dfrac{V\ls{eff}(r)}{\Sigma(r,\theta)^2}
 &= 0
\\
 k^\varphi
 &= \dfrac{1}{\Sigma(r,\theta)}\Bigl[  a X(r) - \dfrac{Y(\theta)}{\sin^2\theta} \Bigr]
 &,&
 &\bigl( k^\theta \bigr)^2
 + \dfrac{U\ls{eff}(\theta)}{\Sigma(r,\theta)^2}
 &= 0 \,,
\end{aligned}
\eqe
where the auxiliary functions are 
$X(r) = (r^2 + a^2 - a b)/\Delta(r)$ and 
$Y(\theta) = a \sin^2\theta - b$, 
and the radial and zenithal effective potentials are
\eqb
\label{eq:formalism.effectivepotential}
 V\ls{eff}(r) = q^2 \Delta(r) - \bigl[ \Delta(r) X(r) \bigr]^2
 \quad,\quad
 U\ls{eff}(\theta) = - q^2 + \dfrac{Y(\theta)^2}{\sin^2\theta} \,.
\eqe
Note that the normic impact parameter $q$ appears only in the squared form, $q^2$. 
This means that the solution of the null geodesic equation, $k^\mu(\eta)$ and $x\ls{ng}^\mu(\eta)$, does not depend on the signature of $q$. 
We set $q \ge 0$ throughout this paper. 
Also, in our numerical calculation, we set $\eta = 0$ at the emission event of light rays: $x\ls{emi}^\mu = x\ls{ng}^\mu(0)$ and $k\ls{emi}^\mu = k^\mu(0)$.

We can recognize from Eq.\eqref{eq:formalism.geodesic} that, given the emission event $x\ls{emi}^\mu$ and the initial null vector $k\ls{emi}^\mu$ at the emission, the values of impact parameters $(b,q)$ corresponding to $x\ls{emi}^\mu$ and $k\ls{emi}^\mu$ are determined by solving Eq.\eqref{eq:formalism.geodesic} algebraically for $(b,q)$. 
In our numerical calculation, we use Eq.\eqref{eq:formalism.geodesic} for the following 2 purposes:
(i) to evaluate the values of impact parameters $(b,q)$ for given $x\ls{emi}^\mu$ and $k\ls{emi}^\mu$, and (ii) to judge whether or not the light ray of given $x\ls{emi}^\mu$ and $k\ls{emi}^\mu$ is to be absorbed eventually by BH. 
The criterion of the judgment is given by the effective potentials \eqref{eq:formalism.effectivepotential} as explained in Appendix~\ref{app:ic}.

If one tries to integrate Eq.\eqref{eq:formalism.geodesic}, then there arises the problem of how to choose the signatures of $k^r = \pm \sqrt{|V\ls{eff}|}/\Sigma$ and $k^\theta = \pm \sqrt{|U\ls{eff}|}/\Sigma$. 
We avoid this problem by using the Hamiltonian formalism of geodesics instead of Eq.\eqref{eq:formalism.geodesic}. 
Define the Hamiltonian of a light ray by
\eqb
\label{eq:formalism.H}
 \mathcal{H}(x\ls{ng},k) \defeq
 \dfrac{1}{2} k_\mu k_\nu g^{\mu\nu}(x\ls{ng}) \,,
\eqe
where we regard the position $x\ls{ng}^\mu(\eta)$ and the null 1-form $k_\mu(\eta)$ as the dynamical variables in the present Hamiltonian formalism. 
In Kerr spacetime, the components of the 1-form in BL coordinates are
\eqb
\label{eq:formalism.k-form}
 k_\mu = (\,-1\,,\,k_r\,,\,k_\theta\,,\,b\,) \,,
\eqe
where the $t$-component and $\varphi$-component are constant because of Eq.\eqref{eq:formalism.constant} and \eqref{eq:formalism.impactparameter}. 
The Hamilton equations, $\diff{x\ls{ng}^\mu(\eta)}/\diff{\eta} = \partial\mathcal{H}/\partial k_\mu$ and $\diff{k_\mu(\eta)}/\diff{\eta} = - \partial\mathcal{H}/\partial x\ls{ng}^\mu$, are reduced to the following 6 simultaneous equations:
\eqb
\label{eq:formalism.geodesic-hamilton}
 \od{k_r}{\eta}
 = -\dfrac{1}{2} k_\mu k_\nu \pd{g^{\mu\nu}(x\ls{ng})}{r\ls{ng}}
 \quad,\quad
 \od{k_\theta}{\eta}
 = -\dfrac{1}{2} k_\mu k_\nu \pd{g^{\mu\nu}(x\ls{ng})}{\theta\ls{ng}}
 \quad,\quad
 \od{x\ls{ng}^\mu}{\eta}
 = k_\nu g^{\mu\nu}(x\ls{ng}) \,.
\eqe
In these equations, no square root appears.

Note that the Euler-Lagrange equations given by the Lagrangian corresponding to the Hamiltonian \eqref{eq:formalism.H} can be arranged to geodesic equations of the form \eqref{eq:formalism.geodesic}. 
Also, note that the value of the Hamiltonian, $\mathcal{H} \propto k^2 \equiv 0$, is conserved along the geodesic. 
Therefore, the null geodesic $x\ls{ng}^\mu(\eta)$ is obtained automatically by solving Eq.\eqref{eq:formalism.geodesic-hamilton} with the initial condition satisfying $k\ls{emi}^2 = 0$. 
In our numerical calculation, the initial null 1-form $k\ls{emi\,\mu}$ is constructed by making use of Eq.\eqref{eq:formalism.geodesic} as explained in Appendix~\ref{app:ic}. 
Then, the null geodesics of the p-ray $x\ls{ng(p)}^\mu(\eta)$ and the s-ray $x\ls{ng(s)}^\mu(\eta)$, which connect the source and observer, are obtained by solving Eq.\eqref{eq:formalism.geodesic-hamilton} numerically, and we can plot those solutions in the $t$-$r$ plane and read out the time delay $\delay$ as shown in Fig.\ref{fig:rays.example}.

\subsection{Total Doppler factor}
\label{sec:formalism.doppler}

Given the observed frequency $\nu\ls{obs}$ of a light ray detected by the observer, the emission frequency $\nu\ls{emi}$ of the ray can be given by
\eqb
\label{eq:formalism.freq-emission}
 \nu\ls{emi} = - K\ls{emi\,\mu} u\ls{emi}^\mu = - \nu\ls{obs}\, k\ls{emi\,\mu} u\ls{emi}^\mu
 = \nu\ls{obs}
   \bigl( u\ls{emi}^t
        - k\ls{emi\,{\it r}} u\ls{emi}^r
        - k\ls{emi\,\theta} u\ls{emi}^\theta
        - b u\ls{emi}^\varphi \bigr) \,,
\eqe
where Eq.\eqref{eq:formalism.k-form} is used. 
When the source is moving near the BH, we cannot identify which term in Eq.\eqref{eq:formalism.freq-emission} is responsible for the gravitational or kinetic Doppler effect. 
Let us define the \emph{total-Doppler factor} $\tdf$ as the ratio of $\nu\ls{obs}$ to $\nu\ls{emi}$,
\eqb
\label{eq:formalism.D}
 \tdf(x\ls{emi},u\ls{emi},k\ls{emi}) \defeq
 \dfrac{\nu\ls{obs}}{\nu\ls{emi}} = -\dfrac{1}{k\ls{emi\,\mu} u\ls{emi}^\mu} =
 -\dfrac{1}{g_{\mu\nu}(x\ls{emi})k\ls{emi}^\mu u\ls{emi}^\nu} \,.
\eqe
This can be regarded as a function of the emission event $x\ls{emi}^\mu$, the velocity of the source $u\ls{emi}^\mu$, and the initial null vector of the light ray $k\ls{emi}^\mu$. 
Thus, the value of $\tdf$ can be calculated numerically once the null geodesic that connects the source and observer is obtained numerically by the formulas shown in Sect.\ref{sec:formalism.delay}. 
The values of $\tdf\ls{(p)}$ of the p-ray and $\tdf\ls{(s)}$ of the s-ray are generally different, since the initial null vectors of them do not coincide ($k\ls{emi(p)}^\mu \neq k\ls{emi(s)}^\mu$), while the other parameters $x\ls{emi}^\mu$ and $u\ls{emi}^\mu$ are shared.

\subsection{Specific flux $F\ls{obs}$ : Jacobi equations}
\label{sec:formalism.flux}

\subsubsection{Primitive form of $F\ls{obs}$}
\label{sec:formalism.flux.primitive}

\begin{figure}[t]
 \begin{center}
 \includegraphics[scale=0.45]{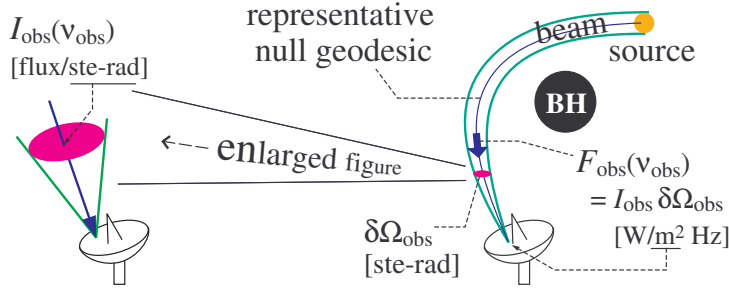}
 \end{center}
\caption{Illustration of the specific intensity $I\ls{obs}(\nu\ls{obs})$, the visible solid-angle $\delta\Omega\ls{obs}$, and the specific flux $F\ls{obs}$. 
The dimensions of them are shown in MKS units, where 1/m$^2$ in the dimensions of $F\ls{obs}$ denotes \emph{per unit area on the telescope}, and 1/ste-rad in the dimensions of $I\ls{obs}$ denotes \emph{per unit solid-angle seen from the telescope}.}
\label{fig:flux}
\end{figure}

In real situations, each winding ray (p-ray, s-ray, and higher winding rays) is detected by the telescope as a beam made of neighboring light rays. 
The 4-dimensional track of the beam in spacetime is a bundle made of neighboring null geodesics of light rays composing the beam. 
In order to calculate the specific flux, which is the \emph{surface density} of the observed power of light rays on the telescope's face, we focus on the beam incident to \emph{1 point} on the telescope's face, as illustrated in Fig.\ref{fig:flux}.

Because we assume that the source of light rays is much smaller than the BH, the beam of light rays is narrow. 
Hence, the orbits (null geodesics) and the spectrum of all light rays in 1 beam can be regarded approximately as the same ones, which are represented by 1 constituent light ray in the beam. 
We call such a ray the \emph{representative light ray}, and its null geodesic the \emph{representative null geodesic}.

Let $I\ls{obs}(\nu\ls{obs})$ denote the observed specific intensity of the representative light ray, and let $\delta\Omega\ls{obs}$ denote the visible solid-angle of the source measured by the observer. 
Then, the observed specific flux of the null geodesic bundle is given by 
\eqb
\label{eq:formalism.Fobs-primitive}
 F\ls{obs}(\nu\ls{obs})
 = \int I\ls{obs} \diff{\Omega}
 \,\simeq\, I\ls{obs}(\nu\ls{obs})\, \delta\Omega\ls{obs} \,,
\eqe
where $\Omega$ is the solid-angle around the telescope.

Our numerical calculation of $F\ls{obs}(\nu\ls{obs})$ is performed under 2 suppositions: (i) the parameters \eqref{eq:principle.parameter-delay} and \eqref{eq:principle.parameter-ratio} are already given, and (ii) the representative null geodesic has already been obtained numerically by the procedure given in Sect.\ref{sec:formalism.delay}. 
Under supposition (i), we derive $I\ls{obs}(\nu\ls{obs})$ from $I\ls{emi}(\nu\ls{emi})$. 
Under supposition (ii), we derive the formula for the visible solid-angle $\delta\Omega\ls{obs}$ in order to calculate it from the narrow null geodesic bundle including the given representative null geodesic (see Fig.\ref{fig:flux}).

\subsubsection{Relation between $I\ls{obs}(\nu\ls{obs})$ and $I\ls{emi}(\nu\ls{emi})$}
\label{sec:formalism.flux.Iobs}

Let us relate the observed specific intensity $I\ls{obs}(\nu\ls{obs})$ to the intrinsic specific intensity $I\ls{emi}(\nu\ls{emi})$ of the source of light rays. 
Since the distribution of light rays in the beam is described by the collisionless Boltzman equation (leading to Liouville's theorem for the distribution of light rays in the beam), the null geodesic bundle of the beam possesses a conserved quantity,
\eqb
\label{eq:formalism.constant-bundle}
 \dfrac{I(\nu)}{\nu^3} = \text{constant inside the null geodesic bundle} \,,
\eqe
where $I(\nu)$ is the specific intensity measured by an arbitrary observer at an arbitrary spacetime point inside the null geodesic bundle~\cite{ref:misner+2.1973}. 
The quantity, $I(\nu)/\nu^3$, keeps the same value from observer to observer and from spacetime point to spacetime point inside the null geodesic bundle. 
Then, we find the relation between $I\ls{obs}(\nu\ls{obs})$ and $I\ls{emi}(\nu\ls{emi})$, 
\eqb
\label{eq:formalism.Iobs}
 I\ls{obs}(\nu\ls{obs}) = \tdf^3 I\ls{emi}\bigl(\nu\ls{obs}/\tdf\bigr) \,,
\eqe
where $\tdf$ is the total-Doppler factor~\eqref{eq:formalism.D}. 
The value of $\tdf$ is uniquely determined by the representative null geodesic.

\subsubsection{Formulation of $\delta\Omega\ls{obs}$}
\label{sec:formalism.flux.deltaOmega}

\begin{figure}[t]
 \begin{center}
 \includegraphics[scale=0.45]{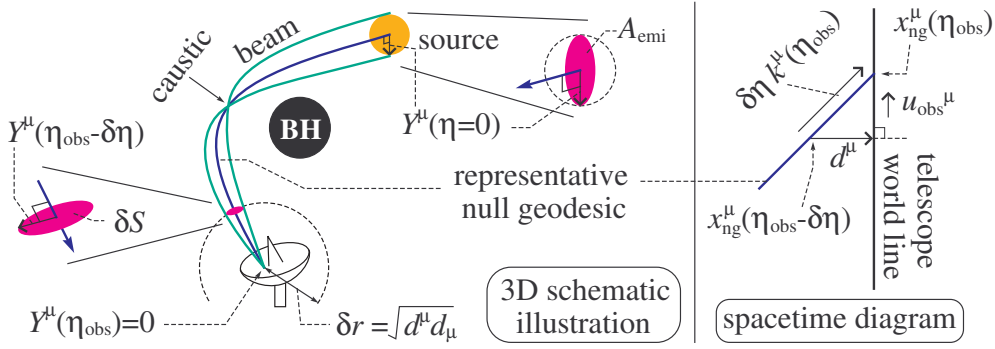}
 \end{center}
\caption{The setup for calculating the visible solid-angle $\delta\Omega\ls{obs}$. 
The left-hand panel is a 3D schematic illustration, and the right-hand panel is a spacetime diagram near the observation event $x\ls{ng}^\mu(\eta\ls{obs})$. 
The quantity $Y^\mu$ is the Jacobi vector field on the representative null geodesic. 
By considering a small celestial sphere around the observation point, the visible solid-angle $\delta\Omega\ls{obs}$ is calculated from the area swept by $Y^\mu$ on the celestial sphere. 
}
\label{fig:deviation}
\end{figure}

Let us construct the formula of visible solid-angle $\delta\Omega\ls{obs}$ so as to be calculated from the narrow null geodesic bundle including the given representative null geodesic. 
The setup of our formulation is shown in Fig.\ref{fig:deviation}. 
Consider a small celestial sphere of radius $\delta r$ centered at the observation point, and let $\delta S$ denote the area on the celestial sphere that is penetrated by the beam of light rays. 
The visible solid-angle is given by
\eqb
\label{eq:formalism.deltaOmega-primitive}
 \delta\Omega\ls{obs} = \dfrac{\delta S}{\delta r^2} \,.
\eqe
We evaluate $\delta r$ and $\delta S$ in following discussions of this Sect.\ref{sec:formalism.flux.deltaOmega}.

For the first, in order to evaluate the radius $\delta r$ of the observer's celestial sphere, let $\eta\ls{obs}$ denote the value of the affine parameter $\eta$ of the representative null geodesic when it reaches the observer, and let $\eta\ls{obs}-\delta\eta$ denote the value of $\eta$ when the representative null geodesic crosses the observer's celestial sphere. 
In terms of the position of representative light ray $x\ls{ng}^\mu(\eta)$, the observation event is $x\ls{ng}^\mu(\eta\ls{obs})$ and the crossing event is $x\ls{ng}^\mu(\eta\ls{obs}-\delta\eta)$. 
Then, the following vector connects these spacetime events (see the right-hand panel of Fig.\ref{fig:deviation}):
\eqb
 x\ls{ng}^\mu(\eta\ls{obs}) - x\ls{ng}^\mu(\eta\ls{obs}-\delta\eta) \,\simeq\,
 \delta\eta \od{x\ls{ng}}{\eta}(\eta\ls{obs})
 = \delta\eta\, k\ls{obs}^\mu \,,
\eqe
where $k\ls{obs}^\mu \defeq k^\mu(\eta\ls{obs})$ is the null tangent vector of the representative null geodesic at the observation event. 
The projection of this vector onto the spacelike hypersurface perpendicular to the observer's velocity $u\ls{obs}^\mu$ is
\eqb
 d^\mu
 = \delta\eta\, \bigl(\, g\ls{obs}^{\lambda\mu}+u\ls{obs}^\lambda u\ls{obs}^\mu \,\bigr)
    k\ls{obs\,\lambda}
 = \delta\eta\, \bigl(\, k\ls{obs}^\mu - u\ls{obs}^\mu \,\bigr) \,,
\eqe
where $g\ls{obs}^{\lambda\mu}+u\ls{obs}^\lambda u\ls{obs}^\mu$ is the projection tensor onto the hypersurface perpendicular to $u\ls{obs}^\mu$, and Eqs.\eqref{eq:formalism.uobs} and \eqref{eq:formalism.k-form} are used in the second equality. 
The radius $\delta r$ measured by the observer is the norm of this spatial vector,
\eqb
\label{eq:formalism.deltar}
 \delta r = \sqrt{d^\mu d_\mu} = \delta\eta \,.
\eqe

Next, in order to evaluate $\delta S$, we make use of the \emph{Jacobi vector field} $Y^\mu(\eta)$ (the \emph{geodesic deviation vector field}) that connects the representative null geodesic to a neighboring null geodesic in the bundle of null geodesics. 
The Jacobi field can be defined so as to be spacelike ($Y^\mu Y_\mu > 0$), perpendicular to the representative null geodesic ($Y^\mu k_\mu = 0$), and invariant under the Lie transport along the geodesic ($\mathcal{L}_k Y^\mu = 0$)~\cite{ref:hawking+1.1973}. 
The Lie-invariant condition, $\mathcal{L}_k Y^\mu = 0$, can be arranged into the \emph{Jacobi equations} (the \emph{geodesic deviation equations}) in terms of covariant derivatives,
\eqb
\label{eq:formalism.jacobi}
 k^\beta \bigl( k^\alpha Y^\mu_{\phantom{\mu};\alpha} \bigr)_{;\beta}
 = - R^\mu_{\phantom{\mu}\alpha \lambda \beta} k^\alpha Y^\lambda k^\beta \,,
\eqe
where $R^\mu_{\phantom{\mu}\alpha \lambda \beta}(\eta)$ is the Riemann curvature tensor at the spacetime point $x\ls{ng}^\mu(\eta)$ on the representative null geodesic. 
Furthermore, since we are now considering the null geodesic bundle converging at the observation event (see Fig.\ref{fig:flux} and the left-hand panel of Fig.\ref{fig:deviation}), the Jacobi equations are to be solved under the ``initial'' condition, 
\eqb
\label{eq:formalism.jacobi-ic}
 Y\ls{obs}^\mu = 0 \quad,\quad
 Y\ls{obs\,;\alpha}^\mu \neq 0 \,,
\eqe
where $Y\ls{obs}^\mu \defeq Y^\mu(\eta\ls{obs})$ and $Y\ls{obs\,;\alpha}^\mu \defeq Y^\mu_{\phantom{\mu};\alpha}(\eta\ls{obs})$ are the Jacobi vector and its covariant derivative at the observation event. 
The value of $Y\ls{obs\,;\alpha}^\mu$ is not specified uniquely. 
Different values of $Y\ls{obs\,;\alpha}^\mu$ correspond to the different Jacobi fields connecting the representative null geodesic to different neighboring null geodesics in the bundle.

The significance of Jacobi fields is that, as illustrated in the left-hand panel of Fig.\ref{fig:deviation} and explained in detail in Appendix \ref{app:area}, the 2-dimensional cross-sectional area $\delta S$ of the narrow null geodesic bundle is regarded as the area swept by the Jacobi vectors surrounding the representative null geodesic at the crossing event $x\ls{ng}^\mu(\eta\ls{obs}-\delta\eta)$. 
Thus, the Jacobi equations \eqref{eq:formalism.jacobi} can be rearranged into a form that describes the evolution of cross-sectional area from the emission event $x\ls{emi}^\mu = x\ls{ng}^\mu(0)$ to the observation event $x\ls{obs}^\mu = x\ls{ng}^\mu(\eta\ls{obs})$ along the representative null geodesic. 
The following paragraphs are the outline of the calculation procedure of $\delta S$, and the details are given in Appendix~\ref{app:area}.

From condition \eqref{eq:formalism.jacobi-ic}, we can set for sufficiently small $\delta\eta$,
\eqb
\label{eq:formalism.deltaY}
 Y^\mu(\eta\ls{obs}-\delta\eta) \,\simeq\,
 - \delta\eta\, k\ls{obs}^\alpha Y\ls{obs\,;\alpha}^\mu \,.
\eqe
Therefore, we find the relation
\eqb
\label{eq:formalism.deltaS}
 \delta S = \delta\eta^2 A\ls{obs} \,,
\eqe
where $A\ls{obs}$ is the spacelike area swept by the spacelike vector $k\ls{obs}^\alpha Y\ls{obs\,;\alpha}^\mu$ which surrounds the representative null geodesic at the observation event $x\ls{obs}^\mu$. 
(Since $Y^\mu$ is spacelike, Eq.\eqref{eq:formalism.deltaY} denotes that $k^\alpha Y_{\phantom{\mu};\alpha}^\mu$ is also spacelike.)

Because the Jacobi equations \eqref{eq:formalism.jacobi} describe the evolution of the Jacobi vector $Y^\mu(\eta)$ along the representative null geodesic, the evolution of the cross-sectional area of the null geodesic bundle along the representative null geodesic is also described by the Jacobi equations \eqref{eq:formalism.jacobi}. 
As explained in detail in Appendix \ref{app:area}, an appropriate rearrangement of Eq.\eqref{eq:formalism.jacobi} provides us with the transformation of the cross-sectional area of the null geodesic bundle at $x\ls{emi}^\mu$, which is denoted by $A\ls{emi}$ hereafter, to the area $A\ls{obs}$ at $x\ls{obs}^\mu$ appearing in Eq.\eqref{eq:formalism.deltaS},
\eqb
\label{eq:formalism.Aobs}
 A\ls{obs} = \atc(x\ls{obs},u\ls{obs},k\ls{obs},x\ls{emi},u\ls{emi},k\ls{emi})\, A\ls{emi} \,,
\eqe
where the precise form of the coefficient $\atc$ is given in Eq.\eqref{eq:area.Aobs-C} of Appendix \ref{app:area}. 
We call this coefficient $\atc$ the \emph{area-transfer coefficient}. 
The point of this relation is that the transformation between $A\ls{obs}$ and $A\ls{emi}$ is a linear relation, and the area-transfer coefficient $\atc$ is determined by the set of quantities, $x\ls{obs}^\mu$, $u\ls{obs}^\mu$, $k\ls{obs}^\mu$, $x\ls{emi}^\mu$, $u\ls{emi}^\mu$ and $k\ls{emi}^\mu$. 
Also, it has to be noted that the area $A\ls{emi}$ can be regarded as the cross-sectional area of the source of light rays seen from the emission direction of the light ray.

Now, we have obtained the ingredients for calculating the visible solid-angle $\delta\Omega\ls{obs}$.
Combining Eqs.\eqref{eq:formalism.deltaOmega-primitive}, \eqref{eq:formalism.deltar}, \eqref{eq:formalism.deltaS}, and \eqref{eq:formalism.Aobs}, we find
\eqb
\label{eq:formalism.deltaOmega}
 \delta\Omega\ls{obs} = \atc A\ls{emi} \,.
\eqe
This is the formula for the visible solid-angle we use in our numerical calculation. 
This formula does not depend on the observation frequency $\nu\ls{obs}$.

\subsubsection{Our formulas for $F\ls{obs}$ and $\ratio$}
\label{sec:formalism.flux.Fobs}

Substituting Eqs.\eqref{eq:formalism.Iobs} and \eqref{eq:formalism.deltaOmega} into Eq.\eqref{eq:formalism.Fobs-primitive}, the formula for the observed specific flux used in our numerical calculation is
\eqb
\label{eq:formalism.Fobs}
 F\ls{obs}(\nu\ls{obs}) =
 \tdf^3 I\ls{emi}\bigl(\nu\ls{obs}/\tdf\bigr)\, \atc A\ls{emi} \,,
\eqe
where the total-Doppler factor $\tdf$ is given in Eq.\eqref{eq:formalism.D}, and the area-transfer coefficient $\atc$ is given in Eq.\eqref{eq:area.Aobs-C} of Appendix \ref{app:area}. 
It is important to specify which factors in Eq.\eqref{eq:formalism.Fobs} include the dependence on the observation frequency $\nu\ls{obs}$ and on the choice of p-ray or s-ray: 
The $\nu\ls{obs}$-dependence of $F\ls{obs}$ arises from only the intrinsic specific intensity of the source, $I\ls{emi}(\nu\ls{obs}/\tdf)$. 
And, the dependence of $F\ls{obs}$ on the choice of p-ray or s-ray (dependence on the winding number around the BH) arises from the factors $\tdf(x\ls{emi},u\ls{emi},k\ls{emi})$ and $\atc(x\ls{obs},u\ls{obs},k\ls{obs},x\ls{emi},u\ls{emi},k\ls{emi})$. 
Here, the arguments $k\ls{emi}^\mu$, $k\ls{obs}^\mu$, and $t\ls{obs} \, (= x\ls{obs}^t)$ depend on the choice of p-ray or s-ray, while the other arguments are shared by the p-ray and the s-ray.

Note that, in our numerical calculation, the cross-sectional area of the source $A\ls{emi}$, which is seen from the emission direction of the ray in the hypersurface perpendicular to $u\ls{emi}^\mu$, is treated as a given parameter. 
For simplicity, we assume that the shape of the source is spherical when it is seen by an observer comoving with the source. 
Then, the value of $A\ls{emi}$ is the same for the p-ray and the s-ray:
\eqb
\label{eq:formalism.assumption-Aemi}
 A\ls{emi(p)}=A\ls{emi(s)} \,(=A\ls{emi}) \,.
\eqe
This is consistent with our model that the source is point-like (see Sect.\ref{sec:principle.assumption}). 
Under this assumption, the ratio of observed specific flux is given by the formula,
\seqb
\eqb
\label{eq:formalism.Robs}
 \ratio =
 \dfrac{\tdf\ls{(s)}^3 I\ls{emi}(\nu\ls{obs(s)}/\tdf\ls{(s)})\, \atc\ls{(s)}}
       {\tdf\ls{(p)}^3 I\ls{emi}(\nu\ls{obs(p)}/\tdf\ls{(p)})\, \atc\ls{(p)}} \,,
\eqe
which is independent of $A\ls{emi}$. 
Then, as explained in Sect.\ref{sec:principle.observable}, the ratio $\ratio\us{(LE)}$ of type LE (line emission) is calculated under the condition $\nu\ls{obs(s)}/\tdf\ls{(s)} = \nu\ls{obs(p)}/\tdf\ls{(p)} \,(\,\Leftrightarrow \nu\ls{emi(s)} = \nu\ls{emi(p)} \,)$, and we find
\eqb
\label{eq:formalism.Robs-LE}
 \ratio\us{(LE)} =
 \dfrac{\tdf\ls{(s)}^3 \atc\ls{(s)}}
       {\tdf\ls{(p)}^3 \atc\ls{(p)}} \,,
\eqe
which is independent of $I\ls{emi}$. 
Also, the ratio $\ratio\us{(LD)}$ of type LD (line detection) is calculated under the condition $\nu\ls{obs(s)} = \nu\ls{obs(p)} \defeqr \nu\ls{obs}\us{(LD)}$, 
\eqb
\label{eq:formalism.Robs-LD}
 \ratio\us{(LD)} =
 \dfrac{\tdf\ls{(s)}^3 I\ls{emi}(\nu\ls{obs}\us{(LD)}/\tdf\ls{(s)})\, \atc\ls{(s)}}
       {\tdf\ls{(p)}^3 I\ls{emi}(\nu\ls{obs}\us{(LD)}/\tdf\ls{(p)})\, \atc\ls{(p)}} \,.
\eqe
\seqe
This $\ratio\us{(LD)}$ depends on $I\ls{emi}(\nu\ls{emi})$ except for the white-noise-type emission, $I\ls{emi}(\nu\ls{emi}) =$ constant. 
For the white-noise emission, the observed flux ratio of type LD and that of type LE are the same and given by Eq.\eqref{eq:formalism.Robs-LE}.

\subsection{Selection rule for the p-ray and the s-ray}
\label{sec:formalism.ray}

Suppose that we have found numerically some null geodesics that connect the given emission event $x\ls{emi}^\mu$ and the given observation position $(r\ls{obs},\theta\ls{obs},\varphi\ls{obs})$. 
This means that we have obtained some numerical solutions of the null geodesic equations \eqref{eq:formalism.geodesic-hamilton} for different values of the initial 1-form $k\ls{emi\,\mu}$. 
Further, suppose that we have not recognized which solutions are the p-ray and the s-ray. 
This is the case that we confront in our numerical calculation, as will be explained in Sect.\ref{sec:formalism.step}. 
The issue in this subsection is how to select the p-ray and the s-ray from the set of numerical solutions of null geodesics.

We should emphasize that, in our numerical calculation, we cannot always regard the numerical solution of the null geodesic possessing the earliest (or second earliest) observation time as the p-ray (or the s-ray). 
The reason is that our numerical search for the solution of null geodesic equations \eqref{eq:formalism.geodesic-hamilton} is the shooting method with discretely varying values of the initial 1-form $k\ls{emi\,\mu}$, and that the initial 1-forms appropriate for the p-ray $k\ls{emi(p)\,\mu}$ and the s-ray $k\ls{emi(p)\,\mu}$ may be omitted in the discrete variation of $k\ls{emi\,\mu}$. 
Therefore, we need the criterion to judge whether the p-ray and the s-ray exist in the set of numerical solutions of null geodesics, and to select the p-ray and the s-ray when they exist in the set of numerical solutions of null geodesics.

To construct the selection rule of the p-ray and the s-ray, let us notice that the light ray can propagate only in the spacetime region where the radial and zenithal effective potentials are non-positive, $V\ls{eff}(r) \le 0$ and $U\ls{eff}(\theta) \le 0$, as indicated by the geodesic equations \eqref{eq:formalism.geodesic}. 
Hence, the $\theta$-coordinate $\theta\ls{ng}(\eta)$ of the null geodesic is confined to the interval $0 < \theta\ls{ng}(\eta) < \pi$ (positions on the spin axis $\theta = 0$ and $\pi$ are excluded) for non-zero toroidal impact parameter $b \neq 0$, because $U\ls{eff} \to +\infty$ as $\theta \to 0$ and $\pi$ for $b \neq 0$. 
Therefore, for the case $b \neq 0$, the winding number of the null geodesic around the BH can be counted by the $\varphi$-coordinate $\varphi\ls{ng}(\eta)$ of the null geodesic. 
We define the winding number $W$ of the null geodesic as the integer given by
\eqb
\label{eq:formalism.W}
 W \defeq 
 \begin{cases}
 \text{Positive Integer}  & \text{for $(2W-1)\pi < \delta\varphi \le (2W+1) \pi$}
 \\
 0  & \text{for $-\pi \le \delta\varphi \le \pi$}
 \\
 \text{Negative Integer} & \text{for $(2W-1) \pi \le \delta\varphi < (2W+1) \pi$}
 \end{cases}\,,
\eqe
where $\delta\varphi \defeq \varphi\ls{ng}(\eta\ls{obs})-\varphi\ls{emi}$. 
Using this winding number, we offer the selection rule for the p-ray and the s-ray as follows: 
\begin{description}
\item[Selection rule for the p-ray and the s-ray: ]
For $b \neq 0$, our selection rule consists of 2 parts.
\begin{itemize}
\item
The p-ray is the null geodesic of the winding number $W = 0$. 
There exists only 1 p-ray, once the emission event $x\ls{emi}^\mu$ and the observation position $(r\ls{obs},\theta\ls{obs},\varphi\ls{obs})$ are specified. 
\item 
Collect the null geodesics of the winding number $W=1$ and $-1$. 
Then, the s-ray is the null geodesic of the earliest observation time among them. 
\end{itemize}
\end{description}
Also, for $b=0$, we can define the winding number $W^\prime$ in the $\theta$-direction by the same form as \eqref{eq:formalism.W}. 
Then, the p-ray may be the null geodesic of $W=0$ and $W^\prime = 0$. 
The s-ray may be the null geodesic of $W = \pm 1$ or $W^\prime = \pm 1$ and the earliest observation time. 
However, since the impact parameters $(b,q)$ in our numerical procedure are not the input parameters but the parameters determined from the initial 1-form as explained in Appendix \ref{app:ic}, the case $b=0$ has not occurred so far in our numerical calculations. 
Hence, we focus on the case $b \neq 0$ in the following discussion.

Let us note again that, in our numerical calculation, the p-ray and/or the s-ray may be omitted in our numerical setup of the initial 1-forms. 
If the true s-ray has not been obtained in the set of numerical solutions for given $x\ls{emi}^\mu$ and $(r\ls{obs},\theta\ls{obs},\varphi\ls{obs})$, then the null geodesic of the earliest observation time among the numerical solutions of winding number $W= \pm 1$ is not the true s-ray. 
Therefore, we need a supplemental rule to check whether such a null geodesic selected by the above-mentioned rule is the true s-ray or not. 
As far as we have searched numerically the solutions of null geodesic equations \eqref{eq:formalism.geodesic-hamilton} and the Jacobi equations \eqref{eq:formalism.jacobi}, we have found the following rule:
\begin{description}
\item[Supplemental rule for selecting the s-ray: ]
Among numerical solutions of the winding number $W = \pm 1$, the s-ray is the ray that passes through only 1 caustic before reaching the observer. 
Here, as explained at the end of Sect.\ref{app:area.evolution} of Appendix \ref{app:area}, the caustic is the spacetime point at which the cross-sectional area of the null geodesic bundle becomes 0. 
\end{description}
Furthermore, we have found numerically that the p-ray passes through no caustic. 
The example of this statement is shown in Fig.\ref{fig:rays.example}. 
We find in Fig.\ref{fig:rays.example} that the p-ray passes through no caustic and the s-ray passes through only 1 caustic.

\subsection{Steps of numerical calculation of $\delay$ and $\ratio$}
\label{sec:formalism.step}

Combining the discussions given so far, our numerical calculation consists of the following steps:
\begin{description}
\item[Step 1 (our calculation): ] 
Specify the values of the BH parameters $M$ and $\chi$. 
Also, specify the emission event $x\ls{emi}^\mu$, the velocity $u\ls{emi}^\mu$, and the intrinsic specific intensity $I\ls{emi}(\nu\ls{emi})$ of the source of light rays. 
\item[Step 2 (our calculation): ]
Consider the sphere of radius $r\ls{obs}$, on which the point is described by $(\theta\ls{obs},\varphi\ls{obs})$. 
Then, create the set of observers (detectors) as the grid points on the sphere. 
We assume the velocity $u\ls{obs}^\mu$ of each observer is given by Eq.\eqref{eq:formalism.uobs}. 
\item[Step 3 (our calculation): ] 
Create the set of the values of the initial direction angles $(\alpha\ls{emi},\beta\ls{emi})$ of $k\ls{emi}^\mu$ as the grid points on the parameter region, $0 \le \alpha\ls{emi} \le \pi$ and $0 \le \beta\ls{emi} < 2 \pi$, where the definition of $(\alpha\ls{emi},\beta\ls{emi})$ is given in Appendix \ref{app:ic}. 
\item[Step 4 (our calculation): ] 
For each value of $(\alpha\ls{emi},\beta\ls{emi})$ created in the previous step, calculate the components of the initial 1-form $k\ls{emi\,\mu}$ by the procedure given in Sect.\ref{app:ic.1form}. 
Further, check whether the light ray of the given initial 1-form is to be absorbed eventually by the BH or not by following the selection rule of the initial 1-form given in Sect.\ref{app:ic.selection}. 
\item[Step 5 (our calculation): ] 
Solve the null geodesic equations \eqref{eq:formalism.geodesic-hamilton} for the initial 1-forms that are not absorbed by the BH. 
Those solutions of the null geodesics arrive at different points on the observation sphere of radius $r\ls{obs}$. 
Then, for each null geodesic, the nearest grid point on the observation sphere, which is created in step~2, is regarded as the position of the observer who detects the null geodesic; 
i.e., the relative location among the BH, source and observer is (approximately) determined for each null geodesic.
\item[Step 6 (our calculation): ]
Count the winding number $W$ by Eq.\eqref{eq:formalism.W} for all null geodesics obtained in the previous step.
Then, for each observer on the observation sphere, there can exist some null geodesics that possess the same winding number $W$. 
Among such null geodesics of the same value of $W$, let us select the one that arrives at the nearest point to the observer on the observation sphere, and delete the others of the same $W$ from the numerical data. 
\item[Step 7 (our calculation): ]
The null geodesics selected in the previous step are regraded as the representative null geodesics of null geodesic bundles. 
Calculate the observed specific flux $F\ls{obs}$ of each null geodesic bundle by the formula \eqref{eq:formalism.Fobs} under the assumption \eqref{eq:formalism.assumption-Aemi}, where the procedure for calculating the area-transfer coefficient $\atc$ is given in Sect.\ref{app:area.step}. 
Also, during the calculation of $F\ls{obs}$, count the number of zeros of $\det \widetilde{J}(\eta)$ given from Eq.\eqref{eq:area.jacobi-matrix} on each null geodesic bundle, which is the number of caustics on the bundle as explained at the end of Sect.\ref{app:area.evolution}. 
\item[Step 8 (our calculation): ]
For each observer on the observation sphere, search the null geodesics detected by the observer for the p-ray and the s-ray by following the selection rules give in Sect.\ref{sec:formalism.ray}. 
Then, if the p-ray and the s-ray are found for the given observer, the time delay $\delay$ and the flux ratio $\ratio$ are obtained by definition \eqref{eq:principle.observable}. 
\end{description}
If the grid points on the parameter plane of $(\alpha\ls{em},\beta\ls{emi})$ are not well prepared in the step~3, then the p-ray and/or s-ray are not found at some observation points in the step~8.

\section{Numerical Results and Potential Detectability of $\delay$ and $\ratio$}
\label{sec:result}

This section is for a discussion of the potential detectability of our observable quantities $\delay$ and $\ratio$. 
After summarizing an estimation of telescope capability, we show some results of our numerical calculations performed by following the procedure given in Sect.\ref{sec:formalism.step}. 
Comparison of telescope capability and the numerical results implies that, if the assumptions introduced in Sect.\ref{sec:principle.assumption} hold, then our observational quantities $\delay$ and $\ratio$ can, in principle, be measured by present or near future telescope capability. 
The cases that modify some of our assumptions will be discussed in other papers.

\subsection{Example of telescope capability}
\label{sec:result.telescope}

As explained in Sect.\ref{sec:principle.assumption}, we assume a transparent environment around the BH at least in the frequency band of observation, and a few candidates for the BH with such a transparent environment are at present recognized by radio observations~\cite{ref:bower+5.2004,ref:doeleman+27.2008,ref:oka+3.2016,ref:shen+4.2005}. 
Then, as an example of a telescope, let us consider the radio telescope of 34\,m diameter operated by the Space-Time Measurement Group at the National Institute of Information and Communications Technology (NICT), Japan. 
In general, the signal-to-noise ratio $R\ls{sn}$ of a radio telescope is given by \cite{ref:thompson+2.2004}
\eqb
\label{eq:result.Rsn}
 R\ls{sn} =
 \dfrac{F\ls{obs}}{F\ls{sefd}}\sqrt{2\, \delta B \,\delta t} \,,
\eqe
where $F\ls{obs}$ is the specific flux of the signal received by the telescope, and $F\ls{sefd}/\sqrt{2\,\delta B\,\delta t}$ is the total noise of the telescope, where $\delta t$ is the duration of the observation time, $\delta B$ is the band-width of observation frequency, and $F\ls{sefd}$ is the \emph{system equivalent flux density}, which measures the system noise of the telescope in the dimensions of specific flux. 
For the radio telescope of NICT, the band-width is $\delta B \simeq 1024$\,MHz, and the system noise is $F\ls{sefd} \simeq 300$\,Jy, where 1\,Jy $= 10^{-26}$\,W/m$^2$Hz is the unit of specific flux.

Here, as an example of a target BH, consider the massive BH candidate at the center of our galaxy, Sgr A$^\ast$ of mass $M\ls{SgrA^\ast} \simeq 4\times 10^6 M_{\odot}$~\cite{ref:bower+5.2004,ref:doeleman+27.2008,ref:shen+4.2005}. 
The Newtonian (Keplerian) dynamical time scale $t\ls{dyn}$ near the horizon radius of Sgr A$^\ast$ is $t\ls{dyn} \defeq \sqrt{r\ls{SgrA^\ast}^3/G M\ls{SgrA^\ast}} \simeq 60$ sec, where $r\ls{SgrA^\ast} = 2 G M\ls{SgrA^\ast}/c^2$. 
Then, let us assume a short duration of emission is $\delta t = t\ls{dyn}/100 \simeq 0.6$ sec, and the criterion for signal detection by telescope is $R\ls{sn} > 5$. 
Note that, because no precise and high resolution observation in the vicinity of the BH horizon has been performed, we do not have relevant observational data for estimating $\delta t$. 
Therefore our assumption $\delta t = t\ls{dyn}/100$ is a simple assumption that should be investigated properly in future studies; however, we expect that 1\% of $t\ls{dyn}$ may not be bad as a short duration emission near the BH horizon. 
On the other hand, the criterion for signal detection, $R\ls{sn} > 5$, is consistent with real astronomical radio observations. 
Then, under the above assumptions, the signal flux $F\ls{obs}$ detectable by the NICT telescope should satisfy
\eqb
\label{eq:result.Fobs-detectable}
 F\ls{obs} >
 \dfrac{5 \times 300}{\sqrt{2\times 1024 \times 10^6 \times 0.6}} \simeq 0.04\,{\rm Jy} \,.
\eqe

Note that a typical observed radio flux $F\ls{SgrA^\ast}$ coming from Sgr A$^\ast$ is $F\ls{SgrA^\ast} \sim 0.1$ Jy~\cite{ref:doeleman+27.2008}, which comes from a region with an approximate size of a few $r\ls{SgrA^\ast}$. 
Then, in order to estimate the value of $\ratio$ that is detectable by the NICT telescope, let us consider 2 cases. 
One is that the flux of the p-ray is stronger than that of the s-ray ($F\ls{obs(p)} > F\ls{obs(s)}$), and the other is the inverse case ($F\ls{obs(p)} < F\ls{obs(s)}$). 
Note that, as mentioned in Sect.\ref{sec:intro}, our setup introduced in Sect.\ref{sec:principle.assumption} is similar to the setup considered by Cunningham and Bardeen~\cite{ref:cunningham+1.1973}, which considered a star on a circular orbit on the equatorial plane of an extreme Kerr BH. 
Cunningham and Bardeen~\cite{ref:cunningham+1.1973} had already shown that the brightness of the primary image can be stronger and weaker than the brightness of the secondary image, due to frame-dragging by the Kerr BH, and the beaming and kinetic Doppler effects on the light rays emitted by the source star. 
Therefore, in our situation where the source is not a star (radiating lights continuously) but a short duration emission, it may be expected that both cases $F\ls{obs(p)} > F\ls{obs(s)}$ and $F\ls{obs(p)} < F\ls{obs(s)}$ can be found. 
Indeed, at the end of this section, it will be shown by our numerical calculation that these 2 cases are possible for a Kerr BH but not for a Schwarzschild BH.

For the case $F\ls{obs(p)} > F\ls{obs(s)}$, we can set $F\ls{obs(p)} = F\ls{SgrA^\ast} \sim 0.1$\,Jy. 
In order to let the s-ray be detectable, its specific flux $F\ls{obs(s)}$ needs to satisfy condition \eqref{eq:result.Fobs-detectable}. 
Then, we obtain the condition on the flux ratio detectable by the NICT telescope,
\seqb
\eqb
 \ratio \defeq \dfrac{F\ls{obs(s)}}{F\ls{obs(p)}} > 0.4 \,.
\eqe
And, for the case $F\ls{obs(p)} < F\ls{obs(s)}$, we can set $F\ls{obs(s)} = F\ls{SgrA^\ast} \sim 0.1$\,Jy. 
In this case, the detectable p-ray needs to satisfy condition \eqref{eq:result.Fobs-detectable}, and we obtain a condition for the detectable flux ratio:
\eqb
 \ratio \defeq \dfrac{F\ls{obs(s)}}{F\ls{obs(p)}} < \dfrac{1}{0.4} = 2.5 \,.
\eqe
\seqe
Hence, although some assumptions are introduced, the above estimation with Sgr~A$^\ast$ seems to suggest that present or near future telescopes can detect the p-ray and the s-ray from Sgr~A$^\ast$, if there occur some emission events near the BH candidate that result in the observed flux ratio in the interval,
\eqb
\label{eq:result.Robs-interval}
 -0.4 < \log_{10} \ratio < 0.4 \,.
\eqe
The interesting issue is whether the general relativity permits the flux ratio $\ratio$ within this interval.

\subsection{Some results of the numerical estimation of $\delay$ and $\ratio$}
\label{sec:result.result}

To discuss whether or not the flux ratio $\ratio$ can take values in the interval~\eqref{eq:result.Robs-interval}, we estimate the values of $\delay$ and $\ratio$ numerically. 
The procedure is given in Sect.\ref{sec:formalism.step}, and our numerical results are obtained with Mathematica version 10. 
The numerical results shown in this section are some typical results obtained with following values of parameters (see Fig.\ref{fig:setup-numerical}):

\begin{figure}[t]
 \begin{center}
 \includegraphics[scale=0.5]{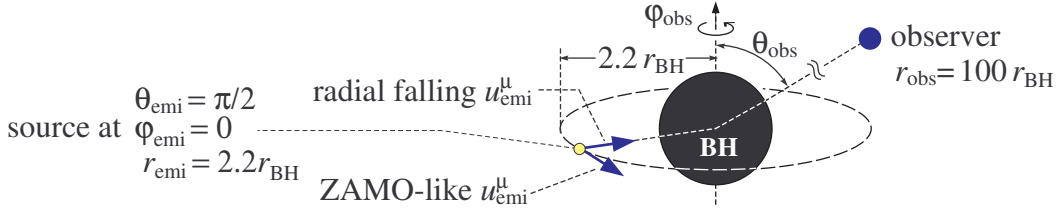}
 \end{center}
\caption{Setup for the numerical results shown in this section.}
\label{fig:setup-numerical}
\end{figure}

\begin{itemize}
\item
The BH's mass is set to unity ($M=1$) and all quantities are calculated with this unit. 
\item
Three cases of the BH's spin parameter are calculated: $\chi = 0$, $0.3$, and $0.8$.
\item
The emission position of the source of light rays is set to $(r\ls{emi},\theta\ls{emi}) = (2.2 r\ls{BH},\pi/2)$. 
This means that the emission event is outside (but near) the ergo-surface, $r\ls{erg}(\pi/2) = 2M$.
\item
Two cases of the source's velocity are calculated, 
the radial falling case $(u\ls{emi}^t,-1,0,0)$ and the ZAMO-like case $u\ls{emi}^\mu = u\ls{zamo}^\mu$, where $u\ls{emi}^t$ is determined by $u\ls{emi}^\mu u\ls{emi \mu} = -1$ and $u\ls{zamo}^\mu$ is given in Eq.\eqref{eq:principle.uzamo}.
\item
The radial coordinate of the observer is $r\ls{obs} = 100\, r\ls{BH}$ (see Sect.\ref{sec:formalism.observer}). 
\item
Two cases of the direction angle of the BH's spin axis are calculated, $\theta\ls{obs} = 4\pi/31$ and $16\pi/31$. 
Also, some cases of the azimuthal position of the observer are calculated: $\varphi\ls{obs} = n (2\pi/N\ls{obs})$, where $n = 0, 1, \cdots , N\ls{obs}$, and $N\ls{obs} = 24$ for $\theta\ls{obs} = 4\pi/31$, and $N\ls{obs} = 60$ for $\theta\ls{obs} = 16\pi/31$.
\item
For simplicity, we suppose the observation type LE, $\ratio\us{(LE)}$, given in Eq.\eqref{eq:formalism.Robs-LE}. 
Alternatively, this can also be understood as type LD with white-noise type emission, $I\ls{emi}(\nu\ls{emi}) =$ constant, as explained at the end of Sect.\ref{sec:formalism.flux.Fobs}. 
\end{itemize}

We do not show all the numerical results of the possible combinations of the values of $\chi$, $u\ls{emi}^\mu$, $\theta\ls{obs}$, and $\varphi\ls{obs}$. 
However, we do show some typical results in order to discuss whether the flux ratio $\ratio$ can take values in the interval \eqref{eq:result.Robs-interval}. 
Further, note that, although we have tried to calculate the observable quantities $\delay$ and $\ratio$ for all values of $\varphi\ls{obs}$, we could not obtain the numerical values of $\delay$ and $\ratio$ for some values of $\varphi\ls{obs}$ because the appropriate initial condition for the geodesic equations could not be created. 
(See the comment at the end of Sect.\ref{sec:formalism.step}.) 
To obtain the values of $\delay$ and $\ratio$ for all given values of the input parameters, we may need a more sophisticated calculation procedure than the present one and a more powerful computer than the author uses at present.

\subsubsection{Radial falling source toward a BH of spin $\chi = 0$}
\label{sec:result.result.falling-0}

\begin{figure}[t]
 \begin{center}
 \includegraphics[scale=0.6]{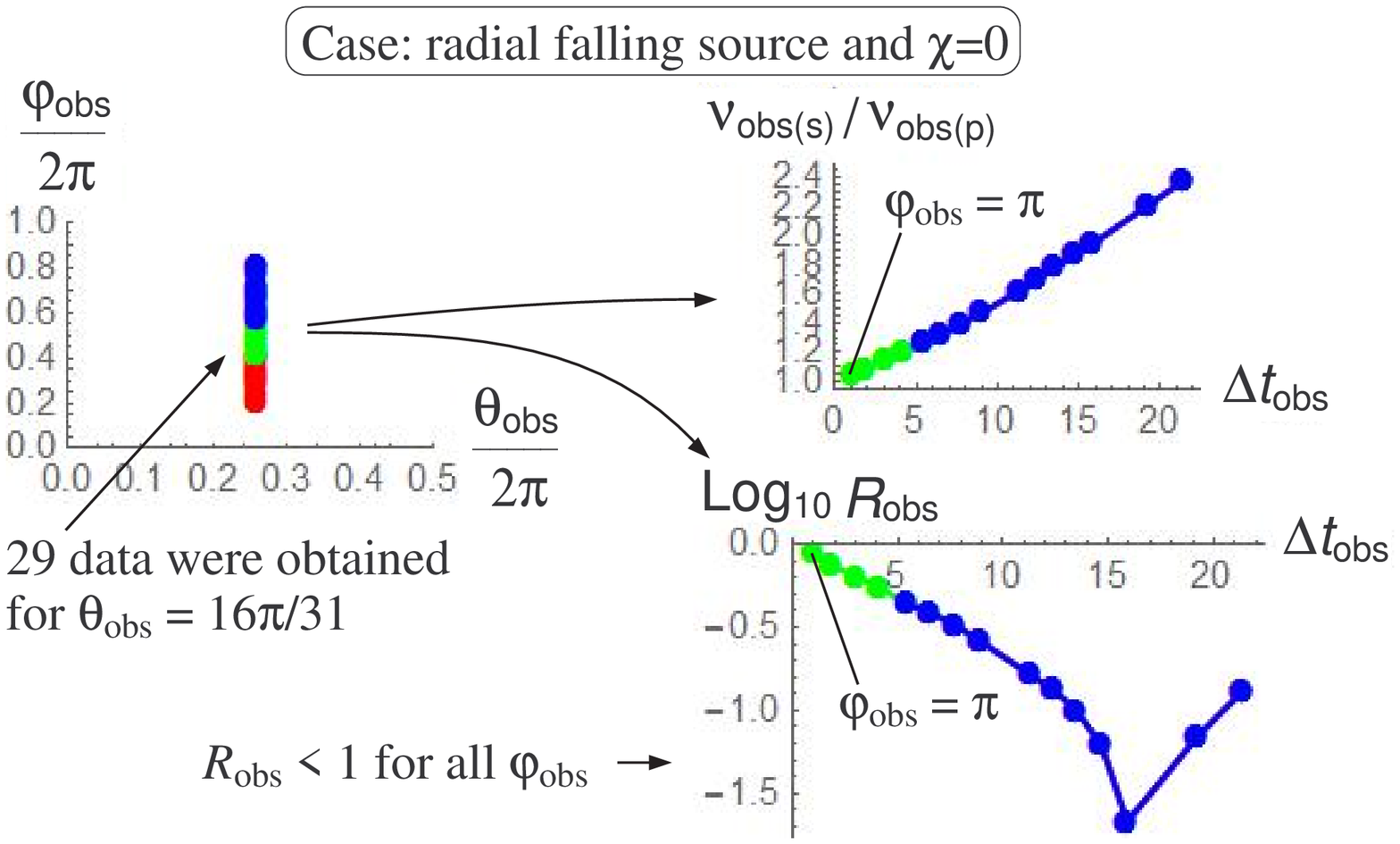}
 \end{center}
\caption{
Numerical results for $\delay$, $\ratio$, and $\nu\ls{obs(s)}/\nu\ls{obs(p)}$ for a radial falling source near a BH with $\chi = 0$. 
The numerical results with $\theta\ls{obs} = 16\pi/31$ are plotted, where $\varphi\ls{obs}$ is varied from $0$ to $2\pi$. 
However, in our numerical program, numerical calculations at some values of $\varphi\ls{obs}$ could not produce an appropriate initial value for the geodesic equations. 
The values of $\varphi\ls{obs}$ for which the numerical calculation was completed are shown in the left-hand panel. 
The colors of the points in each of the panels denote the variation in the value of $\varphi\ls{obs}$. 
In the right-hand panels, the data points corresponding to $\varphi\ls{obs} < \pi$ do not appear since those data points degenerate to the data points corresponding to $\varphi\ls{obs} \ge \pi$ in the case $\chi = 0$.
}
\label{fig:result-1}
\end{figure}

Figure~\ref{fig:result-1} shows numerical results for $\delay$, $\ratio$, and $\nu\ls{obs(s)}/\nu\ls{obs(p)}$ with the parameter values listed in the figure's caption. 
The source is radially falling toward the BH. 
As shown in the left-hand panel, we were able to complete the numerical calculations for 29 values of $\varphi\ls{obs}$ at $\theta\ls{obs} = 16\pi/31 (\simeq 0.516 \pi)$.

Note that (or return to here after reading Sect.\ref{sec:result.result.falling-03}), in comparison with the results in Fig.\ref{fig:result-2}, the time delay $\delay$ should extend up to $\delay \sim 30$. 
However, our numerical calculations for $\delay > 20$ were not successful (see the end of Sect.\ref{sec:formalism.step}). 
Also note that the values of $\delay$, $\ratio$ and $\nu\ls{obs(s)}/\nu\ls{obs(p)}$ for $0 < \varphi\ls{obs} \le \pi$ seem to degenerate to those for $\pi < \varphi\ls{obs} \le 2\pi$. 
This seems to be the typical situation for Schwarzschild BH case.

It can be read from the lower-right panel that the flux ratio $\ratio$ can take values in the interval \eqref{eq:result.Robs-interval}. 
However, note that, if the observed frequency of the p-ray $\nu\ls{obs(p)}$ and that of the s-ray $\nu\ls{obs(s)}$ are different so that the band-width of the telescope does not include both of them, the flux ratio $\ratio$ made of such a p-ray and s-ray cannot be detected even when $\ratio$ takes a detectable value \eqref{eq:result.Robs-interval}. 
Hence, numerical results with $\nu\ls{obs(s)}/\nu\ls{obs(p)} \simeq 1$ are desirable for expecting \emph{safe} detectability of both of the p-ray and the s-ray by 1 telescope. 
From the upper-right panel, we find that a flux ratio $\ratio$ in the interval \eqref{eq:result.Robs-interval}, together with a frequency ratio around unity $\nu\ls{obs(s)}/\nu\ls{obs(p)} \simeq 1$, is realized for $\varphi\ls{obs} \simeq \pi$.

\subsubsection{Radial falling source toward a BH of spin $\chi = 0.3$}
\label{sec:result.result.falling-03}

\begin{figure}[t]
 \begin{center}
 \includegraphics[scale=0.6]{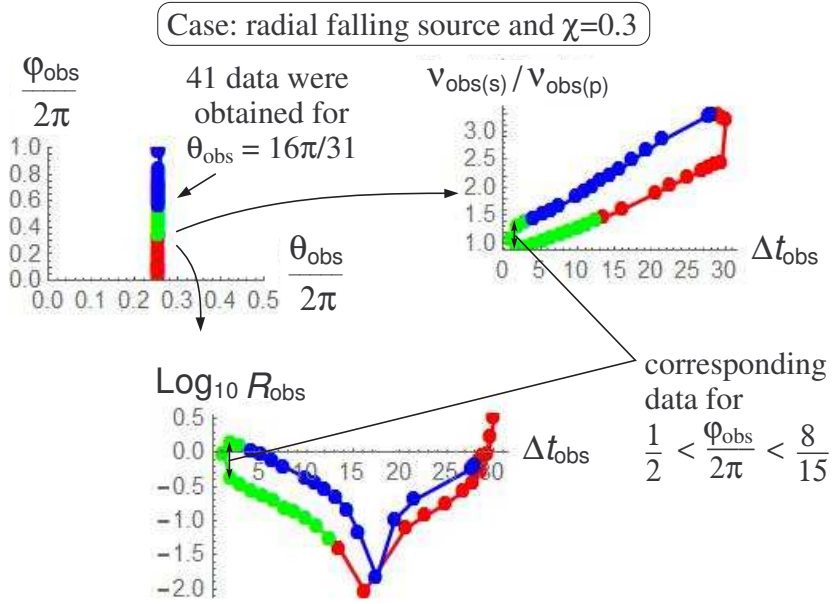}
 \end{center}
\caption{
Numerical results for $\delay$, $\ratio$, and $\nu\ls{obs(s)}/\nu\ls{obs(p)}$ for a radial falling source near a BH with $\chi = 0.3$.
The numerical results with $\theta\ls{obs} = 16\pi/31$ are plotted, where $\varphi\ls{obs}$ is varied from $0$ to $2\pi$, but some values of $\varphi\ls{obs}$ could not produce an appropriate initial value for the geodesic equations. 
The values of $\varphi\ls{obs}$ for which the numerical calculation was completed are shown in the upper-left panel. 
The colors of the points denote the variation of $\varphi\ls{obs}$.
}
\label{fig:result-2}
\end{figure}

Figure~\ref{fig:result-2} shows a modification of Fig.\ref{fig:result-1} by increasing the BH's spin from $\chi = 0$ to $0.3$ while keeping the other parameters fixed at the same value as those in Fig.\ref{fig:result-1}. 
The source is radially falling toward BH. 
As shown in the upper-left-hand panel, we could complete the numerical calculations for 41 values of $\varphi\ls{obs}$ at $\theta\ls{obs} = 16\pi/31 (\simeq 0.516 \pi)$. 
It can be read from the upper-right and lower panels that a flux ratio $\ratio$ in the interval \eqref{eq:result.Robs-interval}, together with a frequency ratio around unity $\nu\ls{obs(s)}/\nu\ls{obs(p)} \simeq 1$, is realized for $\pi < \varphi\ls{obs} < 16\pi/15 (\simeq 1.066 \pi)$.

\subsubsection{Radial falling source toward a BH of spin $\chi = 0.8$}
\label{sec:result.result.falling-08}

\begin{figure}[t]
 \begin{center}
 \includegraphics[scale=0.6]{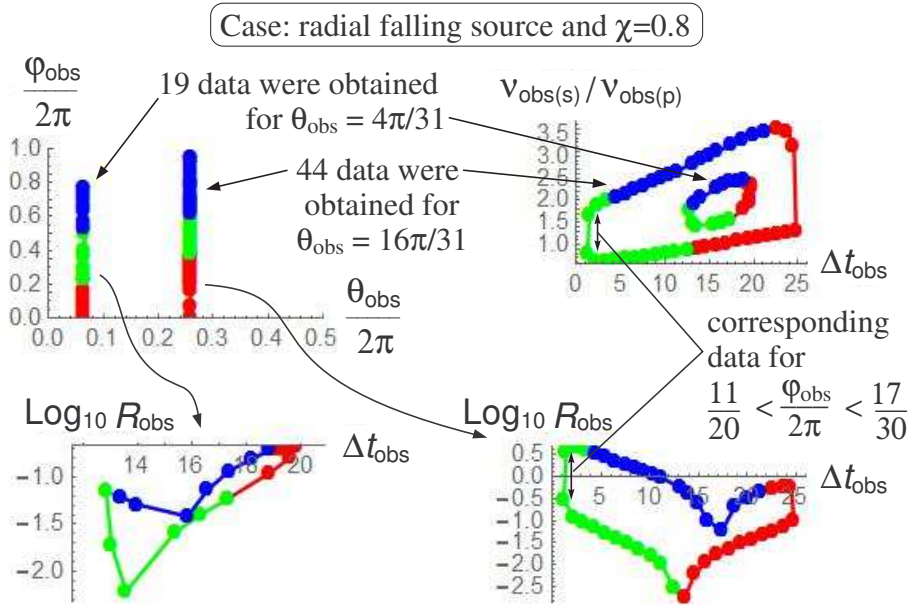}
 \end{center}
\caption{
Numerical results for $\delay$, $\ratio$, and $\nu\ls{obs(s)}/\nu\ls{obs(p)}$ for a radial falling source near a BH with $\chi = 0.8$. 
Two cases, $\theta\ls{obs} = 4\pi/31$ and $16\pi/31$, are plotted. 
For each case, $\varphi\ls{obs}$ is varied from $0$ to $2\pi$, but some values of $\varphi\ls{obs}$ could not produce an appropriate initial value for the geodesic equations. 
The values $(\theta\ls{obs},\varphi\ls{obs})$ for which the numerical calculation was completed are shown in the upper-left panel. 
The colors of the points denote the variation of $\varphi\ls{obs}$.
}
\label{fig:result-3}
\end{figure}

Figure~\ref{fig:result-3} shows a modification of Fig.\ref{fig:result-1} by increasing the BH's spin from $\chi = 0$ to $0.8$ while keeping the other parameters fixed at the same value as those in Fig.\ref{fig:result-1} and adding another parameter value of the direction angle of the BH's spin. 
The source is radially falling toward the BH. 
As shown in the upper-left panel, we were able to complete the numerical calculations for 19 values of $\varphi\ls{obs}$ at $\theta\ls{obs} = 4\pi/31 (\simeq 0.129 \pi)$, and 44 values of $\varphi\ls{obs}$ at $\theta\ls{obs} = 16\pi/31 (\simeq 0.516 \pi)$.
It can be read from the upper-right and lower panels that a flux ratio $\ratio$ in the interval \eqref{eq:result.Robs-interval}, together with a frequency ratio around unity $\nu\ls{obs(s)}/\nu\ls{obs(p)} \simeq 1$, is realized for $\theta\ls{obs} = 16\pi/31 (\simeq 0.516 \pi)$ and $1.1\pi < \varphi\ls{obs} < 17\pi/15 (\simeq 1.133 \pi)$.

Note that, in comparison with the results in Fig.\ref{fig:result-2}, the time delay $\delay$ should extend up to $\delay \sim 30$. 
However, our numerical calculations for $\delay > 25$ were not successful (see the end of Sect.\ref{sec:formalism.step}).

\subsubsection{ZAMO-like source around a BH of $\chi = 0.8$}
\label{sec:result.result.zamo-08}

\begin{figure}[t]
 \begin{center}
 \includegraphics[scale=0.6]{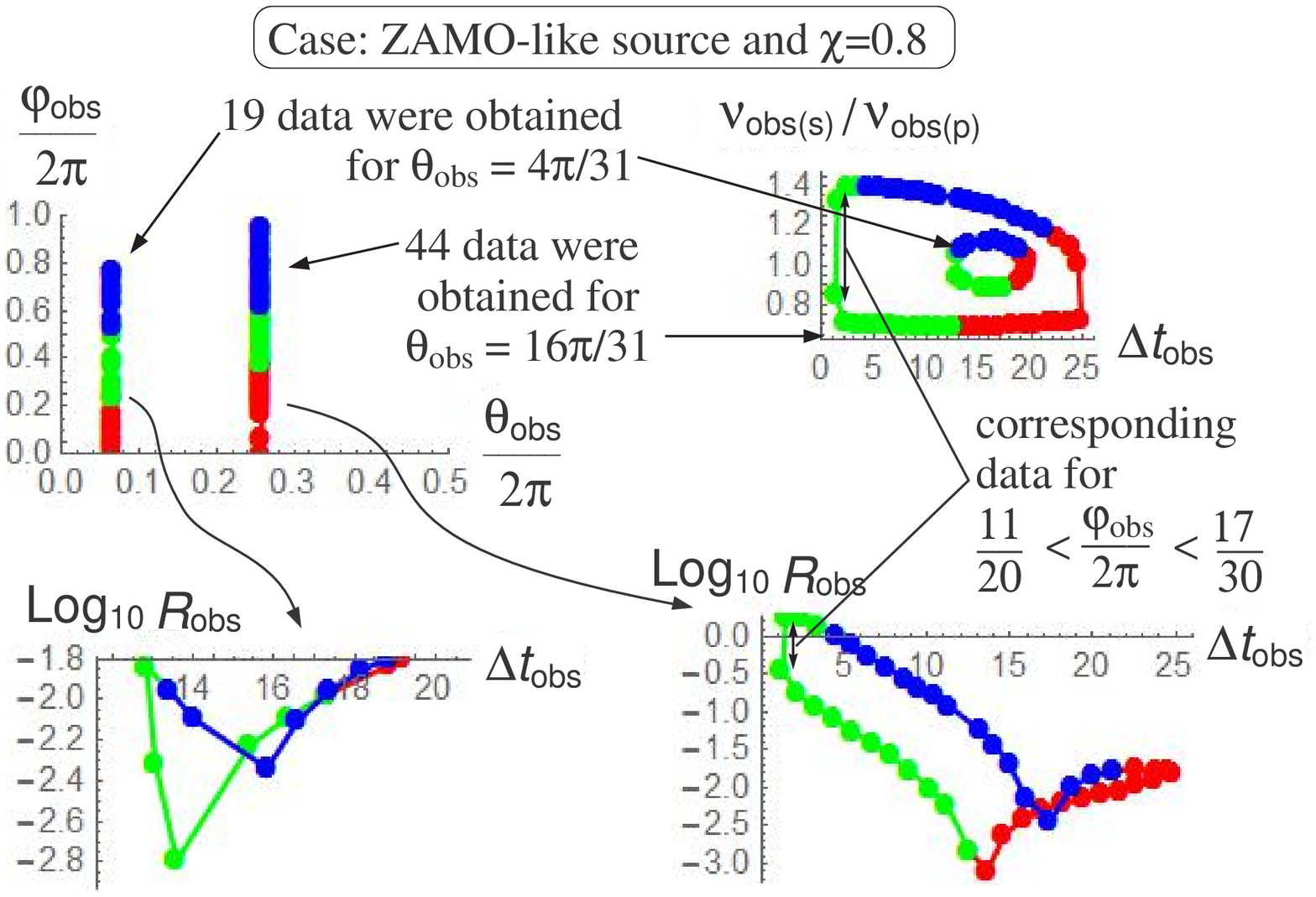}
 \end{center}
\caption{Numerical results for $\delay$, $\ratio$, and $\nu\ls{obs(s)}/\nu\ls{obs(p)}$ for a ZAMO-like source near a BH with $\chi = 0.8$.
The other parameters are the same as those in Fig.\ref{fig:result-3}. 
The values of $\ratio$ and $\nu\ls{obs(s)}/\nu\ls{obs(p)}$ are changed from those in Fig.\ref{fig:result-3} by replacing the source's velocity. 
The colors of the points denote the variation of $\varphi\ls{obs}$.
}
\label{fig:result-4}
\end{figure}

Figure~\ref{fig:result-4} shows a modification of Fig.\ref{fig:result-3} by replacing the source's velocity from the radial falling case to the ZAMO-like case while keeping the other parameters fixed at the same values as those in Fig.\ref{fig:result-3}. 
Since the same numerical data for the solution of the null geodesic equations are used in Fig.\ref{fig:result-3} and~\ref{fig:result-4}, the number of data for each value of $\theta\ls{obs}$ is the same as indicated in Fig.\ref{fig:result-3}. 
But the behaviors of $\ratio$ and $\nu\ls{obs(s)}/\nu\ls{obs(p)}$ in this case are somewhat different from those in Fig.\ref{fig:result-3}, because the beaming effect and the kinetic Doppler effect are different, due to the difference in the source's velocity. 
It can be read from the lower and upper-right panels that a flux ratio $\ratio$ in the interval \eqref{eq:result.Robs-interval}, together with a frequency ratio around unity $\nu\ls{obs(s)}/\nu\ls{obs(p)} \simeq 1$, is realized for $\theta\ls{obs} = 16\pi/31 (\simeq 0.516 \pi)$ and $1.1\pi < \varphi\ls{obs} < 17\pi/15 (\simeq 1.133 \pi)$.

\subsubsection{Accuracy errors in numerical calculations of $\delay$ and $\ratio$}
\label{sec:result.result.error}

We may be able to estimate the errors in our numerical calculations of $\delay$ and $\ratio$ that arise from a numerical uncertainty in the position of the observer in our numerical procedure. 
In our setup, given at the beginning of this Sect.\ref{sec:result.result}, the angular uncertainty $(\delta\ls{obs},\delta\varphi\ls{obs})$ in the position of the observer is $\delta\theta\ls{obs} \simeq \pi/31 \sim 0.1$ and $\delta\varphi\ls{obs} \simeq \pi/30 \sim 0.1$ for $\theta\ls{obs} = 16\pi/31$. 
This uncertainty corresponds to the angular separation between neighboring grid points on the observation sphere of radius $r\ls{obs}$, which are prepared in the step~2 of our numerical procedure given in Sect.\ref{sec:formalism.step}. 
The length size $\delta l$ of the position uncertainty of the observer is $\delta l \sim r\ls{obs} \delta\theta\ls{obs} \simeq 0.1\, r\ls{obs}$.

Let us estimate the error $\delta\delay$ in our numerical calculation of the time delay $\delay$. 
The direction of the observer's position uncertainty $\delta l$ is tangent to the observation sphere of radius $r\ls{obs}$. 
Also, $\delay$ can be roughly estimated by the spatial length of a path on which a light ray propagates, $\delay \sim r\ls{obs}$. 
Hence, the numerical error $\delta\delay$ may be estimated by considering a right triangle whose legs are the sides of length $r\ls{obs}$ and $\delta l$. 
(The angle between these legs is the right angle.) 
Our estimation is $\delta\delay \simeq \sqrt{r\ls{obs}^2 + \delta l^2} - r\ls{obs} \simeq (1/2) (\delta l)^2/r\ls{obs} \sim 0.005\,r\ls{obs}$. 
The relative error is $\delta\delay/\delay \sim 0.005$. 
Further, including another error in our numerical setup estimated in Sect.\ref{sec:formalism.observer}, we expect that the total numerical error in the accuracy of $\delay$ in our numerical calculation may be several percent.

Next, the error $\delta\ratio$ in our numerical calculation of the specific flux ratio $\ratio$ may be estimated by the uncertainty of the curvature tensor $R^\mu_{\nu\alpha\beta}$, because the Jacobi matrix, which constitutes the area-transfer coefficient $\atc$ in Eq.\eqref{eq:area.Aobs-C}, is determined by $R^\mu_{\nu\alpha\beta}$ due to the Jacobi equation~\eqref{eq:area.jacobi-matrix}. 
And note that, since the curvature is roughly the second derivative of the metric tensor, the curvature $R^\mu_{\nu\alpha\beta}$ can be roughly estimated as $R^\mu_{\nu\alpha\beta} \sim r\ls{obs}^{-2}$. 
Then, the numerical error $\delta R^\mu_{\nu\alpha\beta}$ is estimated as $\delta R^\mu_{\nu\alpha\beta} \sim [\,r\ls{obs}^{-2} - (r\ls{obs} + \delta\delay)^{-2}\,] \sim 2\,\delta\delay/r\ls{obs} \sim 0.01$. 
This may be the origin of the numerical error in the specific flux of the p-ray $F\ls{obs(p)}$ and that of the s-ray $F\ls{obs(s)}$. 
Hence, because of $\ratio \defeq F\ls{obs(s)}/F\ls{obs(p)}$, we expect $\delta\ratio \sim \delta R^\mu_{\nu\alpha\beta} \sim 0.01$. 
Further, including another error in our numerical setup estimated in Sect.\ref{sec:formalism.observer}, we expect that the total numerical error in the accuracy of $\ratio$ in our numerical calculation may be several percent.

These errors in $\delay$ and $\ratio$ do not seriously affect our conclusion derived from the numerical results shown in Fig.\ref{fig:result-1} to Fig.\ref{fig:result-4}.

\subsubsection{Summary of numerical results}
\label{sec:result.result.summary}

Numerical results similar to those shown in Fig.\ref{fig:result-1} to Fig.\ref{fig:result-4} have been obtained for various parameters, as far as the author has checked. 
Hence, \emph{it may be reasonable to expect detectability of the observable quantities $\delay$ and $\ratio$ by present or near future telescope capability, at least, for the case that the source of light rays and the observer are located near the equatorial plane of a BH's spin ($\theta\ls{obs} \simeq \theta\ls{emi} = 0.5\pi$) and the emission event is behind the BH seen from the observer ($\varphi\ls{obs} \simeq \pi$)}.

Let us note that both cases $\log_{10}\ratio > 0 \, (F\ls{obs(s)} > F\ls{obs(p)})$ and $\log_{10}\ratio < 0\, (F\ls{obs(s)} < F\ls{obs(p)})$ appear for Kerr BH cases, Fig.\ref{fig:result-2} to Fig.\ref{fig:result-4}. 
As mentioned in Sect.\ref{sec:result.telescope}, because our setup is similar to the setup considered in Cunningham and Bardeen~\cite{ref:cunningham+1.1973}, which predicted that the brightness of the primary image of a star orbiting a Kerr BH can be stronger and weaker than that of the secondary image, it was expected that our numerical results would show both $\ratio > 1$ and $\ratio < 1$ cases. 
This expectation is supported by our numerical results. 
Further note that the case $F\ls{obs(s)} > F\ls{obs(p)}$ does not appear in the Schwarzschild BH case, Fig.\ref{fig:result-1}. 
Therefore, the case $F\ls{obs(s)} > F\ls{obs(p)}$ may be mainly due to the frame-dragging effect of a spinning BH. 
More detailed numerical study will be reported in another paper.

We find from all presented figures, Figs.\ref{fig:result-1} to~\ref{fig:result-4}, that the flux ratio $\ratio$ has local minima about $\delay \sim 15 GM/c^3$ in the plots of $\log_{10}\ratio$ versus $\delay$. 
This may reflect some universal property of BH spacetime, since our numerical results in Figs.\ref{fig:result-1} to~\ref{fig:result-4} include some cases of BH spin $\chi$, its direction angle $\theta\ls{obs}$, and source velocity $u\ls{obs}^\mu$. 
However, we could not specify physical reasons for the appearance of local minima in $\ratio$-$\delay$ relation. 
This behavior of $\ratio$ remains an open issue for future works.

\section{Summary and discussions}
\label{sec:sd}

The main theoretical proposal in this paper is in Sect.\ref{sec:principle.principle}, which is the principle to measure the mass $M$, spin parameter $\chi$, and direction angle $\theta\ls{obs}$ of BHs through observing the time delay $\delay$ and specific flux ratio $\ratio$ created by the p-ray and the s-ray. 
This principle is a method of the direct BH measurement under the definition given in Sect.\ref{sec:intro}, since $\delay$ and $\ratio$ are the quantities created by the Kerr BH lens effect (a general relativistic effect). 
And, following the numerical procedure for calculating $\delay$ and $\ratio$ constructed in Sect.\ref{sec:formalism} and Appendices \ref{app:ic} and \ref{app:area}, we showed in Sect.\ref{sec:result} the potential detectability of $\delay$ and $\ratio$ by present or near future telescope capability. 
However, since our assumptions on the source of light rays are very simple, cases of complicated source emissions are to be studied by appropriately summing the results of this paper.

The conditions assumed in this paper are described in Sect.\ref{sec:principle.assumption}, and the numerical setup is given at the beginning of Sect.\ref{sec:result.result}. 
These assumptions and related issues are summarized as follows:
\begin{description}
\item[Source of light rays: ]
We have assumed that the source of light rays is point-like and emits light rays isotropically in its comoving frame, and also that the emission duration is much shorter than a typical dynamical time scale of the system composed of the source and the BH. 
On the other hand, in astrophysical situations, not only such simple emissions, but also other complicated emissions, would occur. 
Complicated source emissions, such as spatially and temporally continuous or random emissions, will be constructed by summing appropriately some simple emissions. 
Such an extension of the source's structure is a task for future works. 
\item[Environment around the BH: ]
We have assumed a transparent environment around the BH, at least in the frequency band of observation. 
Some observational evidence of such an environment around massive BH candidates in the central region of our galaxy has been reported~\cite{ref:bower+5.2004,ref:doeleman+27.2008,ref:shen+4.2005,ref:oka+3.2016}. 
On the other hand, in astrophysical situations, it is also expected that a BH is surrounded by dense plasmas that form some opaque environment around the BH. 
Inclusion of such opaque effects in our study is also an interesting issue for future works. 
(We may give priority not to the opaque effects but to the complicated source emissions, since some observational evidence of a transparent environment around a BH is already been known.)
\item[Numerical calculation: ]
Under the numerical setup given at the beginning of Sect.\ref{sec:result.result}, our numerical procedure, which is summarized in Sect.\ref{sec:formalism.step}, could not produce some desired null geodesics, that should connect the source and observer, within the numerical error evaluated in Sect.\ref{sec:result.result.error}. 
Therefore, we need a more sophisticated numerical procedure or technique to obtain all the desired null geodesics. 
At present, the author is modifying the numerical procedure in order to obtain all the desired null geodesics. 
After completing the modification, a complicated source case will be reported in a future work. 
\end{description}

We have found 2 by-products of our numerical study. 
The first by-product, explained in Sect.\ref{sec:formalism.ray}, is that the p-ray passes through no caustics and the s-ray passes through only 1 caustic before reaching the observer. 
Note that this statement is nothing but a conjecture based on our numerical calculation, and theoretical exact proof of this statement remains to be constructed. 
The second by-product, explained at the end of Sect.\ref{sec:result}, is that a large flux ratio $\ratio > 1$ seems to occur due to the frame-dragging effect of a spinning BH. 
However, other detailed properties of $\delay$ and $\ratio$, such as the local minima of $\ratio$ at $\delay \sim 15 M$ in the $\ratio$-$\delay$ relation, can hardly be analyzed by our present numerical results shown in Sect.\ref{sec:result}. 
More detailed numerical study of $\delay$ and $\ratio$ will be reported in future works.

Let us make a comment on the significance of our principle of direct BH measurement. 
The existing famous observable quantities of the Kerr BH lens effect may be the BH shadow (see Takahashi~\cite{ref:takahashi.2004} and references therein) and the broadening of the iron line emission by an accretion disk around a BH (see Kojima~\cite{ref:kojima.1991}, Fanton et al.~\cite{ref:fanton+3.1997}, and references therein). 
Although these quantities have not been clearly detected so far, intensive observational approaches are now developing. 
While the imaging of a BH shadow needs many radio telescopes in order to compose a very-long-baseline-interferometer system, our method of measuring $(M,\chi,\theta\ls{obs})$ can be carried out, in principle, by just 1 telescope. 
And, while the broadening of the iron line depends, by definition, on the accretion disk model, our observable quantities $\delay$ and $\ratio$ do not depend so largely on the disk model. 
We expect that the combination of the BH shadow, the broadening of the iron line and our proposal will strengthen the observational study of BHs by astronomical methods.

Finally, from the viewpoint of the general relativity, it is necessary to recognize exactly what we can extract from the observation of the Kerr BH lens effect. 
Each of the BH shadow, the broadening of the iron line, and our proposal observe the light rays emitted by sources moving around the BH. 
Those light rays wind sometimes around the BH before reaching the observer. 
Here, it must be emphasized that, if the source of the light rays is located outside the so-called \emph{photon sphere} (which is a sphere of radius $r = 3M$, for a non-spinning BH, where the peak of the photon's radial potential is located), any light ray entering the inside of the photon sphere can never escape from the photon sphere but will be absorbed by the BH eventually. 
Therefore, any light ray connecting the source and the distant observer can never approach nearer than the radius of the photon sphere before reaching the observer. 
This theoretical fact implies that, even if the BH shadow, the broadening of the iron line, and our observable quantities $\delay$ and $\ratio$ are observed clearly, the direct conclusion of these observational data is never the existence of the BH horizon but the existence of the photon sphere. 
\emph{Hence, whenever we aim to observe the BH through the Kerr BH lens effect, we are faced with the theoretical issue of whether the existence of the photon sphere denotes the existence of the BH horizon.} 
At present, the final answer to this issue has not been obtained. 
However, there is some positive theoretical evidence for this issue, e.g., by Cardoso et al.~\cite{ref:cardoso+4.2014} and Saida et al.~\cite{ref:saida+3.2016} (see also references therein). 
Thus, although we still do not have complete theoretical support for the existence of the BH horizon under the existence of the photon sphere, the existence of the BH horizon seems probable if the existence of the photon sphere is shown by the observation of the Kerr BH lens effect.

\section*{Acknowledegments}

I would like to express my gratitude to some specialists in radio observation, Masato Tsuboi, Makoto Miyoshi and Kazuhiro Takefuji, and also to some specialists in theoretical physics and astronomy, Masaaki Takahashi and Yasusada Nambu.
The discussion with M.Tsuboi and M.Miyoshi was the very beginning of this research, and the estimation of the detectability of our observable quantities depends on K.Takefuji. 
The discussions with M.Takahashi and Y.Nambu gave me many hints for the theoretical formulation and numerical calculation.  I was supported by the Japan Society for the Promotion of Science (JSPS), Grant-in-Aid for Scientific Research (KAKENHI, Exploratory Research, 26610050), and partially supported by Daiko Foundation (Grant no. 9130).

\appendix
\section{Setup of the initial null 1-form $k\ls{emi\,\mu}$ for Eq.\eqref{eq:formalism.geodesic-hamilton}}
\label{app:ic}

\begin{figure}[t]
 \begin{center}
 \includegraphics[scale=0.45]{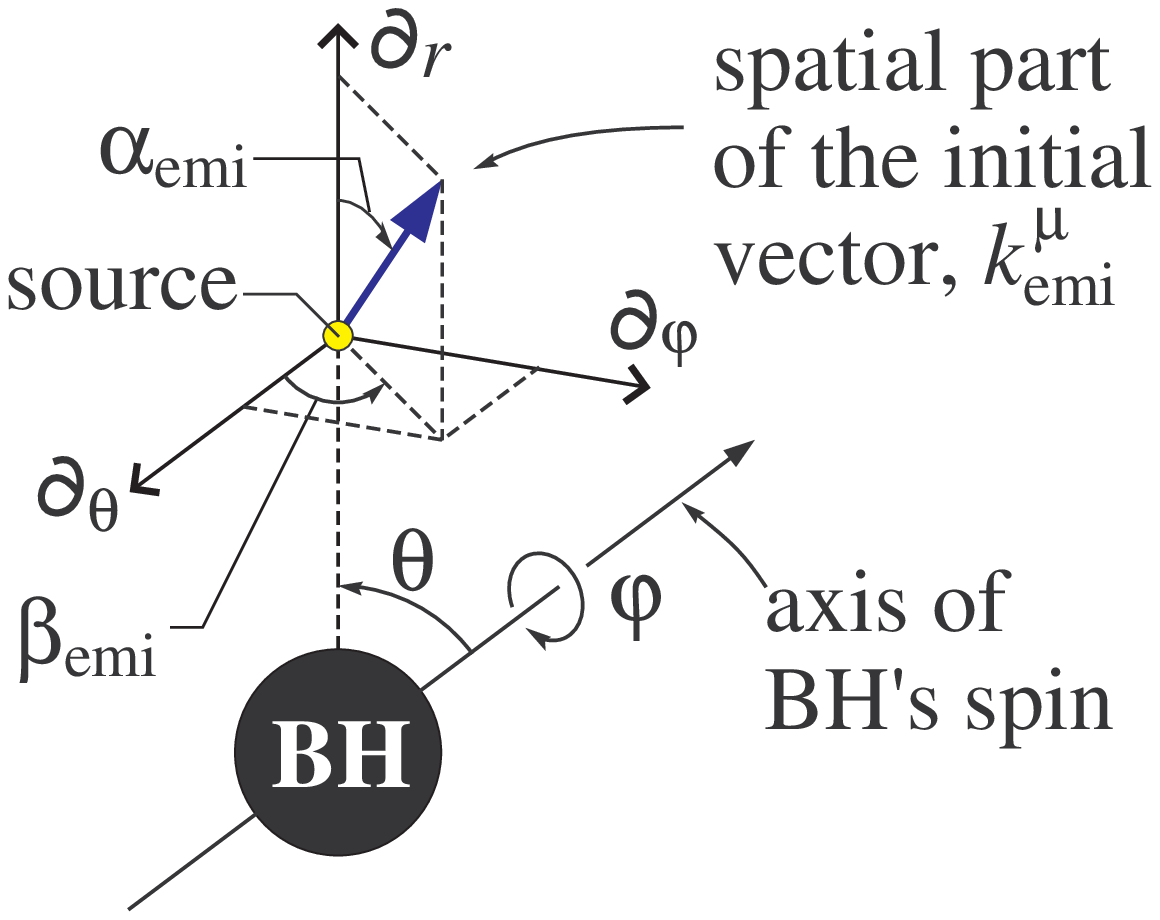}
 \end{center}
\caption{The emission angles, $\alpha\ls{emi}$ and $\beta\ls{emi}$, are specified in the spatial hypersurface $t =$ constant. 
In our numerical calculation, these angles are the input values determining the values of the initial 1-form $k\ls{emi\,\mu}$ and the impact parameters $(b,q)$. 
}
\label{fig:emission}
\end{figure}

In our numerical calculation, we need the initial 1-form $k\ls{emi\,\mu}$ to solve Hamilton's equations of the null geodesic \eqref{eq:formalism.geodesic-hamilton}. 
This Appendix explains our setup of the initial null 1-form $k\ls{emi\,\mu}$. 
The outline of our setup of $k\ls{emi\,\mu}$ consists of the following parts:
\begin{itemize}
\item
Specify the direction angles $(\alpha\ls{emi},\beta\ls{emi})$ of the initial vector $k\ls{emi}^\mu$ seen from the source as shown in Fig.\ref{fig:emission}. 
Here, $\alpha\ls{emi}$ is the angle between the basis vector $\partial_r$ and the initial vector $k\ls{emi}^\mu$, and $\beta\ls{emi}$ is the angle between the basis vector $\partial_\theta$ and the projection of $k\ls{emi}^\mu$ onto the spatial surface spanned by $\partial_\theta$ and $\partial_\varphi$. 
Then, calculate the components of the initial vector $k\ls{emi}^\mu$ from the given $(\alpha\ls{emi},\beta\ls{emi})$, which gives the initial 1-form $k\ls{emi\,\mu} = g\ls{emi\,\mu\nu}k\ls{emi}^\nu$.
\item
Before carrying out the numerical integration of Eq.\eqref{eq:formalism.geodesic-hamilton}, we judge whether the light ray emitted in the direction $(\alpha\ls{emi},\beta\ls{emi})$ is to be absorbed eventually by the BH, where the criterion of judgment is given by the effective potentials \eqref{eq:formalism.effectivepotential}. 
Then, if the light ray of the given $k\ls{emi\,\mu}$ is not to be absorbed by the BH, we solve Eq.\eqref{eq:formalism.geodesic-hamilton} numerically. 
\end{itemize}
The details of these parts are explained in the following subsections. 

\subsection{Calculation of $k\ls{emi\,\mu}$ from $(\alpha\ls{emi},\beta\ls{emi})$}
\label{app:ic.1form}

Given the emission angles $(\alpha\ls{emi},\beta\ls{emi})$, the components of the initial vector in BL coordinates are
\eqb
\label{eq:ic.kemi.vector-1}
 k\ls{emi}^\mu =
 (\, k\ls{emi}^t \,,\, k\ls{sn}\cos\alpha\ls{emi} \,,\,
     k\ls{sn}\sin\alpha\ls{emi}\,\cos\beta\ls{emi} \,,\, k\ls{sn}\sin\alpha\ls{emi}\,\sin\beta\ls{emi} \,) \,,
\eqe
where $0\le\alpha\ls{emi}\le\pi$ and $0\le\beta\ls{emi} < 2\pi$. 
The spatial norm $k\ls{sn}$ and the $t$-component $k\ls{emi}^t$ are determined below.

In order to calculate $k\ls{sn}$ and $k\ls{emi}^t$, let us make use of the relation $k\ls{emi}^\mu = g^{\mu\nu} k\ls{emi\,\nu}$, where the components of the inverse of the metric $g^{\mu\nu}$ in BL coordinates are
\eqb
 g^{\mu\nu} =
 \left[
 \begin{array}{cccc}
 -Z/(\Sigma \Delta) & 0 & 0 & -2Mar/(\Sigma \Delta)
 \\
 & \Delta/\Sigma & 0 & 0
 \\
 && 1/\Sigma & 0
 \\
 \multicolumn{2}{c}{\rm sym.} && -4M^2a^2r^2/(Z \Sigma \Delta)
 \end{array}
 \right] \,.
\eqe
From these components together with $k\ls{emi\,\mu}$ in Eq.\eqref{eq:formalism.k-form}, we find
\eqb
\label{eq:ic.kemi.vector-2}
 k\ls{emi}^\mu = g\ls{emi}^{\mu\nu} k\ls{emi\,\nu} =
 \Bigl(\, -g\ls{emi}^{tt} + b\, g\ls{emi}^{t \varphi} \,,\,
          \dfrac{\Delta\ls{emi}}{\Sigma\ls{emi}}\, k\ls{emi\,{\it r}} \,,\,
          \dfrac{1}{\Sigma\ls{emi}}\, k\ls{emi\,\theta} \,,\,
          -g\ls{emi}^{t \varphi} + b\, g\ls{emi}^{\varphi \varphi}
 \,\Bigr) \,,
\eqe
where the subscript ``emi'' denotes the value at the emission event. 
Comparing the $\varphi$-component $k\ls{emi}^\varphi$ in Eqs.\eqref{eq:ic.kemi.vector-1} and \eqref{eq:ic.kemi.vector-2}, we find the relation,
\eqb
\label{eq:ic.b}
 b = \dfrac{g\ls{emi}^{t \varphi} + k\ls{sn} \sin\alpha\ls{emi}\,\sin\beta\ls{emi}}{g\ls{emi}^{\varphi \varphi}} \,.
\eqe
This relation determines the value of the toroidal impact parameter $b$ after obtaining the value of $k\ls{sn}$. 
Substituting \eqref{eq:ic.b} into the $t$-component $k\ls{emi}^t$ in Eq.\eqref{eq:ic.kemi.vector-2}, we find the relation,
\eqb
\label{eq:ic.kt}
 k\ls{emi}^t =
  - g\ls{emi}^{tt} +
  \dfrac{g\ls{emi}^{t \varphi}}{g\ls{emi}^{\varphi \varphi}}\,
  \bigl(\,g\ls{emi}^{t \varphi} + k\ls{sn} \sin\alpha\ls{emi}\,\sin\beta\ls{emi}\,\bigr)
 \,.
\eqe
By this relation, the unknown quantities $k\ls{emi}^t$ and $k\ls{sn}$ in Eq.\eqref{eq:ic.kemi.vector-1} are reduced to only one unknown quantity $k\ls{sn}$. 
Then, the spatial norm $k\ls{sn}$ is determined by the null condition,
\eqb
\label{eq:ic.nullcondition}
 g\ls{emi\,\mu\nu}\, k\ls{emi}^\mu\, k\ls{emi}^\nu = 0 \,.
\eqe
It must be noted that Eq.\eqref{eq:ic.nullcondition} is a 2nd order algebraic equation in $k\ls{sn}$, and the appropriate solution of $k\ls{sn}$ should satisfy $k\ls{emi}^t > 0$ (future-pointing vector).

From the above discussions, the initial 1-from, $k\ls{emi\,\mu} = (-1,k\ls{emi\,{\it r}},k\ls{emi\,\theta},b)$, is given by the following procedure in our numerical calculation:
\begin{description}
\item[Step 1 ($k\ls{emi\,\mu}$): ]
Specify the value of the emission angles $(\alpha\ls{emi} , \beta\ls{emi})$ as shown in Fig.\ref{fig:emission}. 
\item[Step 2 ($k\ls{emi\,\mu}$): ]
Solve the 2nd order algebraic equation \eqref{eq:ic.nullcondition} for $k\ls{sn}$, and adopt the solution satisfying $k\ls{emi}^t > 0$. 
\item[Step 3 ($k\ls{emi\,\mu}$): ]
Calculate the $\varphi$-component $k\ls{emi\,\varphi} = b$ using Eq.\eqref{eq:ic.b}. 
\item[Step 4 ($k\ls{emi\,\mu}$): ]
Calculate the $r$-component $k\ls{emi\,{\it r}}$ and the $\theta$-component $k\ls{emi\,\theta}$ using the following formulas, given by comparing Eq.\eqref{eq:ic.kemi.vector-1} and \eqref{eq:ic.kemi.vector-2}:
\eqb
\label{eq:ic.kemi.form}
 k\ls{emi\,{\it r}} = \dfrac{\Sigma\ls{emi}}{\Delta\ls{emi}}\,k\ls{sn}\cos\alpha\ls{emi}
 \quad,\quad
 k\ls{emi\,\theta} = \Sigma\ls{emi}\,k\ls{sn}\sin\alpha\ls{emi}\,\cos\beta\ls{emi} \,.
\eqe
\item[Step 5 ($k\ls{emi\,\mu}$): ]
Calculate the normic impact parameter $q$ from the $\theta$-component of Eq.\eqref{eq:formalism.geodesic}:
\eqb
\label{eq:ic.q}
 q = \sqrt{\bigl(k\ls{emi\,\theta}\bigr)^2 + \dfrac{Y(\theta\ls{emi})^2}{\sin^2\theta\ls{emi}}} \,.
\eqe
\end{description}
By these 5 steps, we can calculate the impact parameters $(b,q)$ and the components of the initial 1-form $k\ls{emi\,\mu}$ from given values of the emission angles $(\alpha\ls{emi},\beta\ls{emi})$.

Here, let us comment on Eq.\eqref{eq:ic.q}. 
It is recognized from the $\theta$-component of Eq.\eqref{eq:formalism.geodesic} that the light ray exists in the region where the zenithal effective potential \eqref{eq:formalism.effectivepotential} is non-positive, $U\ls{eff}(\theta) = -(k_\theta)^2 \le 0$. 
Therefore, the impact parameters $(b,q)$ included in the functional form of $U\ls{eff}$ have to satisfy the relation $U\ls{eff}(\theta\ls{emi};b,q) = - (k\ls{emi\,\theta})^2 \le 0$. 
This requirement is guaranteed by Eq.\eqref{eq:ic.q}, because Eq.\eqref{eq:ic.q} is derived from the relation $(k\ls{emi\,\theta})^2 + U\ls{eff}(\theta\ls{emi};b,q) = 0$.

\subsection{Selection rule of the initial direction of the light ray escaping to infinity}
\label{app:ic.selection}

Suppose that the values of the emission angles $(\alpha\ls{emi},\beta\ls{emi})$ are given, and the values of the corresponding impact parameters $(b,q)$ and initial 1-form $k\ls{emi\,\mu}$ are calculated as explained above. 
Then, the next procedure in our numerical calculation is to judge whether the light ray of the given $k\ls{emi\,\mu}$ is absorbed eventually by the BH. 
The criterion for this judgment, i.e., the \emph{selection rule for the initial direction of the light ray escaping to infinity}, is given by the radial effective potential $V\ls{eff}(r)$ defined in Eq.\eqref{eq:formalism.effectivepotential}.

\begin{figure}[t]
 \begin{center}
 \includegraphics[scale=0.60]{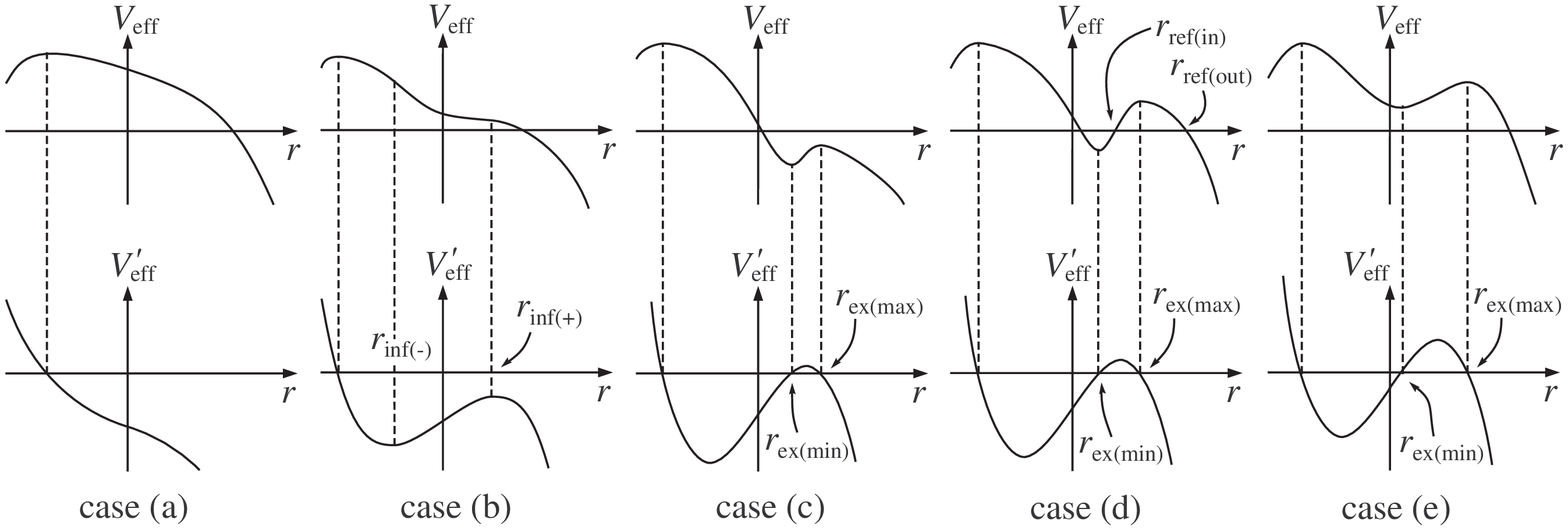}
 \end{center}
\caption{
Typical graphs of the radial effective potential $V\ls{eff}(r)$. 
These graphs are examples under the condition $V\ls{eff}(0) > 0$. 
Another condition, $V\ls{eff}(0) \le 0$, is also possible for cases (a) to (d).
The light ray and BH horizon exist in the region of non-positive potential $V\ls{eff}(r) \le 0$. 
}
\label{fig:Veff}
\end{figure}

Before constructing the selection rule, it is useful to summarize the functional form of $V\ls{eff}(r)$ and its derivatives:
\seqb
\begin{align}
 V\ls{eff}(r) &=
 -r^4 + [q^2 - 2a (a-b)] r^2 - 2M q^2 r + a^2 [q^2 - (a-b)^2]
 \\
 V\ls{eff}'(r) &=
 -4 r^3 + 2 [q^2 - 2a (a-b)] r - 2M q^2
 \\
 V\ls{eff}''(r) &=
 -12 r^2 + 2 [q^2 - 2a (a-b)] \,,
\end{align}
\seqe
where the prime denotes differentiation, $Q' = \diff{Q}/\diff{r}$. 
The facts we need here are that
\begin{itemize}
\item
by the relation $\Delta(r\ls{BH})=0$, we find $V\ls{eff}(r\ls{BH}) = - (2M r\ls{BH} - ab)^2 \le 0$;
\item
obviously, $V\ls{eff}'(0) \le 0$ holds;
\item
if $q^2 - 2a (a-b) \ge 0$ holds, then the inflection points of $V\ls{eff}(r)$ appear at $r = r\ls{inf(\pm)} \defeq \pm \sqrt{[q^2 - 2a (a-b)]/6}$;
\item
typical shapes of the graph of $V\ls{eff}(r)$ are illustrated in Fig.\ref{fig:Veff}, and classified as
\begin{description}
\item[case (a)\,] $q^2 - 2a (a-b) \le 0$.
\item[case (b)\,] $q^2 - 2a (a-b) > 0$ and $V\ls{eff}'(r\ls{inf(+)}) \le 0$.
\item[case (c)\,] $q^2 - 2a (a-b) > 0$, $V\ls{eff}'(r\ls{inf(+)}) > 0$, and $V\ls{eff}(r\ls{ex(max)}) < 0$, where $r\ls{ex(max)}$ is a local maximum point of $V\ls{eff}(r)$ satisfying $V\ls{eff}'(r\ls{ex(max)})=0$.
\item[case (d)\,] $q^2 - 2a (a-b) > 0$, $V\ls{eff}'(r\ls{inf(+)}) > 0$, $V\ls{eff}(r\ls{ex(max)}) \ge 0$, and $V\ls{eff}(r\ls{ex(min)}) \le 0$, where $r\ls{ex(min)}$ is a local minimum point of $V\ls{eff}(r)$ satisfying $V\ls{eff}'(r\ls{ex(min)})=0$.
\item[case (e)\,] $q^2 - 2a (a-b) > 0$, $V\ls{eff}'(r\ls{inf(+)}) > 0$ and $V\ls{eff}(r\ls{ex(min)}) > 0$.
\end{description}
Note that the shapes of the graphs illustrated in Fig.\ref{fig:Veff} are examples under the condition $V\ls{eff}(0) > 0$. 
The other condition, $V\ls{eff}(0) \le 0$, is also possible for cases (a) to (d).
\end{itemize}

Given the above properties of the effective potential $V\ls{eff}(r)$, we can construct a selection rule for the initial direction of the light ray escaping to infinity. 
There are 2 important points for the selection rule; (i) the light ray exists in the region of non-positive potential, $V\ls{eff}(r) \le 0$, because of the relation $V\ls{eff}(r) = - (\Sigma k^r)^2 \le 0$ given by the $r$-component of Eq.\eqref{eq:formalism.geodesic}, and (ii) the BH horizon is also in the region of non-positive potential as indicated by $V\ls{eff}(r\ls{BH}) \le 0$. 
Then, our selection rule consists of 2 parts as follows: 
\begin{description}
\item[Selection rule 1 for $k\ls{emi\,\mu}$: ]
If the potential $V\ls{eff}(r)$ corresponds to 1 of the cases (a), (b), (c) or (e) for the given values of the parameters $(b,q)$, then the light ray escaping to infinity is given by the condition, $k\ls{emi}^r > 0$. 
(The light ray of $k\ls{emi}^r \le 0$ will fall into the BH eventually.) 
Therefore, in this case, we solve Eq.\eqref{eq:formalism.geodesic-hamilton} numerically if $k\ls{emi}^r > 0$ holds. 
\item[Selection rule 2 for $k\ls{emi\,\mu}$: ]
If the potential $V\ls{eff}(r)$ corresponds to case (d) for the given values of the parameters $(b,q)$, then we need to check following 2 sub-rules. 
In these sub-rules, the zero-points of th potential, $V\ls{eff}(r)=0$, are denoted by $r = r\ls{ref(in)}$ and $r\ls{ref(out)}$ in increasing order as shown in Fig.\ref{fig:Veff}. 
(The radial coordinate of the light ray's position, $r\ls{ng}(\eta)$, is reflected by the potential at the zero-points.) 
\begin{description}
\item[Rule 2-1: ]
If the radius of the BH horizon is larger than the larger zero-point of the potential ($r\ls{ref(out)} < r\ls{BH}$), then the same statement with rule~1 is applied. 
That is, we solve Eq.\eqref{eq:formalism.geodesic-hamilton} numerically if $k\ls{emi}^r > 0$ holds.
\item[Rule 2-2: ]
If the radius of the BH horizon is smaller than the smaller zero-point of the potential ($r\ls{BH} \le r\ls{ref(in)}$), then the light ray escaping to infinity is given by the condition, $r\ls{ref(out)} \le r\ls{emi}$. 
(The light ray of $r\ls{BH} < r\ls{emi} \le r\ls{ref(in)}$ will fall into the BH eventually.) 
Therefore, in this case, we solve Eq.\eqref{eq:formalism.geodesic-hamilton} numerically if $r\ls{ref(out)} \le r\ls{emi}$ holds. 
\end{description}
\end{description}

The above procedures are our selection rule for the light ray escaping to infinity. 
In our numerical calculations, the initial 1-form $k\ls{emi\,\mu}$ given by the procedure of Sect.\ref{app:ic.1form} is filtered by this selection rule. 
Then, we carry out the numerical integration of Eq.\eqref{eq:formalism.geodesic-hamilton} only with those $k\ls{emi\,\mu}$ that pass our selection rule. 
Using this method of selecting appropriate $k\ls{emi\,\mu}$, we perform numerically a shooting search of the p-ray and the s-ray.

\section{Cross-sectional area of the null geodesic bundle}
\label{app:area}

This Appendix derives the detailed form of the area-transfer coefficient $\atc$ appearing in Eq.\eqref{eq:formalism.Aobs} by making use of the Jacobi equation \eqref{eq:formalism.jacobi}.

For the preparation for the derivation of $\atc$, we need to clarify some geometrical setup in the null geodesic bundle. 
Remember that the value of $Y\ls{obs\,;\alpha}^\mu$ in the ``initial'' condition \eqref{eq:formalism.jacobi-ic} is not uniquely determined. 
The different values of $Y\ls{obs\,;\alpha}^\mu$ correspond to the different Jacobi vector fields that connect the representative null geodesic to different neighboring null geodesics in the narrow null geodesic bundle. 
Then, in the neighborhood of the spacetime point $x\ls{ng}^\mu(\eta)$ on the representative null geodesic, the perpendicular condition of the Jacobi vector to the representative null geodesic ($Y^\mu k_\mu =0$) denotes that all Jacobi vectors at $x\ls{ng}^\mu(\eta)$ that are distinguished by the value of $Y\ls{obs\,;\alpha}^\mu$ compose a 3-dimensional region perpendicular to $k^\mu(\eta)$.
In such a 3-dimensional region in the neighborhood of $x\ls{ng}^\mu(\eta)$, let us introduce a \emph{reference-2D-surface} as a 2-dimensional \emph{spacelike} surface inside the 3-dimensional region. 
Obviously, there can be infinitely many reference-2D-surfaces in the neighborhood of $x\ls{ng}^\mu(\eta)$, since we have not specified the normal direction to the reference-2D-surface in the 3-dimensional region perpendicular to $k^\mu(\eta)$.

Once a reference-2D-surface is specified in the neighborhood of $x\ls{ng}^\mu(\eta)$, we can define the cross-sectional area of the null geodesic bundle measured on the reference-2D-surface at $x\ls{ng}^\mu(\eta)$ as the intersection area of the reference-2D-surface with the null geodesic bundle. 
Under this definition, the value of the cross-sectional area changes as the normal direction to the reference-2D-surface changes in the neighborhood of $x\ls{ng}^\mu(\eta)$. 
This change in the value of the cross-sectional area can be understood as the Lorentz transformation of the cross-sectional area in the neighborhood of $x\ls{ng}^\mu(\eta)$.

\begin{figure}[t]
 \begin{center}
 \includegraphics[scale=0.45]{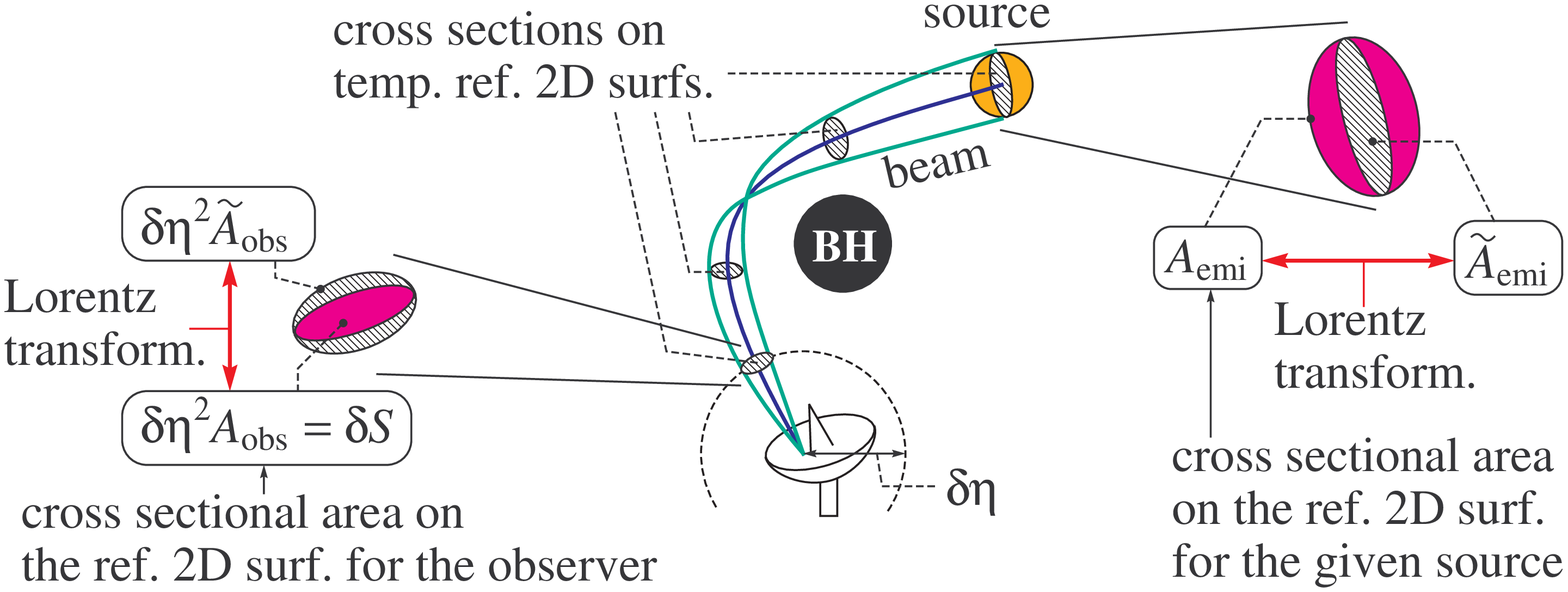}
 \end{center}
\caption{
The evolution of the cross-section is described by the Jacobi equations. 
The areas $\widetilde{A}\ls{obs}$ and $\widetilde{A}\ls{emi}$ are measured on the temporal reference-2D-surface. 
Our desired areas $A\ls{obs}$ and $A\ls{emi}$ are obtained by an appropriate Lorentz transformation of $\widetilde{A}\ls{obs}$ and $\widetilde{A}\ls{emi}$.
}
\label{fig:section}
\end{figure}

Given the definitions of the reference-2D-surface and the cross-sectional area of the null geodesic bundle, our derivation of the area-transfer coefficient $\atc$ consists of the following parts:
\begin{description}
\item[Part 1 of $\atc$: ]
We focus on the cross-sectional area of null geodesic bundle measured on a \emph{temporal} reference-2D-surface which is useful for our numerical calculation and defined exactly in Sect.\ref{app:area.evolution}. 
Then, by making use of the Jacobi equations, we calculate the relation between 2 cross-sectional areas: (i) the cross-sectional area $\delta\widetilde{S}$ on the celestial sphere around the observer, and (ii) the cross-sectional area $\widetilde{A}\ls{emi}$ of the source of light rays. 
Here the tilde, such as in $\widetilde{A}$, denotes the value evaluated on the temporal reference-2D-surface. 
\item[Part 2 of $\atc$: ]
We construct the Lorentz transformation from the temporal reference-2D-surface to the \emph{appropriate} reference-2D-surface which is defined exactly in Sect.\ref{app:area.transformation} so that it provides us with the value of $\delay$ and $\ratio$ that our telescope measures. 
Such a Lorentz transformation lets us obtain the detailed form of $\atc$. 
\end{description}
An illustrative summary of these parts is shown in Fig.\ref{fig:section}, and the details of these parts are explained in the following subsections.

Some items in this Appendix may already be well known to readers familiar with the application of general relativity to astrophysics and astronomy. 
However, we describe the detail of the derivation of Eq.\eqref{eq:formalism.Aobs} so that it becomes accessible by as many researchers as possible. 
Also, a detailed explanation of the formula \eqref{eq:formalism.Aobs} may be useful for future improvements of the numerical calculation of observable quantities $\delay$ and $\ratio$.

\subsection{Part 1 of $\atc$: Evolution of the cross-section in the temporal reference-2D-surface}
\label{app:area.evolution}

For part~1 of deriving $\atc$, suppose that the representative null geodesic $x\ls{ng}^\mu(\eta)$ is given. 
Then, in order to derive the details of the formula \eqref{eq:formalism.Aobs}, it is useful to rearrange the Jacobi equations \eqref{eq:formalism.jacobi} into some simultaneous ordinary differential equations by using appropriate tetrad components.

In our numerical calculation, we construct double null tetrad basis vectors on the representative null geodesic, $\{ k^\mu(\eta)$~, $l^\mu(\eta)$~, $e_{(1)}^\mu(\eta)$~, $e_{(2)}^\mu(\eta) \}$, which are parallel transported along the representative null geodesic,
\seqb
\label{eq:area.tetradbasis}
\eqb
\label{eq:area.tetradbasis-evolution}
 k^\alpha l^\mu_{\phantom{\mu};\alpha} = 0 \,,\,
 k^\alpha e_{(i);\alpha}^\mu = 0 \,,
\eqe
where there should hold the orthonormal condition,
\eqb
\label{eq:area.tetradbasis-orthonormal.null}
 l^\mu k_\mu = -1 \,,\, 
 l^\mu l_\mu = 0 \,,\,
 e_{(i)}^\mu e_{(j)\mu}^{\phantom{(i)}} = \delta_{(i) (j)} \,,\,
 e_{(i)}^\mu k_\mu = e_{(i)}^\mu l_\mu =  0 \,.
\eqe
\seqe
Here, the spacelike tetrad index is denoted with parentheses, $(i) = (1) , (2)$ and $(j) = (1) , (2)$. 
The vector $k^\mu$ is the given null vector tangent to the representative null geodesic, and the null vector $l^\mu$ and 2 spacelike vectors $e_{(i)}^\mu$ are the other basis vectors used in our numerical calculation. 
Our double null tetrad basis is constructed by solving Eq.\eqref{eq:area.tetradbasis-evolution} under the initial condition satisfying \eqref{eq:area.tetradbasis-orthonormal.null} at the emission event $x\ls{ng}^\mu(0) = x\ls{emi}^\mu$.\footnote{
When the initial condition satisfies the orthonormal condition \eqref{eq:area.tetradbasis-orthonormal.null}, the orthonormality of the tetrad basis is automatically preserved at all spacetime points on the representative null geodesic, because the inner product of any 2 tetrad basis vectors is invariant under parallel transport along the representative null geodesic due to Eq.\eqref{eq:area.tetradbasis-evolution}. 
} 
Note that the initial condition is not completely specified by the orthonormal condition \eqref{eq:area.tetradbasis-orthonormal.null} alone, because there are only 9 constraints in \eqref{eq:area.tetradbasis-orthonormal.null} for the 12 components of $l^\mu$ and $e_{(i)}^\mu$ at $x\ls{emi}^\mu$. 
The remaining 3 freedoms in the specification of the initial values correspond to the freedoms for choosing the reference-2D-surface at $x\ls{emi}^\mu$. 
In our numerical calculation, we adopt the following ansatz for the initial condition for simplicity:
\eqb
\label{eq:area.tetradbasis-ic.temporal}
\begin{aligned}
 \tilde{l}^\mu(\eta=0)
 &= (\tilde{l}\ls{emi}^t \,,\, \tilde{l}\ls{emi}^r \,,\,
     \tilde{l}\ls{emi}^\theta \,,\, \tilde{l}\ls{emi}^\varphi)
 \\
 \tilde{e}_{(1)}^\mu(\eta=0)
 &= (0 \,,\, \tilde{e}\ls{emi (1)}^r \,,\,
     \tilde{e}\ls{emi (1)}^\theta \,,\, \tilde{e}\ls{emi (1)}^\varphi)
 \\
 \tilde{e}_{(2)}^\mu(\eta=0)
 &= (0 \,,\, 0 \,,\, \tilde{e}\ls{emi (2)}^\theta \,,\, \tilde{e}\ls{emi (2)}^\varphi) \,,
\end{aligned}
\eqe
where the tilde, such as in $\tilde{e}$, denotes the value evaluated with this ansatz, and the non-zero components are determined by the orthonormal condition \eqref{eq:area.tetradbasis-orthonormal.null}.

Let the \emph{temporal tetrad basis}, $\{ k^\mu(\eta)$~, $\tilde{l}^\mu(\eta)$~, $\tilde{e}_{(1)}^\mu(\eta)$~, $\tilde{e}_{(2)}^\mu(\eta) \}$, denote the ones constructed from the ansatz \eqref{eq:area.tetradbasis-ic.temporal}. 
We assign the tilde, such as in $\widetilde{Q}$, to the value of quantity $Q$ when it is evaluated on the temporal tetrad basis. 
And, given the temporal tetrad basis, we can introduce the spacelike 2-dimensional surface spanned by the 2 spacelike vectors $\tilde{e}_{(i)}^\mu(\eta)$ in the neighborhood of $x\ls{ng}^\mu(\eta)$. 
We call this surface the \emph{temporal reference-2D-surface}. 
Note that every spacetime point $x\ls{ng}^\mu(\eta)$ $(0\le\eta\le\eta\ls{obs})$ on the representative null geodesic possesses one temporal reference-2D-surface in its neighborhood, and, due to Eq.\eqref{eq:area.tetradbasis-evolution}, all such temporal reference-2D-surfaces are generated by transporting parallel the temporal reference-2D-surface at $x\ls{emi}^\mu$ along the representative null geodesic. 
Therefore, the cross-sectional area of the null geodesic bundle $\delta\widetilde{S}$ ($= \delta\eta^2 \widetilde{A}\ls{obs}$ due to Eq.\eqref{eq:formalism.deltaS}\,) measured on the temporal reference-2D-surface at $x\ls{obs}^\mu$ and the cross-sectional area of the bundle $\widetilde{A}\ls{emi}$ measured on the temporal reference-2D-surface at $x\ls{emi}^\mu$ can be related by tracing the parallel transport of area $\widetilde{A}\ls{obs}\,(= \delta\widetilde{S}/\delta\eta^2)$ from $x\ls{obs}^\mu$ back to $x\ls{emi}^\mu$. 
Such a relation of areas can be calculated by making use of the Jacobi equations \eqref{eq:formalism.jacobi}.

In order to rearrange the Jacobi equations \eqref{eq:formalism.jacobi}, we decompose the Jacobi vector by the temporal tetrad basis,
$Y^\mu(\eta) = \widetilde{Y}^{(k)}(\eta)\, k^\mu(\eta) +
 \widetilde{Y}^{(l)}(\eta)\, \tilde{l}^\mu(\eta) +
 \widetilde{Y}^{(i)}(\eta)\, \tilde{e}_{(i)}^\mu(\eta)
$, 
where the tetrad components are the scalar quantities calculated from the orthonormal condition \eqref{eq:area.tetradbasis-orthonormal.null} as
\eqb
\label{eq:area.jacobivector-tetradcomponent}
 \widetilde{Y}^{(k)} = -\tilde{l}^\mu Y_\mu \quad,\quad
 \widetilde{Y}^{(l)} = -k^\mu Y_\mu \quad,\quad
 \widetilde{Y}^{(i)} = \tilde{e}_{(i)}^\mu Y_\mu \,.
\eqe
Here, the component $\widetilde{Y}^{(l)}$ is constant along the representative null geodesic as indicated by 
$\diff{\widetilde{Y}^{(l)}}/\diff{\eta} = - k^\lambda \bigl( k^\mu Y_\mu \bigr)_{;\lambda} = 0$, 
where the last equality is obtained from the geodesic equations $k^\lambda k^\mu_{\phantom{\mu};\lambda} =0$, the original form of the Jacobi equations ${\mathcal L}_k Y^\mu = 0 \,( \Leftrightarrow k^\lambda Y_{\phantom{\mu};\lambda}^\mu - Y^\lambda k_{\phantom{\mu};\lambda}^\mu = 0 )$ and the constancy of the norm $k^\mu k_\mu = \it{const}$. 
Hence, without loss of generality, we require the simplification,
\eqb
\label{eq:area.jacobivector-simplification}
 \widetilde{Y}^{(l)}(\eta) \equiv 0 \quad(\Leftrightarrow k^\mu Y_\mu \equiv 0) \,.
\eqe
Then, the decomposition of $Y^\mu$ by the temporal tetrad basis becomes
\eqb
\label{eq:area.jacobivector-decomposition}
 Y^\mu(\eta) =
 \widetilde{Y}^{(k)}(\eta)\, k^\mu(\eta) +
 \widetilde{Y}^{(i)}(\eta)\, \tilde{e}_{(i)}^\mu(\eta) \,.
\eqe
This decomposition guarantees $Y^\mu$ to be spacelike, since $Y^\mu Y_\mu = (\widetilde{Y}^{(1)})^2 + (\widetilde{Y}^{(2)})^2 \ge 0$, where the equality $Y^\mu Y_\mu = 0$ holds if and only if the spacelike components vanish, $Y^{(i)} = 0$.

Substituting the decomposition \eqref{eq:area.jacobivector-decomposition} into the Jacobi equations \eqref{eq:formalism.jacobi}, we obtain
\eqb
   \od{^2 \widetilde{Y}^{(k)}}{\eta^2} k^\mu
 + \od{^2 \widetilde{Y}^{(i)}}{\eta^2} e_{(i)}^\mu
 = - R^\mu_{\phantom{\mu}\alpha \lambda \beta} k^\alpha Y^\lambda k^\beta \,,
\eqe
where Eqs.\eqref{eq:area.tetradbasis} are used in deriving this expression. 
The inner products of this equation with $l^\mu$ and $e_{(i)}^\mu$ give, respectively, the following simultaneous ordinary differential equations,
\eqb
\label{eq:area.jacobi-tetradcomponent.k}
 \od{\widetilde{Y}^{(k)}}{\eta} = \widetilde{Z}^{(k)}
 \quad,\quad
 \od{\widetilde{Z}^{(k)}}{\eta} =
 \widetilde{R}^{(l)}_{\phantom{(l)}(k) (j) (k)} \widetilde{Y}^{(j)} \,,
\eqe
and
\eqb
\label{eq:area.jacobi-tetradcomponent.spacelike}
 \od{\widetilde{Y}^{(i)}}{\eta} = \widetilde{Z}^{(i)}
 \quad,\quad
 \od{\widetilde{Z}^{(i)}}{\eta} =
 - \widetilde{R}^{(i)}_{\phantom{(i)}(k) (j) (k)} \widetilde{Y}^{(j)} \,,
\eqe
where 
$\widetilde{R}^{(l)}_{\phantom{(l)}(k) (j) (k)} \defeq
 R^\lambda_{\phantom{\lambda}\alpha\mu\beta}l_\lambda k^\alpha \tilde{e}_{(j)}^\mu k^\beta$ 
and 
$\widetilde{R}^{(i)}_{\phantom{(i)}(k) (j) (k)} \defeq
 R^\lambda_{\phantom{\lambda}\alpha\mu\beta}
 \tilde{e}_{(i)\lambda} k^\alpha \tilde{e}_{(j)}^\mu k^\beta$ 
are the tetrad components of the Riemann tensor. 
Here, since the tetrad indices $(i)$ and $(j)$ take only the spacelike values $(1)$ and $(2)$, the latter \eqref{eq:area.jacobi-tetradcomponent.spacelike} is independent of the former \eqref{eq:area.jacobi-tetradcomponent.k}. 
Hence, in order to calculate the cross-sectional area on the temporal reference-2D-surface, we focus on the projection of the Jacobi vector \eqref{eq:area.jacobivector-decomposition} onto the temporal reference-2D-surface,
\eqb
\label{eq:area.jacobivector-projection}
 \widetilde{Y}^\mu(\eta) = \widetilde{Y}^{(i)}(\eta)\, \tilde{e}_{(i)}^\mu(\eta) \,,
\eqe
and this $\widetilde{Y}^\mu(\eta)$ is determined by the simultaneous equations \eqref{eq:area.jacobi-tetradcomponent.spacelike} without being affected by the remaining tetrad component $\widetilde{Y}^{(k)}$.

The ``initial'' condition of Eq.\eqref{eq:area.jacobi-tetradcomponent.spacelike} at $x\ls{obs}^\mu$ is read from condition \eqref{eq:formalism.jacobi-ic} as
\eqb
\label{eq:area.jacobi-tetradcomponent.ic}
 \widetilde{Y}^{(i)}(\eta\ls{obs}) = 0 \quad,\quad
 \widetilde{Z}^{(i)}(\eta\ls{obs}) \neq 0 \,,
\eqe
where the latter condition is obtained from the relation 
$\widetilde{Z}^{(i)} =
 k^\alpha (Y^\lambda \tilde{e}_{(i)\lambda} )_{;\alpha} = 
 \tilde{e}_{(i)\lambda} k^\alpha Y_{\phantom{\lambda};\alpha}^\lambda$. 
The latter condition, $\widetilde{Z}^{(i)}(\eta\ls{obs}) \neq 0$, guarantees that the area $\widetilde{A}\ls{obs}$ is non-zero, since $\widetilde{A}\ls{obs}$ is swept by the non-zero vector 
$\widetilde{Z}\ls{obs}^\mu \defeq
 k\ls{obs}^\alpha \widetilde{Y}\ls{obs\,;\alpha}^\mu =
 \widetilde{Z}^{(i)} \tilde{e}_{(i)}^\mu|_{\eta=\eta\ls{obs}} \neq 0$, 
as implied by Eq.\eqref{eq:formalism.deltaS}.

In order to calculate the area $\widetilde{A}\ls{obs}$ as it is swept by $\widetilde{Z}\ls{obs}^\mu$, we transform the Jacobi equations \eqref{eq:area.jacobi-tetradcomponent.spacelike} into the other form. 
The appropriate transformation of Eq.\eqref{eq:area.jacobi-tetradcomponent.spacelike} is given by the well-known theory of simultaneous 1st order ordinary differential equations. 
According to the theory, the general solution of the simultaneous equations \eqref{eq:area.jacobi-tetradcomponent.spacelike} under the condition \eqref{eq:area.jacobi-tetradcomponent.ic} is expressed as
\eqb
\label{eq:area.jacobi-tetradcomponent.solution}
 \widetilde{Y}^{(i)}(\eta) =
 \widetilde{J}^{(i)}_{\phantom{(i)}(j)}(\eta)\, \widetilde{Z}^{(j)}(\eta\ls{obs}) \,,
\eqe
where $\widetilde{J}^{(i)}_{\phantom{(i)}(j)}(\eta)$ is the \emph{Jacobi matrix}. 
The evolution equation of the Jacobi matrix along the given representative null geodesic is obtained by substituting Eq.\eqref{eq:area.jacobi-tetradcomponent.solution} into Eq.\eqref{eq:area.jacobi-tetradcomponent.spacelike},
\seqb
\label{eq:area.jacobi-matrix}
\eqb
\label{eq:area.jacobi-matrix.evolution}
 \od{^2\widetilde{J}^{(i)}_{\phantom{(i)}(j)}}{\eta^2} =
 - \widetilde{R}^{(i)}_{\phantom{(i)}(k) (m) (k)} \widetilde{J}^{(m)}_{\phantom{(m)}(j)} \,,
\eqe
and the ``initial'' condition is given by substituting Eq.\eqref{eq:area.jacobi-tetradcomponent.solution} into Eq.\eqref{eq:area.jacobi-tetradcomponent.ic},
\eqb
\label{eq:area.jacobi-matrix.ic}
 \widetilde{J}^{(i)}_{\phantom{(i)}(j)}(\eta\ls{obs}) = 0 \quad,\quad
 \od{\widetilde{J}^{(i)}_{\phantom{(i)}(j)}}{\eta}(\eta\ls{obs})
 = \delta^{(i)}_{\phantom{(i)}(j)} \,.
\eqe
\seqe
Once these equations are solved, we obtain the relation between the tetrad components of the Jacobi vector at $x\ls{obs}^\mu$ and those at $x\ls{emi}^\mu$,
\eqb
\label{eq:area.transformation-vector}
 \widetilde{Y}^{(i)}(0) = \widetilde{J}\ls{emi\it (j)}^{(i)} \widetilde{Z}^{(j)}(\eta\ls{obs}) \,,
\eqe
where $\widetilde{J}\ls{emi} = \widetilde{J}(\eta=0)$ is the Jacobi matrix at $x\ls{emi}^\mu$. 
Then, by supposing the value of $\widetilde{Z}^{(i)}$ is sufficiently small, we can understand that the matrix $\widetilde{J}\ls{emi}$ transforms the areal element on the temporal reference-2D-surface at $x\ls{obs}^\mu$ into that at $x\ls{emi}^\mu$. 
Here let us remember that the area $\widetilde{A}\ls{obs}$ is swept by the vector $\widetilde{Z}\ls{obs}^\mu = \widetilde{Z}^{(i)} \tilde{e}_{(i)}^\mu\bigr|_{\eta=\eta\ls{obs}}$ and that the area $\widetilde{A}\ls{emi}$ is swept by the vector $\widetilde{Y}^\mu(0) = \widetilde{Y}^{(i)} \tilde{e}_{(i)}^\mu\bigr|_{\eta=0}$. 
This fact, together with the relation \eqref{eq:area.transformation-vector}, implies the relation of areas
\eqb
\label{eq:area.transformation-area}
 \widetilde{A}\ls{emi} = (\det\widetilde{J}\ls{emi})\,\widetilde{A}\ls{obs} \,.
\eqe

It has to be emphasized here that, although the relation \eqref{eq:area.transformation-area} is derived by using the temporal double null tetrad basis given by condition \eqref{eq:area.tetradbasis-ic.temporal}, the value of the coefficient $\det\widetilde{J}\ls{emi}$ is invariant under the change of the condition \eqref{eq:area.tetradbasis-ic.temporal}. 
To show this invariance, let us consider the Lorentz transformation from the present tetrad basis $\{ k^\mu$~, $\tilde{l}^\mu$~, $\tilde{e}_{(i)}^\mu \}$ to the new tetrad basis $\{ k^\mu$~, $\hat{l}^\mu$~, $\hat{e}_{(i)}^\mu \}$, which is determined by Eq.\eqref{eq:area.tetradbasis} under an initial condition different from \eqref{eq:area.tetradbasis-ic.temporal}. 
The Lorentz transformation is expressed as
\eqb
\label{eq:area.transformation-tetradbasis}
 \hat{e}_{(a)}^\mu
 = \Lambda_{(a)}^{(b)} \tilde{e}_{(b)}^\mu \,,
\eqe
where $(a) , (b) = (k) , (l) , (1) , (2)$ and $\hat{e}_{(l)}^\mu = \hat{l}^\mu$, $\tilde{e}_{(l)}^\mu = \tilde{l}^\mu$, $\hat{e}_{(k)}^\mu = \tilde{e}_{(k)}^\mu = k^\mu$. 
This transformation \eqref{eq:area.transformation-tetradbasis} is equivalent to the change of condition \eqref{eq:area.tetradbasis-ic.temporal}. 
Since the transformation matrix $\Lambda_{(a)}^{(b)}$ does not depend on the affine parameter $\eta$, the transformation \eqref{eq:area.transformation-tetradbasis} holds at both $x\ls{emi}^\mu$ and $x\ls{obs}^\mu$. 
Therefore, the relation \eqref{eq:area.transformation-area} is transformed to the form
\eqb
\label{eq:area.transformation-area.hat.primitive}
 \widehat{A}\ls{emi}
 = (\det\widehat{J}\ls{emi})\,\widehat{A}\ls{obs}
 = (\det\Lambda\ls{(2D)}) (\det[\Lambda^{-1}]\ls{(2D)}) (\det\widetilde{J}\ls{emi})\,\widehat{A}\ls{obs} \,.
\eqe
Here, $\widehat{A}\ls{emi}$ is the area measured on the new reference-2D-surface spanned by $\{ \hat{e}_{(1)}^\mu$~, $\hat{e}_{(2)}^\mu \}$ at $x\ls{emi}^\mu$, and $\widehat{A}\ls{obs}$ is the parallel transport of $\widehat{A}\ls{emi}$ from $x\ls{emi}^\mu$ to $x\ls{obs}^\mu$. 
Also, $\widehat{J}\ls{emi}$ is the Jacobi matrix evaluated on the new reference-2D-surface. 
Further, $\Lambda\ls{(2D)} = \Lambda_{(i)}^{(j)}$ ($(i) , (j) = (1) , (2)$) is the $2\times 2$ part of the $4\times 4$ matrix $\Lambda_{(a)}^{(b)}$, and $[\Lambda^{-1}]\ls{(2D)} = [\Lambda^{-1}]_{(i)}^{(j)}$ is the $2\times 2$ part of the $4\times 4$ inverse matrix $[\Lambda^{-1}]_{(a)}^{(b)}$. 
Here note that the orthonormal conditions \eqref{eq:area.tetradbasis-orthonormal.null} for $\tilde{e}_{(a)}^\mu$ and $\hat{e}_{(a)}^\mu$ determine the values of some elements of the transformation matrix, $\Lambda_{(a)}^{(k)} = 1$ and $\Lambda_{(i)}^{(l)} = 0$, and these values of elements prove the relation, $\det[\Lambda^{-1}]\ls{(2D)} = (\det\Lambda\ls{(2D)})^{-1}$. 
Thus we find from \eqref{eq:area.transformation-area.hat.primitive},
\eqb
\label{eq:area.transformation-area.hat}
 \widehat{A}\ls{emi}
 = (\det\widehat{J}\ls{emi})\,\widehat{A}\ls{obs}
 = (\det\widetilde{J}\ls{emi})\,\widehat{A}\ls{obs} \,.
\eqe
This relation is used in the next subsection.

Here, let us make a comment on the \emph{caustic}. 
The caustic on the null geodesic bundle is the spacetime point $x\ls{ng}^\mu(\eta\ls{c})$ where $\det\widetilde{J}(\eta\ls{c}) = 0$ holds. 
Equation~\eqref{eq:area.jacobi-tetradcomponent.solution} indicates that the cross-sectional area vanishes at caustics. 
Further, by condition \eqref{eq:area.jacobi-tetradcomponent.ic}, the observation event $x\ls{obs}^\mu$ is interpreted as a caustic of the null geodesic bundle.

\subsection{Part 2 of $\atc$: Lorentz transformation to the observational value $A\ls{obs}$}
\label{app:area.transformation}

The area-transfer coefficient $\atc$ is given by formula \eqref{eq:formalism.Aobs}, which is similar to Eq.\eqref{eq:area.transformation-area}. 
Our desired area $A\ls{emi}$, which appears in Eq.\eqref{eq:formalism.Aobs}, is the cross-sectional area of the source measured on the appropriate reference-2D-surface that is perpendicular to $u\ls{emi}^\mu$ at $x\ls{emi}$. 
Another desired area $A\ls{obs}$, which appears in Eq.\eqref{eq:formalism.Aobs}, is the area measured on the appropriate reference-2D-surface perpendicular to $u\ls{obs}^\mu$ at $x\ls{obs}$. 
On the other hand, the areas $\widetilde{A}\ls{obs}$ and $\widetilde{A}\ls{emi}$ in Eq.\eqref{eq:area.transformation-area} are the areas evaluated on the temporal reference-2D-surface. 
Therefore, we need some appropriate Lorentz transformations in order to obtain the detailed form of Eq.\eqref{eq:formalism.Aobs}. 
Our procedure for calculating the appropriate Lorentz transformations consists of 3 subparts:
\begin{description}
\item[Part 2-1 of $\atc$: ]
For the first, at the emission event $x\ls{emi}^\mu$, determine the Lorentz transformation matrix \eqref{eq:area.transformation-tetradbasis} so that the temporal double null tetrad basis 
$\{ k\ls{emi}^\mu$~, $\tilde{l}\ls{emi}^\mu$~, $\tilde{e}\ls{{\it (i)} emi}^\mu \}$ 
is transformed to the new double null tetrad basis 
$\{ k\ls{emi}^\mu$~, $\hat{l}\ls{emi}^\mu$~, $\hat{e}\ls{{\it (i)} emi}^\mu \}$ 
and the new spacelike basis $\hat{e}\ls{{\it (i)} emi}^\mu$ is perpendicular to the source's velocity,
\eqb
\label{eq:area.e.hat}
 u\ls{emi\,\mu} \hat{e}\ls{{\it (i)} emi}^\mu = 0 \,.
\eqe
The cross-sectional area of the source $\widehat{A}\ls{emi}$ measured on the reference-2D-surface spanned by $\hat{e}\ls{{\it (i)} emi}^\mu$ is our desired area, $\widehat{A}\ls{emi} = A\ls{emi}$, which appears in the right-hand side of Eq.\eqref{eq:formalism.Aobs}. 
\item[Part 2-2 of $\atc$: ]
Next, at the observation event $x\ls{obs}^\mu$, transform the temporal double null basis $\{ k\ls{obs}^\mu$~, $\tilde{l}\ls{obs}^\mu$~, $\tilde{e}\ls{{\it (i)}obs}^\mu \}$ to the new basis $\{ k\ls{obs}^\mu$~, $\hat{l}\ls{obs}^\mu$~, $\hat{e}\ls{{\it (i)} obs}^\mu \}$ by applying the Lorentz transformation obtained in the previous step. 
This new basis $\{ k\ls{obs}^\mu$~, $\hat{l}\ls{obs}^\mu$~, $\hat{e}\ls{{\it (i)} obs}^\mu \}$ is the parallel transport of $\{ k\ls{emi}^\mu$~, $\hat{l}\ls{emi}^\mu$~, $\hat{e}\ls{{\it (i)} emi}^\mu \}$ along the representative null geodesic whose initial tangent vector is $k\ls{emi}^\mu$. 
The area $\widehat{A}\ls{obs}$ measured on the new reference-2D-surface at $x\ls{obs}^\mu$ spanned by $\hat{e}\ls{{\it (i)}obs}^\mu$ is given by Eq.\eqref{eq:area.transformation-area.hat}. 
\item[Part 2-3 of $\atc$: ]
Finally, calculate the Lorentz transformation of the area $\widehat{A}\ls{obs}$ to the desired area $A\ls{obs}$ at $x\ls{obs}^\mu$. 
Then, we will arrive at the detailed form of Eq.\eqref{eq:formalism.Aobs}, which is in Eq.\eqref{eq:area.Aobs}. 
\end{description}

Hereafter we carry out these 3 subparts. 
Part 2-1 of deriving $\atc$ is the calculation of the appropriate transformation matrix $\Lambda_{(a)}^{(b)}$ at $x\ls{emi}^\mu$. 
If the new double null tetrad basis $\{ k\ls{emi}^\mu$~, $\hat{l}\ls{emi}^\mu$~, $\hat{e}\ls{{\it (i)} emi}^\mu \}$ at $x\ls{emi}^\mu$ satisfying the condition \eqref{eq:area.e.hat} is given, then the transformation matrix is calculated by making use of the orthonormal condition \eqref{eq:area.tetradbasis-orthonormal.null} which holds for $\{ k\ls{emi}^\mu$~, $\tilde{l}\ls{emi}^\mu$~, $\tilde{e}\ls{{\it (i)} emi}^\mu \}$ and $\{ k\ls{emi}^\mu$~, $\hat{l}\ls{emi}^\mu$~, $\hat{e}\ls{{\it (i)} emi}^\mu \}$,
\eqb
\label{eq:area.Lambda}
 \Lambda_{(a)}^{(b)} =
 \left[
 \begin{array}{cccc}
 1 & 0 & 0 & 0
 \\
 - \hat{l}\ls{emi}^\mu \tilde{l}\ls{emi \,\mu} & 1 &
 \hat{l}\ls{emi}^\mu \tilde{e}\ls{(1) emi \,\mu} &
 \hat{l}\ls{emi}^\mu \tilde{e}\ls{(2) emi \,\mu}
 \\
 - \hat{e}\ls{(1) emi}^\mu \tilde{l}\ls{emi \,\mu} & 0 &
 \hat{e}\ls{(1) emi}^\mu \tilde{e}\ls{(1) emi \,\mu} &
 \hat{e}\ls{(1) emi}^\mu \tilde{e}\ls{(2) emi \,\mu}
 \\
 - \hat{e}\ls{(2) emi}^\mu \tilde{l}\ls{emi \,\mu} & 0 &
 \hat{e}\ls{(2) emi}^\mu \tilde{e}\ls{(1) emi \,\mu} &
 \hat{e}\ls{(2) emi}^\mu \tilde{e}\ls{(2) emi \,\mu}
 \end{array}
 \right] \,,
\eqe
where $(a)$ and $(b)$ are the indices distinguishing, respectively, a row and column of this matrix. 
Here, the tetrad basis $\{ k\ls{emi}^\mu$~, $\hat{l}\ls{emi}^\mu$~, $\hat{e}\ls{{\it (i)} emi}^\mu \}$ satisfying Eq.\eqref{eq:area.e.hat} is constructed from $k\ls{emi}^\mu$ and $u\ls{emi}^\mu$ as follows. 
Let the spacelike unit vector $p\ls{emi}^\mu$ be pointing in the propagation direction of the light ray in the 3-dimensional hypersurface perpendicular to $u\ls{emi}^\mu$, 
\seqb
\label{eq:area.tetradbasis-desired.emi}
\eqb
\label{eq:area.pemi}
 p\ls{emi}^\mu =
 \tdf\, (g\ls{emi}^{\mu\nu} + u\ls{emi}^\mu u\ls{emi}^\nu) \,k\ls{emi\, \nu}
 = \tdf k\ls{emi}^\mu - u\ls{emi}^\mu \,,
\eqe
where the total-Doppler factor $\tdf$ is required for the normalization, $p\ls{emi \,\mu} p\ls{emi}^\mu = 1$. 
Then, the desired double null tetrad basis $\{ k\ls{emi}^\mu$~, $\hat{l}\ls{emi}^\mu$~, $\hat{e}\ls{{\it (i)} emi}^\mu \}$, which must satisfy the orthonormal condition \eqref{eq:area.tetradbasis-orthonormal.null}, is given by
\eqb
\label{eq:area.tetradbasis-hatlemi}
 \hat{l}\ls{emi}^\mu = \dfrac{\tdf}{2} ( u\ls{emi}^\mu - p\ls{emi}^\mu ) \,,
\eqe
and $\hat{e}\ls{{\it (i)} emi}^\mu$ ($i = 1, 2$) whose components are determined so as to satisfy the orthonormal condition \eqref{eq:area.tetradbasis-orthonormal.null}. 
Note that the spacelike basis $\hat{e}\ls{{\it (i)} emi}^\mu$ satisfying condition \eqref{eq:area.tetradbasis-orthonormal.null} also automatically satisfies condition \eqref{eq:area.e.hat} due to the construction of $\hat{l}\ls{emi}^\mu$ in Eq.\eqref{eq:area.tetradbasis-hatlemi}. 
Furthermore, for numerical calculation, it should be noted that, given the null basis vectors $k\ls{emi}^\mu$ and $\hat{l}\ls{emi}^\mu$, condition \eqref{eq:area.tetradbasis-orthonormal.null} provides 7 constraints for the 8 components of the 2 vectors $\hat{e}\ls{{\it (i)} emi}^\mu$. 
The remaining 1 degree of freedom corresponds to the rotational degree of freedom of $\hat{e}\ls{{\it (i)} obs}^\mu$ in the 2-dimensional spacelike surface perpendicular to $k\ls{emi}^\mu$ and $\hat{l}\ls{emi}^\mu$. 
In our numerical calculation, we adopt the following ansatz for simplicity:
\eqb
\label{eq:area.hateemi-ansatz}
\begin{aligned}
 \hat{e}\ls{(1) emi}^\mu &=
 (\, \hat{e}\ls{(1) emi}^t \,,\, \hat{e}\ls{(1) emi}^r \,,\,
     \hat{e}\ls{(1) emi}^\theta \,,\, \hat{e}\ls{(1) emi}^\varphi \,)
\\
 \hat{e}\ls{(2) emi}^\mu &=
 (\, \hat{e}\ls{(2) emi}^t \,,\, \hat{e}\ls{(2) emi}^r \,,\,
     \hat{e}\ls{(2) emi}^\theta \,,\, 0 \,) \,.
\end{aligned}
\eqe
\seqe
Substituting this new tetrad basis $\{ k\ls{emi}^\mu$~, $\hat{l}\ls{emi}^\mu$~, $\hat{e}\ls{{\it (i)} emi}^\mu \}$ and the temporal tetrad basis $\{ k\ls{emi}^\mu$~, $\tilde{l}\ls{emi}^\mu$~, $\tilde{e}\ls{{\it (i)} emi}^\mu \}$ into Eq.\eqref{eq:area.Lambda}, the appropriate Lorentz transformation $\Lambda_{(a)}^{(b)}$ at $x\ls{emi}^\mu$ is obtained.

Part 2-2 of deriving $\atc$ is the calculation of the Lorentz transformation \eqref{eq:area.transformation-tetradbasis} at $x\ls{obs}^\mu$. 
Since the transformation matrix $\Lambda_{(a)}^{(b)}$ is already given in Eq.\eqref{eq:area.Lambda}, we obtain $\{ k\ls{obs}^\mu$~, $\hat{l}\ls{obs}^\mu$~, $\hat{e}\ls{{\it (i)} obs}^\mu \}$, which is the parallel transport of $\{ k\ls{emi}^\mu$~, $\hat{l}\ls{emi}^\mu$~, $\hat{e}\ls{{\it (i)} emi}^\mu \}$. 
Also, we find from \eqref{eq:area.transformation-area.hat},
\eqb
 A\ls{emi} = (\det\widetilde{J}\ls{emi})\,\widehat{A}\ls{obs} \,,
\eqe
where our desired area $A\ls{emi} = \widehat{A}\ls{emi}$ by the construction of $\Lambda_{(a)}^{(b)}$.

Part 2-3 of deriving $\atc$ is the calculation of the Lorentz transformation from the area $\widehat{A}\ls{obs}$ measured on the reference-2D-surface spanned by $\hat{e}\ls{{\it (i)} obs}^\mu$ to our desired area $A\ls{obs}$ measured on the appropriate reference-2D-surface that is perpendicular to $u\ls{obs}^\mu$. 
The rest part of this subsection is for this step.

For the preparation of this Lorentz transformation, we introduce the imaginary observer at $x\ls{obs}^\mu$ whose velocity $\hat{u}\ls{obs} (\neq u\ls{obs}^\mu)$ is perpendicular to $\hat{e}\ls{{\it (i)} obs}^\mu$, 
\seqb
\label{eq:area.tetradbasis-hat.obs}
\eqb
\label{eq:area.hatu}
 \hat{u}\ls{obs}^\mu \defeq \frac{1}{\sqrt{2}} ( k\ls{obs}^\mu + \hat{l}\ls{obs}^\mu ) \,.
\eqe
The unity of norm $\hat{u}\ls{obs \,\mu} \hat{u}\ls{obs}^\mu = -1$ holds by this definition. 
In the spacelike 3-dimensional hypersurface perpendicular to $\hat{u}\ls{obs}^\mu$, the imaginary observer of velocity $\hat{u}\ls{obs}^\mu$ recognizes that the beam of light rays moves in a direction along the spacelike vector $(g\ls{obs}^{\mu\nu} + \hat{u}\ls{obs}^\mu \hat{u}\ls{obs}^\nu) k\ls{obs \,\nu} = (k\ls{obs}^\mu - \hat{l}\ls{obs}^\mu)/2$. 
The unit vector pointing in this direction is
\eqb
\label{eq:area.hatp}
 \hat{p}\ls{obs}^\mu = \frac{1}{\sqrt{2}} ( k^\mu - \hat{l}^\mu ) \,.
\eqe
Then, the set of vectors $\{ \hat{u}\ls{obs}^\mu$~, $\hat{p}\ls{obs}^\mu$~, $\hat{e}\ls{{\it (i)} obs}^\mu \}$ can be a tetrad basis, because the vectors in this set satisfy the orthonormal condition due to \eqref{eq:area.tetradbasis-orthonormal.null},
\eqb
\begin{aligned}
 &\hat{u}\ls{obs \,\mu} \hat{p}\ls{obs}^\mu =
  \hat{u}\ls{obs \,\mu} \hat{e}\ls{{\it (i)} obs}^\mu =
  \hat{p}\ls{obs \,\mu} \hat{e}\ls{{\it (i)} obs}^\mu = 0
 \\
 &\hat{e}\ls{{\it(j)} obs\,\mu} \hat{e}\ls{{\it (i)} obs}^\mu = \delta_{(i)(j)} \quad,\quad
  -\hat{u}\ls{obs \,\mu} \hat{u}\ls{obs}^\mu =
  \hat{p}\ls{obs \,\mu} \hat{p}\ls{obs}^\mu = 1 \,.
\end{aligned}
\eqe
\seqe
This tetrad basis is equivalent to the double null tetrad basis $\{ k\ls{obs}^\mu$~, $\hat{l}\ls{obs}^\mu$~, $\hat{e}\ls{{\it (i)} obs}^\mu \}$ in the sense that both tetrad bases include the same spacelike basis vectors $\hat{e}\ls{{\it (i)} obs}^\mu$ that span the reference-2D-surface on which the area $\widehat{A}\ls{obs}$ is measured.

Remember that our desired area $A\ls{obs}$ is related to the cross-sectional area of the null geodesic bundle as $\delta S = \delta\eta^2 A\ls{obs}$ (see Eq.\eqref{eq:formalism.deltaS}\,), which are measured at the intersection event $x\ls{ng}^\mu(\eta\ls{obs}-\delta\eta)$ of the representative null geodesic with the celestial sphere of radius $\delta\eta$ (see Eq.\eqref{eq:formalism.deltar} and Fig.\ref{fig:deviation}). 
Also, the area $\widehat{A}\ls{obs}$ should satisfy the relation $\delta\widehat{S} = \delta\eta^2 \widehat{A}\ls{obs}$, where $\delta\widehat{S}$ is the cross-sectional area measured on the reference-2D-surface spanned by the parallel transform of $\hat{e}\ls{{\it (i)} obs}^\mu$ from $x\ls{obs}^\mu$ to $x\ls{ng}^\mu(\eta\ls{obs}-\delta\eta)$. 
Here note that, for a sufficiently small $\delta\eta$, any reference-2D-surface at $x\ls{ng}^\mu(\eta\ls{obs}-\delta\eta)$ can be regarded as the reference-2D-surface at $x\ls{obs}^\mu$.\footnote{
In the other words, for a sufficiently small $\delta\eta$, both the reference-2D-surface at $x\ls{ng}^\mu(\eta\ls{obs}-\delta\eta)$ and that at $x\ls{obs}^\mu$ are in the intersection region of the neighborhood of $x\ls{ng}^\mu(\eta\ls{obs}-\delta\eta)$ with that of $x\ls{obs}^\mu$.
}
Hence, the construction of the Lorentz transformation between $\delta S$ and $\delta\widehat{S}$ can be carried out at $x\ls{obs}^\mu$.

\begin{figure}[t]
 \begin{center}
 \includegraphics[scale=0.45]{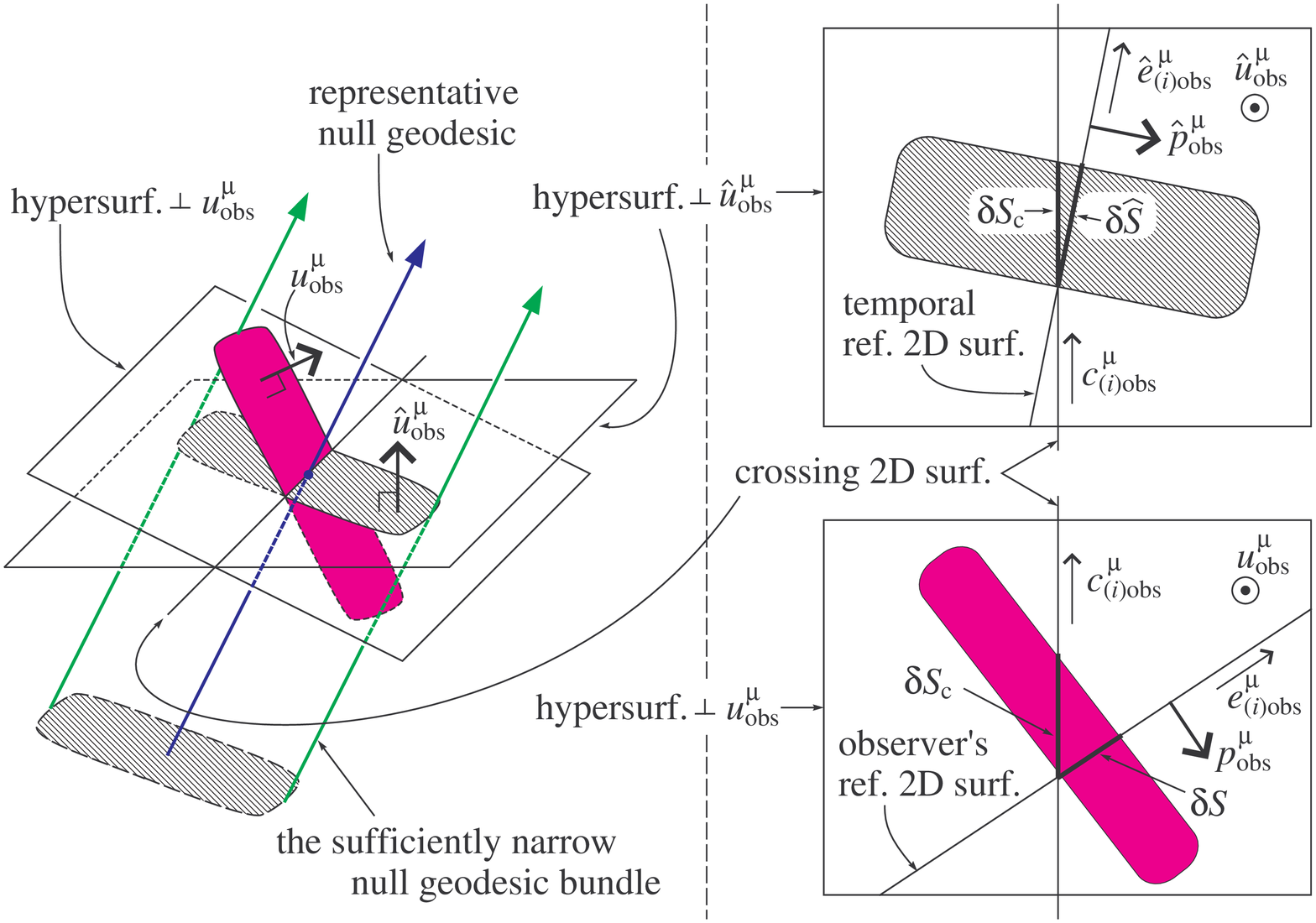}
 \end{center}
\caption{
Geometrical illustration of the relation between the cross-sectional areas of the narrow null geodesic bundle, $\delta\widehat{S}\, (= \delta\eta^2 \widehat{A}\ls{obs})$ and $\delta S\, (= \delta\eta^2 A\ls{obs})$. 
The null geodesic bundle is illustrated as a quadrangular prism, but only 2 null geodesics are illustrated on the side faces of the prism. 
}
\label{fig:area}
\end{figure}

Let $\{ u\ls{obs}^\mu$~, $p\ls{obs}^\mu$~, $e\ls{(1) obs}^\mu$~, $e\ls{(2) obs}^\mu \}$ denote the tetrad basis at $x\ls{obs}^\mu$ so that the the desired area $A\ls{obs}$ is measured on the 2-dimensional spacelike surface spanned by $e\ls{{\it (i)} obs}^\mu$. 
In this basis, $p\ls{obs}^\mu$ has to be the unit spacelike vector pointing in the propagation direction of the light rays in the 3-dimensional hypersurface perpendicular to $u\ls{obs}^\mu$,
\seqb
\label{eq:area.tetradbasis-desired.obs}
\eqb
\label{eq:area.pobs}
 p\ls{obs}^\mu \defeq (g\ls{obs}^{\mu\nu} + u\ls{obs}^\mu u\ls{obs}^\nu)\, k\ls{obs\,\nu}
 = k\ls{obs}^\mu - u\ls{obs}^\mu \,,
\eqe
where the normalization \eqref{eq:formalism.normalize} of the null vector $k^\mu$ and our observer's velocity \eqref{eq:formalism.uobs} are used in the second equality, and the unity of norm $p\ls{obs\,\mu} p\ls{obs}^\mu = 1$ holds. 
The other spacelike basis vectors $e\ls{{\it (i)} obs}^\mu$ are determined by the orthonormal condition,
\eqb
\label{eq:area.eobs}
 e\ls{{\it (i)} obs\,\mu}\, e\ls{{\it (j)} obs}^\mu = \delta_{(i) (j)}
 \quad,\quad
 e\ls{{\it (i)} obs\,\mu}\, u\ls{obs}^\mu =
 e\ls{{\it (i)} obs\,\mu}\, p\ls{obs}^\mu = 0 \,.
\eqe
This condition provides 7 constraints for the 8 components of the 2 vectors $e\ls{{\it (i)} obs}^\mu$. 
The remaining one degree of freedom corresponds to the rotational degree of freedom of $e\ls{{\it (i)} obs}^\mu$ in the 2-dimensional surface perpendicular to $u\ls{obs}^\mu$ and $p\ls{obs}^\mu$. 
In our numerical calculation, we adopt the following ansatz for simplicity,
\eqb
\label{eq:area.eobs-ansatz}
\begin{aligned}
 e\ls{(1) obs}^\mu &=
 (\, e\ls{(1) obs}^t \,,\, e\ls{(1) obs}^r \,,\,
     e\ls{(1) obs}^\theta \,,\, e\ls{(1) obs}^\varphi \,)
\\
 e\ls{(2) obs}^\mu &=
 (\, e\ls{(2) obs}^t \,,\, e\ls{(2) obs}^r \,,\,
     e\ls{(2) obs}^\theta \,,\, 0 \,) \,.
\end{aligned}
\eqe
\seqe
Note that the 2-dimensional spacelike surface spanned by $e\ls{{\it (i)} obs}^\mu$ is the appropriate reference-2D-surface on which our desired area $A\ls{obs}$ is measured. 
Also note that, if the other ansatz $e\ls{(2) obs}^\mu = (\, 0 \,,\, e\ls{(2) obs}^r \,,\, e\ls{(2) obs}^\theta \,,\, e\ls{(2) obs}^\varphi \,)$ was adopted, the condition $e\ls{(2) obs\,\mu}\, u\ls{obs}^\mu = 0$ becomes a trivial one $0=0$ under the setup of $u\ls{obs}^\mu$ in Eq.\eqref{eq:formalism.uobs}. 
We have to avoid such a trivial ansatz.

We have just constructed two 3-dimensional hypersurfaces in the neighborhood of $x\ls{obs}^\mu$.
One of them is perpendicular to $\hat{u}\ls{obs}^\mu$ and spanned by $\{ \hat{p}\ls{obs}^\mu$~, $\hat{e}\ls{{\it (i)} obs}^\mu \}$, and the other hypersurface is perpendicular to $u\ls{obs}^\mu$ and spanned by $\{ p\ls{obs}^\mu$~, $e\ls{{\it (i)} obs}^\mu \}$. 
The intersection of these 2 hypersurfaces is a 2-dimensional spacelike surface that we call the \emph{crossing-2D-surface}. 
Let $\{ c\ls{(1) obs}^\mu$~, $c\ls{(2) obs}^\mu \}$ denote the spacelike orthonormal basis that spans the crossing-2D-surface. 
These basis vectors are determined by
\seqb
\label{eq:area.cobs}
\eqb
 c\ls{{\it (i)} obs\,\mu}\, c\ls{{\it (j)} obs}^\mu = \delta_{(i) (j)}
 \quad,\quad
 c\ls{{\it (i)} obs\,\mu}\, u\ls{obs}^\mu =
 c\ls{{\it (i)} obs\,\mu}\, \hat{u}\ls{obs}^\mu = 0 \,.
\eqe
This condition provides 7 constraints for the 8 components of the 2 vectors $c\ls{{\it (i)} obs}^\mu$. 
The remaining 1 degree of freedom corresponds to the rotational degree of freedom of $c\ls{{\it (i)} obs}^\mu$ in the crossing-2D-surface. 
In our numerical calculation, we adopt the following ansatz for simplicity,
\eqb
\label{eq:area.cobs-ansatz}
\begin{aligned}
 c\ls{(1) obs}^\mu &=
 (\, c\ls{(1) obs}^t \,,\, c\ls{(1) obs}^r \,,\,
     c\ls{(1) obs}^\theta \,,\, c\ls{(1) obs}^\varphi \,)
\\
 c\ls{(2) obs}^\mu &=
 (\, c\ls{(2) obs}^t \,,\, c\ls{(2) obs}^r \,,\,
     c\ls{(2) obs}^\theta \,,\, 0 \,) \,.
\end{aligned}
\eqe
\seqe
The same notice as given for Eq.\eqref{eq:area.eobs-ansatz}, at the end of the previous paragraph, also holds here. 
The geometrical situation introduced so far is illustrated in Fig.\ref{fig:area}.

Given the above preparation, we find 2 different cross-sections of the null geodesic bundle in each of the 2 hypersurfaces, as indicated in Fig.\ref{fig:area}. 
In the hypersurface perpendicular to $u\ls{obs}^\mu$, 1 of the 2 cross-sectional areas is $\delta S$ measured on the reference-2D-surface spanned by $e\ls{{\it (i)} obs}^\mu$, and the other area is $\delta S\ls{c}$ measured on the crossing-2D-surface. 
Also, in the hypersurface perpendicular to $\hat{u}\ls{obs}^\mu$, the area $\delta \widehat{S}$ is measured on the reference-2D-surface spanned by $\hat{e}\ls{{\it (i)} obs}^\mu$, and the other area $\delta S\ls{c}$ is measured on the crossing-2D-surface. 
The area $\delta S\ls{c}$ is shared by the 2 hypersurfaces.

Here it should be noted that the shape of the beam of light rays appearing on the hypersurface in Fig.\ref{fig:area}, which is shown by the shaded or colored area, need not necessarily express the real shape determined by the real distribution of photons in the hypersurface. 
The point is that, in the hypersurface perpendicular to $u\ls{obs}^\mu$, the area $\delta S$ has to be the observable value of the cross-sectional area of the beam measured by the observer of velocity $u\ls{obs}^\mu$. 
And, because the other areas $\delta\widehat{S}$ and $\delta S\ls{c}$ are not the observable quantities in our setup, we can define those areas so that they give the observable value $\delta S$ through the formula \eqref{eq:area.Aobs} derived below. 
To do so, we consider the imaginary distribution of the null geodesic bundle that is illustrated as a quadrangular prism in Fig.\ref{fig:area}. 
As explained below, given such an imaginary distribution of null geodesics, the areas $\delta\widehat{S}$ and $\delta S\ls{c}$ will be determined automatically once the reference-2D-surface spanned by $\hat{e}\ls{{\it (i)} obs}^\mu$ and the crossing-2D-surface are specified.

\begin{figure}[t]
 \begin{center}
 \includegraphics[scale=0.45]{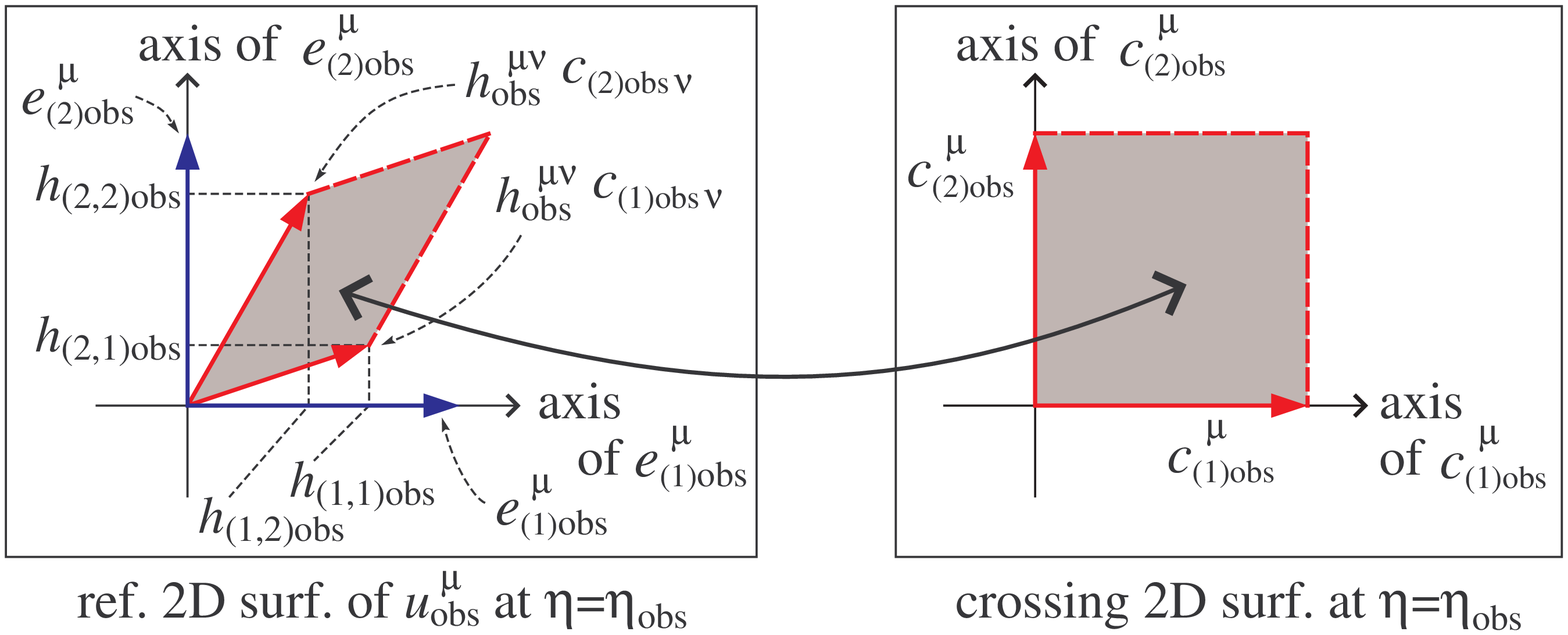}
 \end{center}
\caption{
Projection from the crossing-2D-surface spanned by $c\ls{{\it (i)} obs}^\mu$ into the reference-2D-surface spanned by $e\ls{{\it (i)} obs}^\mu$. 
}
\label{fig:projection}
\end{figure}

The relation between the areas $\delta S\ls{c}$ and $\delta S$ in the hypersurface perpendicular to $u\ls{obs}^\mu$ can be derived by considering the projection of the area $\delta S\ls{c}$, which is measured on the crossing-2D-surface, into the appropriate reference-2D-surface, which is spanned by $e\ls{{\it (i)} obs}^\mu$. 
The projection tensor $h\ls{obs}^{\mu\nu}$ into the appropriate reference-2D-surface is given by
$h\ls{obs}^{\mu\nu} \defeq
 g\ls{obs}^{\mu\nu} - p\ls{obs}^\mu p\ls{obs}^\nu + u\ls{obs}^\mu u\ls{obs}^\nu$. 
For any vector $v^\mu$, the components perpendicular to $p\ls{obs}^\mu$ and $u\ls{obs}^\mu$ are deleted by operating this tensor as $h\ls{obs}^{\mu\nu}v_\nu$. 
Thus, the projection of basis vectors $c\ls{{\it (i)} obs}^\mu$ into the appropriate reference-2D-surface is given by (see Fig.\ref{fig:projection})
\eqb
\label{eq:area.hcobs-primitive}
 hc\ls{{\it (i)} obs}^\mu \defeq h\ls{obs}^{\mu\nu} c\ls{{\it (i)} obs\,\nu}
 = c\ls{{\it (i)} obs}^\mu - (p\ls{obs}^\nu c\ls{{\it (i)}obs\,\nu}) p\ls{obs}^\mu \,,
\eqe
where condition \eqref{eq:area.cobs} is used in the last equality. 
Since this $hc\ls{{\it (i)} obs}^\mu$ is the vector on the appropriate reference-2D-surface, we can expand it as
\label{eq:area.hcobs}
\eqb
 hc\ls{{\it (i) }obs}^\mu = \sum_{j=1,2}h\ls{{\it (i,j)} obs} \, e\ls{{\it (j)} obs}^\mu \,,
\eqe
where the expansion coefficient is given by
\eqb
\label{eq:area.hobs}
 h\ls{{\it (i,j)} obs} = e\ls{{\it (i)} obs\,\mu}\, hc\ls{{\it (j) }obs}^\mu =
 e\ls{{\it (i)} obs\, \mu}\, c\ls{{\it (j)} obs}^\mu \,,
\eqe
where Eq.\eqref{eq:area.hcobs-primitive} and condition \eqref{eq:area.eobs} are used in the last equality. 
As indicated in Fig.\ref{fig:projection}, the parallelogram of area $\det h\ls{obs}$ on the appropriate reference-2D-surface spanned by $e\ls{{\it (i)} obs}^\mu$ corresponds to the unit square on the crossing-2D-surface, where $\det h\ls{obs}$ is the determinant of the matrix made of the expansion coefficients \eqref{eq:area.hobs}. 
This indicates the relation, $\delta S\ls{c}/\delta S = 1/\det h\ls{obs}$, which denotes
\eqb
\label{eq:area.deltaS-deltaSc}
 \delta S\ls{c} = (\det h\ls{obs})\,\delta S \,.
\eqe
Further, by replacing the reference-2D-surface spanned by $e\ls{{\it (i)} obs}^\mu$ with the reference-2D-surface spanned by $\hat{e}\ls{{\it (i)} obs}^\mu$ in the derivation of Eq.\eqref{eq:area.deltaS-deltaSc}, we can derive the relation between the areas $\delta S\ls{c}$ and $\delta\widehat{S}$ in the hypersurface perpendicular to $\hat{u}\ls{obs}^\mu$. 
The resultant relation is
\eqb
\label{eq:area.deltahatS-deltaSc}
 \delta S\ls{c} = (\det \hat{h}\ls{obs})\,\delta\widehat{S} \,,
\eqe
where $\det\hat{h}\ls{obs}$ is the determinant of the matrix,
\eqb
\label{eq:area.hathobs}
 \hat{h}\ls{{\it (i,j)} obs} = \hat{e}\ls{{\it (i)} obs\, \mu}\, c\ls{{\it (j)} obs}^\mu \,.
\eqe
Given the relations \eqref{eq:area.deltaS-deltaSc} and \eqref{eq:area.deltahatS-deltaSc}, we obtain the Lorentz transformation between the areas $A\ls{obs}\, (= \delta S/\delta\eta^2)$ and $\widehat{A}\ls{obs}\, (= \delta\widehat{S}/\delta\eta^2)$,
\eqb
\label{eq:area.lorentztransformation-Aobs}
 A\ls{obs} = \frac{\det \hat{h}\ls{obs}}{\det h\ls{obs}} \widehat{A}\ls{obs} \,.
\eqe

Finally, substituting the Lorentz transformation of area \eqref{eq:area.lorentztransformation-Aobs} into the relation \eqref{eq:area.transformation-area.hat}, we obtain the detailed form of formula \eqref{eq:formalism.Aobs},
\seqb
\label{eq:area.Aobs}
\eqb
 A\ls{obs} = \atc A\ls{emi} \,,
\eqe
where the area-transfer-coefficient is
\eqb
\label{eq:area.Aobs-C}
 \atc =
 \frac{\det \hat{h}\ls{obs}}{\det h\ls{obs}}
 \frac{1}{\det\widetilde{J}\ls{emi}} \,,
\eqe
\seqe
where the matrices $h\ls{{\it (i,j)} obs}$ and $\hat{h}\ls{{\it (i,j)} obs}$ are, respectively, in Eq.\eqref{eq:area.hobs} and \eqref{eq:area.hathobs}. 
We can recognize that $\atc$ is determined by $x\ls{obs}^\mu$, $u\ls{obs}^\mu$, $k\ls{obs}^\mu$, $x\ls{emi}^\mu$, $u\ls{emi}^\mu$, and $k\ls{emi}^\mu$.

\subsection{Steps for numerical calculation of $A\ls{obs}$}
\label{app:area.step}

In our numerical calculation after the solution of the null geodesic equations $x\ls{ng}^\mu(\eta)$ has been obtained, the procedure for calculating the area $A\ls{obs}$ is as follows:
\begin{description}
\item[Step 1 ($A\ls{obs}$): ]
Construct the double null temporal tetrad basis $\{ k^\mu(\eta) , \tilde{l}^\mu(\eta) , \tilde{e}_{(1)}^\mu(\eta) , \tilde{e}_{(2)}^\mu(\eta) \}$ along the given representative null geodesic by solving equations \eqref{eq:area.tetradbasis-evolution} under the initial condition \eqref{eq:area.tetradbasis-ic.temporal} whose non-zero values are determined by the algebraic equations \eqref{eq:area.tetradbasis-orthonormal.null}.
\item[Step 2 ($A\ls{obs}$): ]
Solve the evolution equations of the Jacobi matrix \eqref{eq:area.jacobi-matrix}. 
Then, we obtain the Jacobi matrix at the emission event $\widetilde{J}\ls{emi \it (j)}^{(i)}$ whose components are evaluated on the temporal reference-2D-surface.
\item[Step 3 ($A\ls{obs}$): ]
At the emission event $x\ls{emi}$, calculate the Lorentz transformation matrix $\Lambda_{(a)}^{(b)}$ by formula \eqref{eq:area.Lambda}, where the new basis $\{ k\ls{emi}^\mu$~, $\hat{l}\ls{emi}^\mu$~, $\hat{e}\ls{{\it (i)} emi}^\mu \}$ (\,$(i) = (1) , (2)$\,) is given by Eq.\eqref{eq:area.tetradbasis-desired.emi}. 
\item[Step 4 ($A\ls{obs}$): ]
At the observation event $x\ls{obs}^\mu$, construct the following 3 bases:
\begin{itemize}
\item
Construct the new tetrad basis $\{ k\ls{obs}^\mu$~, $\hat{l}\ls{obs}^\mu$~, $\hat{e}\ls{{\it (i)} obs}^\mu \}$ by the Lorentz transformation \eqref{eq:area.transformation-tetradbasis}, where the transformation matrix $\Lambda_{(a)}^{(b)}$ is obtained in the previous step. 
Then, reform this basis to $\{ \hat{u}\ls{obs}^\mu$~, $\hat{p}\ls{obs}^\mu$~, $\hat{e}\ls{{\it (i)} obs}^\mu \}$ using Eq.\eqref{eq:area.tetradbasis-hat.obs}. 
\item
Construct the appropriate tetrad basis $\{ u\ls{obs}^\mu$~, $p\ls{obs}^\mu$~, $e\ls{{\it (i)} obs}^\mu \}$ using Eq.\eqref{eq:area.tetradbasis-desired.obs}. 
\item
Given the above 2 bases, construct the basis vectors $\{ c\ls{(1) obs}^\mu$~, $c\ls{(2) obs}^\mu \}$ of the crossing-2D-surface using Eq.\eqref{eq:area.cobs}. 
\end{itemize}
\item[Step 5 ($A\ls{obs}$): ]
Given the above 3 bases, the Lorentz transformation \eqref{eq:area.lorentztransformation-Aobs} from $\widehat{A}\ls{obs}$ to $A\ls{obs}$ is calculated, where the matrices $h\ls{{\it (i,j)} obs}$ and $\hat{h}\ls{{\it (i,j)} obs}$ are given by Eqs.\eqref{eq:area.hobs} and \eqref{eq:area.hathobs}.
\item[Step 6 ($A\ls{obs}$): ]
Finally, collecting the results of steps 2 and 5, the relation \eqref{eq:area.Aobs} between $A\ls{obs}$ and $A\ls{emi}$ is calculated, where the value of $A\ls{emi}$ is assumed to satisfy condition \eqref{eq:formalism.assumption-Aemi} in our numerical calculation.
\end{description}



\begin{thebibliography}{}
%
\bibitem{ref:abbott+many.2016}
B.P. Abbott, et al. (LIGO Scientific Collaboration \& Virgo Collaboration), 
{\it Observation of Gravitational Waves from a Binary Black Hole Merger}, 
\PRL{116,061102,2016}
%
\bibitem{ref:boehle+13.2016}
A. Boehle, et al. (13 collaborators),
{\it An Improved Distance and Mass Estimate for Sgr A$^\ast$ from a Multistar Orbit Analysis},
\AJ{830,id.17(23pp.),2016}
%
\bibitem{ref:bower+5.2004}
G.C. Bower, H. Falcke, R.M. Herrnstein, J. Zhao, W.M. Goss \& D.C. Backer, 
{\it Detection of the Intrinsic Size of Sagittarius A$^\ast$ Through Closure Amplitude Imaging}, 
Science, {\bf 304}, 704-708 (2004)
%
\bibitem{ref:bozza+1.2004}
V. Bozza \& L. Mancini,
{\it Gravitational Lensing by Black Holes: a comprehensive treatment and the case of the star S2}, 
\AJ{611,1045-1053,2004}
%
\bibitem{ref:cardoso+4.2014}
V. Cardoso, L.C.B. Crispino, C.F.B. Macedo, H. Okawa, \& P. Pani, 
{\it Light rings as observational evidence for event horizons: Long-lived modes, ergoregions and nonlinear instabilities of ultracompact objects}, 
arXiv:1406.5510 [gr-qc]
%
\bibitem{ref:cunningham+1.1973}
C.T. Cunningham \& J.M. Bardeen, 
{\it The Optical Appearance of a Star Orbiting an Extreme Kerr Black Hole}, 
\AJ{183,237-264,1973}
%
\bibitem{ref:doeleman+27.2008}
S.S. Doeleman, et al.(27 collaborators), 
{\it Event-horizon-scale structure in the supermassive black hole candidate at the Galactic Center},
Nature {\bf 455}, 78 (2008).
%
\bibitem{ref:fanton+3.1997}
C. Fanton, M. Calvani, F. deDelice \& A. \`{C}ade\`{z}, 
{\it Detecting Accretion Disks in Active Galactic Nuclei}, 
Publ. Astrophys. Soc. Japan, {\bf 49}, 159-169 (1997).
%
\bibitem{ref:frolov+1.1998}
V.P. Frolov \& I.D. Novikov, 
{\it Black Hole Physics}, 
Kluwer Academic Publishers (Dordrecht, Netherland), 1998, chap.6
%
\bibitem{ref:fukumura+1.2008}
K. Fukumura \& D. Kazanas, 
{\it Light Echoes in Kerr Geometry: A Source of High-Frequency QPOs from Random X-Ray Bursts}, 
\AJ{679,1413-1421,2008}
%
\bibitem{ref:fukumura+2.2009}
K. Fukumura, D. Kazanas \& G. Stephenson, 
{\it Quasi-Periodic Oscillations from Random X-Ray Bursts around Rotating Black Holes}, 
\AJ{695,1199-1209,2009}
%
\bibitem{ref:hawking+1.1973}
S.W. Hawking \& G.F.R. Ellis, 
{\it The large scale structure of space-time}, 
Cambridge Univ. Press (1973), chap.4
%
\bibitem{ref:holz+1.2002}
D.E. Holz \& J.A. Wheeler, 
{\it Retro-Machos: $\pi$ in the sky?}, 
\AJ{578,330-334,2002}
%
\bibitem{ref:horak+1.2006}
J. Ho\'{r}ak \& V. Karas, 
{\it On the role of strong gravity in polarization from scattering of light in relativistic flows},
Mon. Not. Roy. Aastron. Soc. {\bf 365}, 813-826 (2006).
%
\bibitem{ref:james+3.2015}
O. James, E. von Tunzelmann, P. Franklin \& K.S. Thorne, 
{\it Gravitational Lensing by Spinning Black Holes in Astrophysics, and in the Movie Interstellar}, 
Class. Quant. Grav, {\bf 32}, 065001 (2015).
%
\bibitem{ref:kojima.1991}
Y. Kojima, 
{\it The Effects of Black Hole Rotation on Line Profiles from Accretion Discs}, 
Mon. Not. Roy. Aastron. Soc. {\bf 250}, 629-632 (1991).
%
\bibitem{ref:luminet.1979}
J.P. Luminet, 
{\it Image of a Spherical Black Hole with Thin Accretion Disk}, 
Astron. Astrophys. {\bf 75}, 228-235 (1979).
%
\bibitem{ref:misner+2.1973}
C.W. Misner, K.S. Thorne \& J.A. Wheeler, 
{\it Gravitation}, 
W.H. Freeman and Company (New York, USA), 1973, p.588
%
\bibitem{ref:oka+3.2016}
T. Oka, R. Mizuno, K. Miura \& S. Takekawa,
{\it Signature of an Intermediate-Mass Black Hole in the Central Molecular Zone of Our Galaxy},
\AJ{816,L7,2016}
%
\bibitem{ref:saida+3.2016}
H. Saida, A. Fujisawa, C. Yoo \& Y. Nambu,
{\it Spherical polytropic balls cannot mimic black holes}, 
Prog. Theor. Exp. Phys., {\bf 2016}, 043E02 (2016).
%
\bibitem{ref:shen+4.2005}
Z.-Q. Shen, K.Y. Lo, M.-C. Liang, P.T.P. Ho \& J.-H. Zhao,
{\it A size of $\sim 1$ AU for the radio source Sgr A$^\ast$ at the center of the Milky Way},
Nature, {\bf 438}, 62 (2005)
%
\bibitem{ref:takahashi.2004}
R. Takahashi, 
{\it Shapes and Positions of Black Hole Shadows in Accretion Disks and Spin Parameters of Black Holes}, 
\AJ{611,996-1004,2004}
%
\bibitem{ref:thompson+2.2004}
A.R. Thompson, J.M. Moran \& G.W. Swenson Jr.,
{\it Interferometry and Synthesis in Radio Astronomy (2nd edition)}, 
WILEY-VCH Verlag GmbH \& Co.KGaA (Weinheim, Germany), 2004, Sect.6.2
%
\end{thebibliography}
\end{document}